\documentclass[journal]{IEEEtran}
\usepackage{cite}
\ifCLASSINFOpdf
  \usepackage[pdftex]{graphicx}
  \graphicspath{{../pdf/}{../jpeg/}}
   \DeclareGraphicsExtensions{.pdf,.jpeg,.png}
\else
  \usepackage[dvips]{graphicx}
  \graphicspath{{../eps/}}
   \DeclareGraphicsExtensions{.eps}
\fi
\usepackage[cmex10]{amsmath}
\usepackage{array}
\ifCLASSOPTIONcompsoc
  \usepackage[caption=false,font=normalsize,labelfont=sf,textfont=sf]{subfig}
\else
 \usepackage[caption=false,font=footnotesize]{subfig}
\fi
\usepackage{amsmath,graphicx}
\usepackage{amssymb}
\usepackage{amscd}
\usepackage{algorithm}
\usepackage[noend]{algpseudocode}
\usepackage{epstopdf}

\usepackage{float}

\begin{document}

\title{iMAP Beamforming for High Quality High Frame Rate Imaging}

\author{\IEEEauthorblockN{Tanya Chernyakova~\IEEEmembership{Student Member,~IEEE},
Dan Cohen,
Meged Shoham, and
Yonina C. Eldar~\IEEEmembership{Fellow,~IEEE}\\
\thanks{\textsuperscript{*}~This project has received funding from the European Union's Horizon 2020 research and innovation program under grant agreement No. 646804-ERC-COG-BNYQ.}}
}

\maketitle

\begin{abstract}
We present a statistical interpretation of beamforming to overcome limitations of standard delay-and-sum (DAS) processing. Both the interference and the signal of interest are viewed as random variables and the distribution of the signal of interest is exploited to maximize the a-posteriori distribution of  the aperture signals.
In this formulation the beamformer output is a maximum-a-posteriori (MAP) estimator of the signal of interest. 
We provide a closed form expression for the MAP beamformer and estimate the unknown distribution parameters from the available aperture data using an empirical Bayes approach. We propose a simple scheme that iterates between estimation of distribution parameters and computation of the MAP estimator of the signal of interest, leading to an iterative MAP (iMAP) beamformer. 
This results in a significant improvement of the contrast compared to DAS without severe increase in computational complexity or need for fine-tuning of parameters. By implementing iMAP on both simulated and experimental data, we show that only 13 transmissions are required to obtain contrast comparable to DAS with 75 plane-waves. The proposed method is compared to other interference suppression techniques such as coherence factor and scaled Wiener processing and shows improved contrast and better preserved speckle pattern.
\end{abstract}

\IEEEpeerreviewmaketitle

\setkeys{Gin}{draft=false}

\section{Introduction}
\label{sec:intro}
B-mode ultrasound imaging can be performed using numerous insonification strategies varying from focused beams to plane-wave and diverging-wave transmissions.
Narrow transmissions improve signal-to-noise ratio (SNR) and spatial resolution, while broad unfocused insonification accelerates image acquisition, leading to ultrafast imaging methods \cite{tanter2014ultrafast}.
The signals detected by the transducer elements after transmission include reflections from the entire insonified area.
To determine the value of a b-mode image at a certain point, focusing delays 
are applied on the signals in attempt to isolate an on-axis echo originating from this point. 
In the most common delay-and-sum (DAS) beamforming, the delayed signals are simply averaged to yield the signal of interest. 
Due to finite aperture size, the off-axis echoes are not entirely suppressed by averaging. As e result, DAS beampattern is characterized by the main lobe width and levels of side-lobes.
Another source of clutter is phase aberrations originating from sound speed inhomogeneities. 
Together with reverberation echoes, produced by impedance mismatches, they further reduce the contrast of a DAS image. These limitations are especially prominent for wide transmissions, where due to lack of transmit focus, clutter originates from a much wider insonified area.
Several approaches have been proposed to overcome these  limitations of DAS.

\subsection{Related Work} \label{ssec:related work}
The side-lobe level of a DAS beamformer can be controlled by weighting the aperture elements. When constant, i.e. signal independent, weights are applied, this process is referred to as apodization and reduces side-lobes at the expense of a wider main lobe, degrading the lateral resolution. 
This trade-off can be avoided by using adaptive weights computed based on the detected signals. 

Minimum variance (MV) beamforming \cite{capon1969high} improves the resolution without sacrificing contrast by allowing higher side-lobes in the directions with no received energy. 
 Its application to ultrasound imaging was studied extensively  over the last two decades \cite{mann2002constrained,sasso2005medical,synnevag2007adaptive,vignon2008capon,austeng2011coherent,nguyen2017minimum}. 
The need to estimate and invert the data covariance matrix per each image point makes MV beamforming computationally intensive and creates a bottleneck for real time implementation.

Another approach for clutter suppression assumes that the signal originating at the focal point is highly coherent over the aperture elements while the interference is not. Based on this assumption a coherence factor (CF) \cite{mallart1994adaptive,hollman1999coherence}  is calculated as 
a ratio of coherent to incoherent sums across the signals detected by the aperture to multiply the beamformer output \cite{jeong2000fourier,li2003adaptive,wang2007performance,asl2009minimum,xu2012adaptive,wang2009mvdr}. 
Despite the advantages of CF combined with either DAS or MV beamforming, the resulting images may suffer from reduced image brightness and degraded speckle pattern, 
especially in low SNR. A detailed study of this phenomenon is presented in \cite{nilsen2010wiener}, where CF is analyzed in the context of Wiener beamforming  \cite{van2004detection} and its calculation is interpreted in terms of estimation of signal and noise power. Several alternatives including a scaled Wiener postfilter are proposed by Nilsen and Holm in \cite{nilsen2010wiener} to provide a trade-off between contrast improvement and robustness to low SNR, and are studied for focused acquisition. 

Methods based on unfocused insonification, such as plane-wave imaging, can potentially highly benefit from CF processing to improve contrast.
In this mode, the quality of an image is inherently reduced due to low SNR and  the lack of transmit focus. The latter implies that the clutter originates from a much wider area and severely reduces the contrast of the image.  
By coherent compounding of images obtained by transmitting several tilted plane-waves, the image quality is improved and is proportional to the number of transmissions \cite{montaldo2009coherent}. The latter should be kept as low as possible to maintain high frame rate and reduce the number of computations leading to a trade-off between image quality and frame rate.

The application of CF processing for broad insonification is challenging due to decreased accuracy in coherent sum estimation \cite{wang2009mvdr}. Several works propose to apply CF based methods coupled with variations of MV  beamforming  to overcome this problem \cite{wang2009mvdr,zhao2016subarray}. However, the implementation of MV processing for ultrafast imaging is prohibitive computationally.  

\subsection{Contributions} \label{ssec:contributions}
In this work we present a statistical interpretation of beamforming leading to an alternative framework for processing the detected signals, applicable for both focused and broad transmissions. 
In the proposed approach, both the interference and the signal of interest are viewed as random variables. This allows exploiting the distribution of the signal of interest in order to maximize the a-posteriori distribution of  signals detected by the aperture.
In this formulation the beamformer output is a maximum-a-posteriori (MAP) estimator of the signal of interest. 
We provide a closed form expression for the MAP beamformer which depends on the signal of interest and the interference distribution parameters. Since the latter are unknown, we  estimate them from the available aperture data using an empirical Bayes approach \cite{carlin2000bayes}. We propose a simple scheme that iterates between estimation of distribution parameters using maximum likelihood (ML)  and computation of the MAP estimator, leading to an iterative MAP \mbox{(iMAP)} beamformer.

To put the proposed method in context, we review Wiener beamforming and Wiener postfilter,  CF and scaled Wiener (ScW) processing and the relationship between them \cite{nilsen2010wiener,li2003adaptive,wang2007performance}. We also discuss different strategies for statistics estimation used by these methods and compare them to the ML approach used by iMAP.  

We consider simulated, experimental and \textit{in-vivo} datasets provided on the PICMUS website \cite{liebgott2016plane} to show that iMAP beamforming with one and two iterations, referred to as iMAP1 and iMAP2, provides significant improvement in contrast compared to DAS without affecting the resolution for any number of transmissions. In addition, iMAP1 requires only 13 plane-waves to obtain contrast comparable to that of DAS with 75 transmissions. 
We also compare iMAP to Wiener postfilter, CF and ScW with statistics estimation suggested in \cite{nilsen2010wiener}. The results show that the proposed method provides improved contrast for simulated, experimental and \textit{in-vivo} datasets. 
Finally, the effect of iMAP and other interference suppression methods on speckle are assessed by evaluating the statistics of the resulting images. Our tests show the improved preservation of speckle patterns by our approach.

The rest of the paper is organized as follows. We introduce the signal model and present iMAP beamforming in Sections \ref{sec:signal model} and \ref{sec:MAP}, respectively. Section \ref{sec:related methods} reviews CF, Wiener and ScW processing and their connection to the proposed method. The experimental setup and results are presented in Sections \ref{sec:experiments} and \ref{sec:results}. Discussion and conclusions appear in Section \ref{sec:discussion}.

\section{Signal Model and Standard Processing }
\label{sec:signal model}
We consider a linear array of $M$ elements.
The delays are applied to the signals received by the elements after transmission, either focused or broad, to focus the array to a point of interest $(r, \theta)$ upon reception. These delayed signals are referred to as aperture data. The analysis is performed for each given image point, therefore, the indexes $r$ and $\theta$ are omitted in the following derivations.
The delayed signal at the $m$-th element is 
\begin{equation} \label{eq:y_m}
y_m = x+n_m,
\end{equation}
where $x$ is a signal of interest which represents the reflectivity of the underlying image at the focal point $(r, \theta)$ and $n_m$ stands for additive interference including the off-axis reflections, aberration and reverberation clutter and system noise. 
In vector notation \eqref{eq:y_m} becomes
\begin{equation} \label{eq:y vector}
\mathbf{y} = x\boldsymbol{1} + \mathbf{n}.
\end{equation}
Here the vector $\mathbf{y}$ corresponds to the aperture data,  $\boldsymbol{1} $ is a vector of ones and $\mathbf{n}$ is the interference vector. All the vectors are of length $M$.   

The two most common beamforming techniques for array processing are DAS and MV.
Standard DAS beamforming maximizes the output energy in the direction of interest, namely the foresight.
It reduces to recovering the signal of interest by averaging the aperture data, i.e.
\begin{equation} \label{eq:x DAS}
x_{\text{DAS}}  = \frac{1}{M} \sum_{m=1}^M y_m = \frac{1}{M} \boldsymbol{1}^H \mathbf{y} .
\end{equation}
MV beamforming \cite{capon1969high} maintains unity gain in the foresight, while minimizing the energy received from other directions:
\begin{align} \label{eq:MVproblem}
\min_{\boldsymbol{w}}{E\left[\left | \boldsymbol{w}^H\mathbf{y}\right |^2\right ]} ~~~\text{s.t.}~~~\boldsymbol{w}^H\boldsymbol{1}=1.
\end{align}
Assuming the data vector $\mathbf{y}$ has a correlation matrix \mbox{$ \mathbf{R}_{y}=E[\mathbf{y} \mathbf{y} ^H]$}, the optimal weights are given by
\begin{align} \label{eq:MV beamformer}
 \boldsymbol{w}_{\text{MV}}  = \frac{\mathbf{R}_{y}^{-1}\boldsymbol{1}}{\boldsymbol{1}^H \mathbf{R}_{y}^{-1} \boldsymbol{1} }.
\end{align}
%
Implementation of the MV beamformer requires estimation and inversion of the data correlation matrix per each image point \cite{synnevag2007adaptive,vignon2008capon}. As a result, computational complexity is increased and is often prohibitive for real time implementation. 
In addition, image quality is sensitive to estimation errors so that strategies such as spatial smoothing and diagonal loading are required to improve robustness. This introduces parameters that have to be tuned for every specific application.

The formulation in \eqref{eq:y vector} implies that the aperture data can be viewed as $M$ noisy measurements of the signal of interest $x$. We assume that the interference at the elements is spatially white, namely, i.i.d.  Gaussian. Explicitly, 
\begin{equation} \label{eq:noise dist}
\mathbf{n} \sim N(\mathbf{0},\sigma_n^2\mathbf{I}),
\end{equation}
where $\mathbf{I}$ is the $M\times M$ identity matrix and $\sigma_n^2$ is the noise variance.
Further assumptions on $x$ imply the processing that should be applied on the measurement vector $\mathbf{y}$ to recover the signal of interest. 

Suppose first that $x$ is a deterministic unknown parameter.  The ML estimate of $x$ is then given by
\begin{align}\label{eq:x ML}
x_{\text{ML} }=&\arg\max_x{p(\mathbf{y};x)} \\ \nonumber
               =&\frac{1}{M} \boldsymbol{1}^H \mathbf{y}  = \frac{1}{M} \sum_{m=1}^M y_m,
\end{align}
where $(\cdot)^H$ stands for the Hermitian operator.  Thus, under the assumption of uncorrelated Gaussian noise at the aperture elements,
the ML estimator corresponds to the output of standard DAS beamforming given in \eqref{eq:x DAS}.

To allow for more involved processing we consider $x$ to be a random variable and use MAP estimation to recover it from the noisy measurements. 

\section{MAP beamforming}
\label{sec:MAP}
\subsection{MAP estimator}
\label{ssec:MAP estimation}
Suppose next that the signal of interest is a Gaussian random variable with variance $\sigma_x^2$,  $ x\sim N(0,\sigma_x^2)$, uncorrelated with the noise. We can then consider a MAP estimator
\begin{equation} \label{eq:x MAP general}
x_{\text{MAP}} = \arg\max_{x}{p(\mathbf{y}|x)p(x)}.
\end{equation}
%
Since $\mathbf{y}|x\sim N(x\boldsymbol{1},\sigma_n^2\mathbf{I})$ and $x\sim N(0,\sigma_x^2)$, the MAP estimator of $x$ is given by
\begin{equation} \label{eq:x MAP}
x_{\text{MAP}} =  \frac{\sigma_x^2}{\sigma_n^2+M\sigma_x^2} \boldsymbol{1}^H\mathbf{y} = \frac{M\sigma_x^2}{\sigma_n^2+M\sigma_x^2}x_{\text{DAS}} .
\end{equation}
The beamformer weights are correspondingly 
\begin{equation} \label{eq:w MAP}
\boldsymbol{w}_{\text{MAP}} = \frac{\sigma_x^2}{\sigma_n^2+M\sigma_x^2} \boldsymbol{1}.
\end{equation}

The expression in \eqref{eq:x MAP} implies that the summation of the aperture data in MAP beamforming  is performed using weights defined by the parameters of the prior distribution of the signal of interest and the interference.
In most realistic scenarios the parameters are unknown and therefore need to be estimated from the  data at hand. This leads to the iMAP beamformer, which we introduce next.

\subsection{Prior distribution parameter estimation}
\label{ssec:prior estimation}
To estimate the unknown parameters $\sigma_x^2$ and $\sigma_n^2$ we use an empirical Bayes approach \cite{carlin2000bayes}. Specifically, given $\sigma_x^2$ and $\sigma_n^2$, $x$ can be estimated using \eqref{eq:x MAP}. On the other hand, for a given realization of $x$ and $\mathbf{y}$, the unknown parameters can be estimated using an ML approach
\begin{align} \label{eq:parameters ML general}
\{\hat{\sigma}_x^2,\hat{\sigma}_n^2 \}_{\text{ML}}  =&  \arg\max_{\sigma_x^2,\sigma_n^2} p(\mathbf{y},x;\sigma_x^2,\sigma_n^2) ,
\end{align}
with
\begin{align}\label{eq:dist for iMAP}
 p(\mathbf{y},x;\sigma_x^2,\sigma_n^2) = & \left( 2\pi \sigma_x^2\right)^{-\frac{1}{2}}  \exp{\left\{-\frac{x^2}{2\sigma_x^2} \right\}} \\ \nonumber
 &\left( 2\pi \sigma_n^2\right)^{-\frac{M}{2}}   \exp{\left\{-\frac{1}{2\sigma_n^2}|| \mathbf{y} - x  \boldsymbol{1}||^2 \right\}}.
\end{align}
This results in
\begin{align} \label{eq:parameters ML}
\hat{\sigma}_x^2 =& x^2 \\ \nonumber
\hat{\sigma}_n^2 = &\frac{1}{M} || \mathbf{y} - x  \boldsymbol{1}||^2.
\end{align}
We thus suggest iterating between \eqref{eq:x MAP} and \eqref{eq:parameters ML} so as to use the signal of interest to improve the estimation of  the distribution parameters and vice versa. The resulting beamformer is presented in Algorithm \ref{alg:Algo}.
The only parameters required by the proposed method are the initialization of $x$ and a stopping criteria.  

In Algorithm \ref{alg:Algo} we use the DAS beamformer as an initial guess for $x$. The number of iterations can be chosen, e.g., according to the desired contrast value. However, defining such a value is practical only for predefined phantoms, where the true values of the underlying image are known. Thus, a more practical approach is to define the number iterations empirically. 
We found experimentally that the second iteration of \mbox{iMAP} leads to about 80-100 dB noise suppression inside the cyst regions. Since the typical dynamic range of medical ultrasound varies from 60 to 80 dB, such noise suppression is sufficient and further iterations are unnecessary. We therefore considered only two iterations in all the examples below.

\begin{algorithm}[h!]
\caption{iMAP beamforming}\label{alg:Algo}
\begin{algorithmic}[1]
\State {Initialize $\hat{x}_{(0)} =  \frac{1}{M} \boldsymbol{1}^H \mathbf{y}.$} 
\State \label{it:param}For current estimate $\hat{x}_{(t)}$ use \eqref{eq:parameters ML} to compute  $$\{\hat{\sigma}_x^2,\hat{\sigma}_n^2 \}_{(t)}=\left\{\hat{x}_{(t)}^2,\frac{1}{M} \left\| \mathbf{y} -\hat{x}_{(t)}  \boldsymbol{1}\right\|^2\right\}.$$
\State \label{it:x}Given $\{\hat{\sigma}_x^2,\hat{\sigma}_n^2 \}_{(t)}$ use \eqref{eq:x MAP} to update $$\hat{x}_{(t+1)} =\frac{\hat{\sigma}^2_{x,(t)}}{\hat{\sigma}_{n,(t)}^2+M\hat{\sigma}_{x,(t)}^2} \boldsymbol{1}^H\mathbf{y}.$$
\State Iterate \ref{it:param} and \ref{it:x} until stopping criterion is met.
\end{algorithmic}
\end{algorithm}

\section{Comparison to Related Methods}
\label{sec:related methods}
In this section we review CF scaling, Wiener beamforming and Wiener postfilter and focus on the relationship between them following derivations in \cite{nilsen2010wiener}. We then compare ScW, a generalization of Wiener processing proposed in \cite{nilsen2010wiener}, to \mbox{iMAP} beamforming.

\subsection{Coherence Factor  }
\label{ssec:CF} 
The coherence factor (CF) \cite{mallart1994adaptive,hollman1999coherence}  is computed as a ratio of coherent to incoherent sums across the aperture signals. For a given range, CF is defined by
\begin{equation} \label{eq:CF}
CF =  \frac{\left |\sum_{m=1}^M y_m \right |^2}{M\sum_{m=1}^M \left | y_m \right |^2} = \frac{\left | \boldsymbol{1}^H \mathbf{y}\right |^2}{M  \mathbf{y}^H \mathbf{y}}.
\end{equation}
In \cite{li2003adaptive} and \cite{wang2007performance}, CF is used as a scaling factor multiplying the output of DAS. 
 Assuming that the signal originating at the focal point is highly coherent over the aperture elements while the interference is not, multiplication by CF suppresses the off-axis echoes and improves contrast.
The resulting value of the b-mode image at the point of interest is 
\begin{equation}
x_{\text{CF}} =  CF\cdot x_{\text{DAS}}, 
\end{equation}
with $x_{\text{DAS}}$ given by \eqref{eq:x DAS}. Significant contrast improvement is reported in \cite{li2003adaptive} also for cases with phase distortions stemming from focusing imperfections of different kinds. 
Despite the advantages of CF, the resulting images may suffer from reduced image brightness and degraded speckle pattern,
especially in the low SNR regime.

\subsection{Wiener processing}
\label{ssec:Wiener processing}
For the Gaussian model assumed in this work, MAP estimation is equivalent to minimizing the mean-squared error (MSE). Explicitly, $\boldsymbol{w}_{\text{MAP}}$ is also the solution to the problem
\begin{equation} \label{eq:MMSE opt problem}
\min_{\boldsymbol{w}}{E\left[ \left |x -\boldsymbol{w}^H \mathbf{y} \right |^2\right]},
\end{equation}
and, therefore, can be referred to as Wiener beamforming~\cite{van2004detection}.
Nilsen and Holm in \cite{nilsen2010wiener} studied Wiener processing to provide a theoretical basis for performance and limitations of CF. In their work, they present a derivation of Wiener beamforming and propose the concept of a Wiener postfilter for any distortionless beamformer. The signal of interest is assumed to be independent of additive zero-mean interference, however, no assumptions are made on the statistics of the interference. In particular, it is not assumed to be spatially white.

For this general case, the Wiener beamformer that solves \eqref{eq:MMSE opt problem} is given by \cite{nilsen2010wiener}
\begin{align} \label{eq:wiener full}
\boldsymbol{w}_{\text{wiener}} = \frac{E[x^2]}{E[x^2]+\boldsymbol{w}_{\text{MV}}^H\mathbf{R}_{n} \boldsymbol{w}_{\text{MV}}} \boldsymbol{w}_{\text{MV}}, \\ \nonumber
\end{align}
%
 where $\mathbf{R}_{n}=E[\mathbf{n} \mathbf{n} ^H]$ and  $\boldsymbol{w}_{\text{MV}} $ is the MV beamformer of~\eqref{eq:MV beamformer}.
Thus, implementation of the general Wiener beamformer has the same drawbacks as MV beamforming discussed in Section \ref{sec:signal model}.

It can be seen from \eqref{eq:wiener full} that  the Wiener beamformer is a MV beamformer scaled by a factor defined by the ratio of the signal power to the total power of the MV beamformer output. 
Based on this, Nilsen and Holm define the Wiener postfilter as a scaling factor that can be found for any beamformer $\boldsymbol{w}$ satisfying  $\boldsymbol{w}^H\boldsymbol{1}=1$, to minimize the MSE:

\begin{align} \label{eq:Wiener post opt problem}
H_{\text{wiener}} &= \arg\min_{H}{E\left[ \left |x - H\boldsymbol{w}^H \mathbf{y} \right |^2\right]}  \\ \nonumber
&= \frac{E[x^2]}{E[x^2]+\boldsymbol{w}^H \mathbf{R}_{n} \boldsymbol{w}}.
\end{align}
Implementing $H_{\text{wiener}} $ requires estimation of $E[x^2]$ and $\mathbf{R}_{n}$, and does not involve matrix inversion.
To implement a Wiener postfilter for a DAS beamformer, namely, for \mbox{$\boldsymbol{w} =\boldsymbol{w}_{\text{DAS}}$}, the following estimators are proposed  in \cite{nilsen2010wiener} :
\begin{align} \label{eq:estimators for Wiener postfilter}
\hat{E}[x^2]&= x_{\text{DAS}}^2, \\ \nonumber
\hat{\mathbf{R}}_{n} &= \frac{1}{K} \sum_{k=1}^{K}\left(\mathbf{y}_k-x_{\text{DAS}}\boldsymbol{1} \right) \left(\mathbf{y}_k-x_{\text{DAS}} \boldsymbol{1}\right)^H,
\end{align}
with $\mathbf{y}_k = [y_k,..., y_{k+L-1}]$. The estimator of  $\mathbf{R}_{n}$, therefore,  is computed by applying spatial smoothing, i.e. by dividing the total aperture of length $M$ to overlapping subarrays of length $L$ and averaging their covariance matrices.

Under the assumption of spatially white interference in \eqref{eq:noise dist}, i.e. $\mathbf{R}_{n}=\sigma_n^2\mathbf{I}$, the Wiener beamformer in \eqref{eq:wiener full} coincides with the Wiener postfilter for a DAS beamformer and is given by
\begin{align} \label{eq:WIener BF postfilter}
\boldsymbol{w}_{\text{wiener}}& =  H_{\text{wiener}}\boldsymbol{w}_{\text{DAS}}   \\ \nonumber
&=\frac{E[x^2]}{E[x^2]+\frac{1}{M}\sigma_n^2}\frac{1}{M} \boldsymbol{1} \\ \nonumber
&= \frac{E[x^2]}{M\cdot E[x^2]+\sigma_n^2}\boldsymbol{1}.
\end{align}
For $x\sim N(0,\sigma_x^2)$, \eqref{eq:WIener BF postfilter} becomes
\begin{align} \label{eq:WIener BF postfilter 2}
\boldsymbol{w}_{\text{wiener}}=  H_{\text{wiener}}\boldsymbol{w}_{\text{DAS}} = \frac{\sigma_x^2}{M\sigma_x^2+\sigma_n^2}\boldsymbol{1}
\end{align}
and is equivalent to the MAP beamformer. 
\subsection{Scaled Wiener Postfilter and iMAP}
\label{ssec:ScW and iMAP}
To pinpoint the relationship between the CF and Wiener beamforming, Nilsen and Holm \cite{nilsen2010wiener} rewrite \eqref{eq:CF}  as
\begin{align} \label{eq:CF 2}
CF &=  \frac{\left |\sum_{m=1}^M y_m \right |^2}{M\sum_{m=1}^M \left | y_m \right |^2} \\ \nonumber
& = \frac{\left|\frac{1}{M}\boldsymbol{1}^H \mathbf{y}\right|^2}{\left|\frac{1}{M}\boldsymbol{1}^H \mathbf{y}\right|^2+\frac{1}{M} || \mathbf{y} - \frac{1}{M}\boldsymbol{1}^H \mathbf{y}  \boldsymbol{1}||^2} \\ \nonumber
& = \frac{\left|x_{\text{DAS}}\right|^2}{\left|x_{\text{DAS}}\right|^2+\frac{1}{M} || \mathbf{y} - x_{\text{DAS}}\boldsymbol{1}||^2}.
\end{align}
For the case of spatially white interference, $\left|x_{\text{DAS}}\right|^2$ and $\frac{1}{M} || \mathbf{y} - x_{\text{DAS}}  \boldsymbol{1}||^2$ can be viewed as the estimators of the signal's and noise power, i.e. $\hat{\sigma}_x^2$ and $\hat{\sigma}_n^2$, respectively. This allows for the following interpretation of the CF output
\begin{align}\label{eq:CF 3}
x_{\text{CF}}=\frac{CF}{M}\boldsymbol{1}^H \mathbf{y} = \frac{\sigma_x^2}{M\sigma_n^2+M\sigma_x^2}\boldsymbol{1}^H \mathbf{y}.
\end{align}
Comparing \eqref{eq:CF 3} to \eqref{eq:WIener BF postfilter 2},
CF may be interpreted as a Wiener postfilter for a DAS beamformer with overestimated noise variance \cite{nilsen2010wiener}.
The overestimation of noise results in more severe suppression of the beamformer output, leading on the one hand to improved contrast, but on the other, to reduced image brightness and degraded speckle patterns.   

To combine contrast improvement obtained by CF with the robustness of Wiener processing, preserving image brightness and speckle, the following heuristic solution, referred to as a scaled Wiener postfilter, is proposed in \cite{nilsen2010wiener}:
\begin{align}\label{eq:ScW}
x_{\text{ScW}} = \frac{\sigma_x^2}{\alpha\sigma_n^2+M\sigma_x^2}\boldsymbol{1}^H \mathbf{y}.
\end{align}
Choosing $\alpha$ in the range from $1$ to $M$, one can gradually move from Wiener to CF performance. 
To implement the ScW in \cite{nilsen2010wiener}, Nilsen and Holm use the following estimators:
\begin{align} \label{eq:estimators for ScW}
\hat{\sigma}_x^2 & = \left|x_{\text{DAS}}\right|^2, \\ \nonumber
\hat{\sigma}_n^2 &= \frac{1}{M} || \mathbf{y} - x_{\text{DAS}} \boldsymbol{1}||^2,
\end{align}
which correspond to the ML estimators in \eqref{eq:parameters ML} used for the first iteration of iMAP. We note that for $\alpha = 1$ and estimators in \eqref{eq:estimators for ScW}, ScW is equivalent to the first iteration of iMAP. However, ScW with $\alpha = 1$ was not studied and verified experimentally in \cite{nilsen2010wiener}. To the best of our knowledge, this work is the first to provide such results, referred in the context of this paper as iMAP1.   

We note that theoretically Wiener processing should provide an optimal solution in terms of MSE.
The fact that CF yields improved contrast compared to Wiener beamforming and Wiener postfilter, as reported in \cite{nilsen2010wiener}, can be explained by poor estimation of $\sigma_x^2$ and $\sigma_n^2$.  
The estimator of $\sigma_x^2$  used for Wiener processing,
\begin{equation} \label{eq}
\hat{\sigma}_x^2  = \left|x_{\text{DAS}}\right|^2,
\end{equation}
overestimates the energy of the true signal since it includes all the interference that is not suppressed by DAS beamforming. This, in turn, leads to underestimated noise power, because of the subtraction of the estimated signal of interest:
\begin{equation} \label{eq}
\hat{\sigma}_{n}^2  =\frac{1}{M} || \mathbf{y} -x_{\text{DAS}}  \boldsymbol{1}||^2.
\end{equation}
The introduction of $\alpha$ in \eqref{eq:ScW}, may be viewed as an attempt to compensate for the estimation flaws in a heuristic way. 

The iterative approach proposed in Section \ref{ssec:prior estimation} deals with the same problem of parameter estimation. Instead of arbitrary compensation for limited estimation accuracy, it uses the available estimator of the signal of interest to  improve the  estimation of the parameters and vice versa. Thus, we expect it to provide better performance, as we will demonstrate in Section \ref{sec:results}.

\section{Experiments}
\label{sec:experiments}
 
To verify the performance of the proposed method we used PICMUS datasets \cite{liebgott2016plane}. The above include simulated and experimental data as well as \textit{in-vivo} acquisitions and are provided at the PICMUS website\footnote{https://www.creatis.insa-lyon.fr/Challenge/IEEE\_IUS\_2016/}. The availability of the data together with the detailed description of the proposed method make the results easily reproducible. 

The performance of iMAP1 and iMAP2 is compared to the following methods:
\begin{enumerate}
\item DAS beamforming defined in \eqref{eq:x DAS}.
\item Wiener postfilter for DAS beamformer according to \eqref{eq:Wiener post opt problem} using the estimators in \eqref{eq:estimators for Wiener postfilter}.
\item ScW processing according to \eqref{eq:ScW} with estimators in \eqref{eq:estimators for ScW} and $\alpha = M^{1/2}$ as proposed in \cite{nilsen2010wiener}.
\item CF using the definition in \eqref{eq:CF}.
\end{enumerate}
 
 The difference between the Wiener postfilter as it is defined in \eqref{eq:Wiener post opt problem} and iMAP1 is in the way they estimate the interference covariance matrix. The iMAP1 relies on the assumption of spatially white interference which is not exploited in \eqref{eq:estimators for Wiener postfilter}. The comparison of iMAP1 and Wiener postfilter may, thus, provide valuable insight regarding the noise statistics. As mentioned in Section \ref{ssec:ScW and iMAP}, ScW can be viewed as a heuristic way to compensate for the statistical estimation flaws. It is compared to iMAP2 which threats the same problem in an iterative fashion.

\subsection{Description of the Datasets}
\label{ssec:desc of the datasets}
To study the performance of the proposed method we use the following datasets.  
\begin{itemize}
\item \textbf{Simulated images}: We start with two acquisitions, simulated with Field II \cite{jensen1992calculation,jensen1996field}. First, we evaluate resolution using a point-reflector phantom with isolated scatterers distributed in an anechoic background. Then we proceed to contrast evaluation using a simulated phantom of a number of anechoic cysts embedded in speckle.
\item \textbf{Experimental dataset}: Both resolution and contrast are next evaluated experimentally using a recorded dataset of a CIRS Multi-Purpose Ultrasound Phantom (Model 040GSE). The scan region contains anechoic cysts and a point-reflector.
\item \textbf{\textit{In-vivo} acquisitions}: Finally we apply the proposed method on two \textit{in-vivo} datasets of a carotid artery.  This allows for qualitative assessment of the proposed technique. 
\end{itemize}

\subsection{Experimental and Evaluation Setup}
\label{ssec:exp setup}
Both simulations and experimental acquisitions are performed with a $128$ element probe with $\lambda$ spacing and central frequency of $5.2$ MHz. More details are available on the PICMUS website. Each image acquisition (both simulated and experimental) includes transmission of $75$ plane-waves with steering angles spaced uniformly between $-16^{\circ}$ and $16^{\circ}$.  The results obtained by each one of 75 transmissions are stored and allow  for evaluation of an image comprised of an arbitrary number of plane-waves. 
The acquisition is performed with an \mbox{f-number} of 1.75, and no apodization is applied upon reception.

The resolution is evaluated by computing the full width at half maximum of the point spread function in the axial and lateral directions.  To quantify contrast we measure contrast-to-noise ratio (CNR) as defined in \cite{li2003adaptive}:
\begin{equation} \label{eq:CNR}
CNR = \frac{\mu_{c}-\mu_{b}}{\sigma_{b}}.
\end{equation}
Here $\mu_c$ and $\mu_b$ stand for the mean log compressed intensity of the anechoic cyst region and the background, respectively, and $\sigma_b$ is the standard deviation of the background, also in dB.

\begin{figure*}[t]
\begin{minipage}[b]{0.16\linewidth}
  \centering
  \centerline{\includegraphics[width=3.5cm]{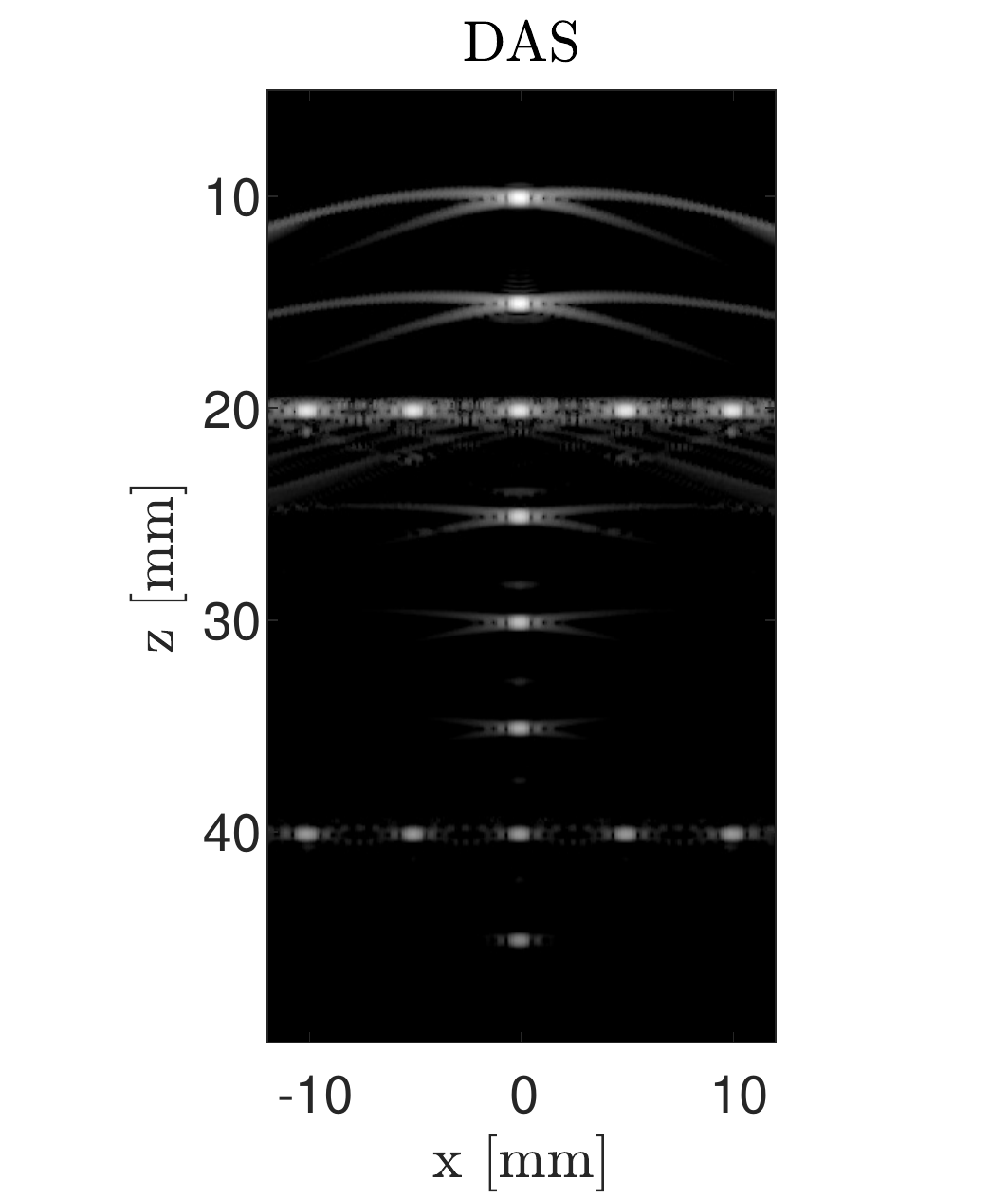}}%
  \centerline{(a)}\medskip
\end{minipage}
\begin{minipage}[b]{0.16\linewidth}
  \centering
  \centerline{\includegraphics[width=3.5cm]{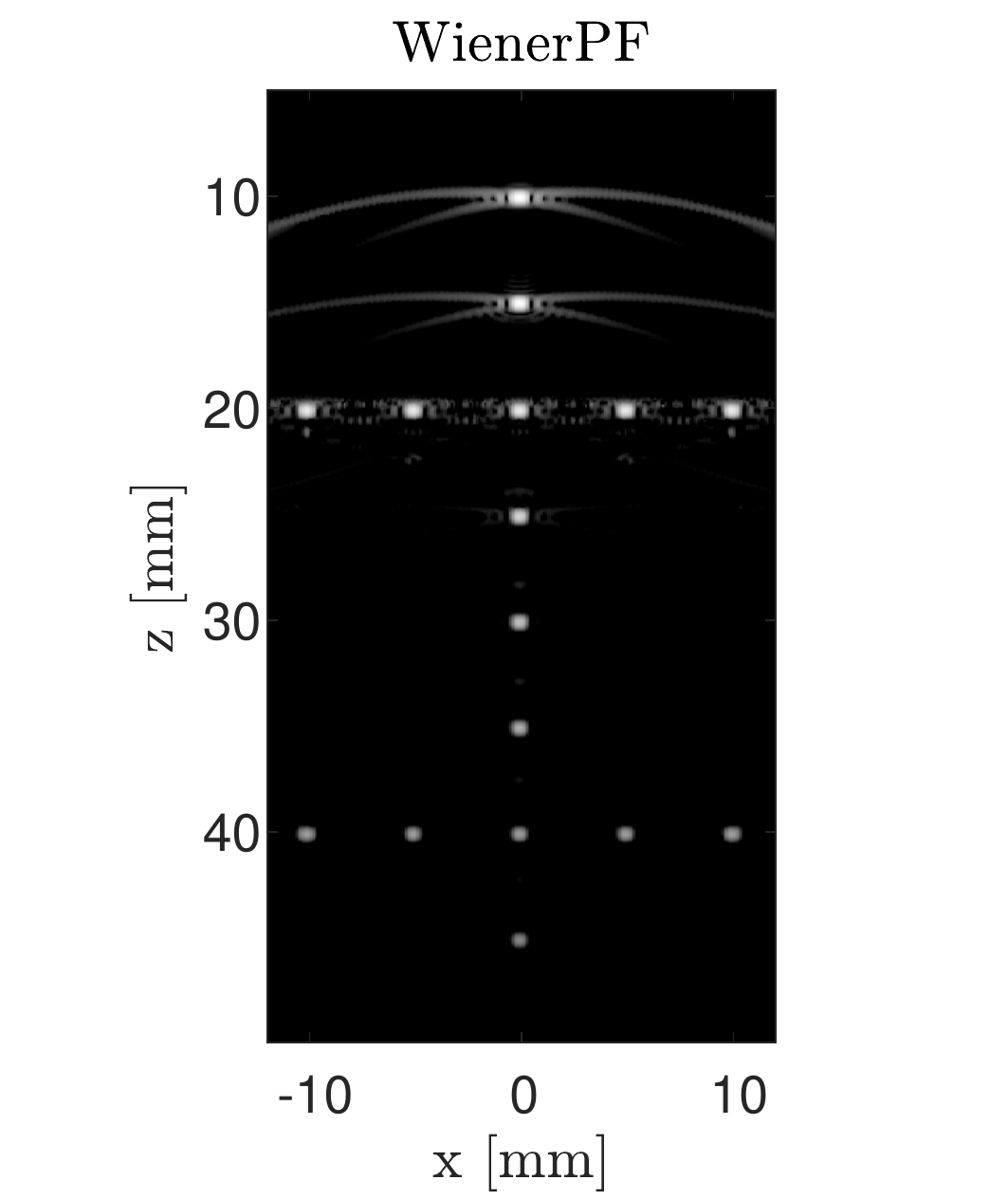}}%
  \centerline{(b)}\medskip
\end{minipage}
\begin{minipage}[b]{0.16\linewidth}
  \centering
  \centerline{\includegraphics[width=3.5cm]{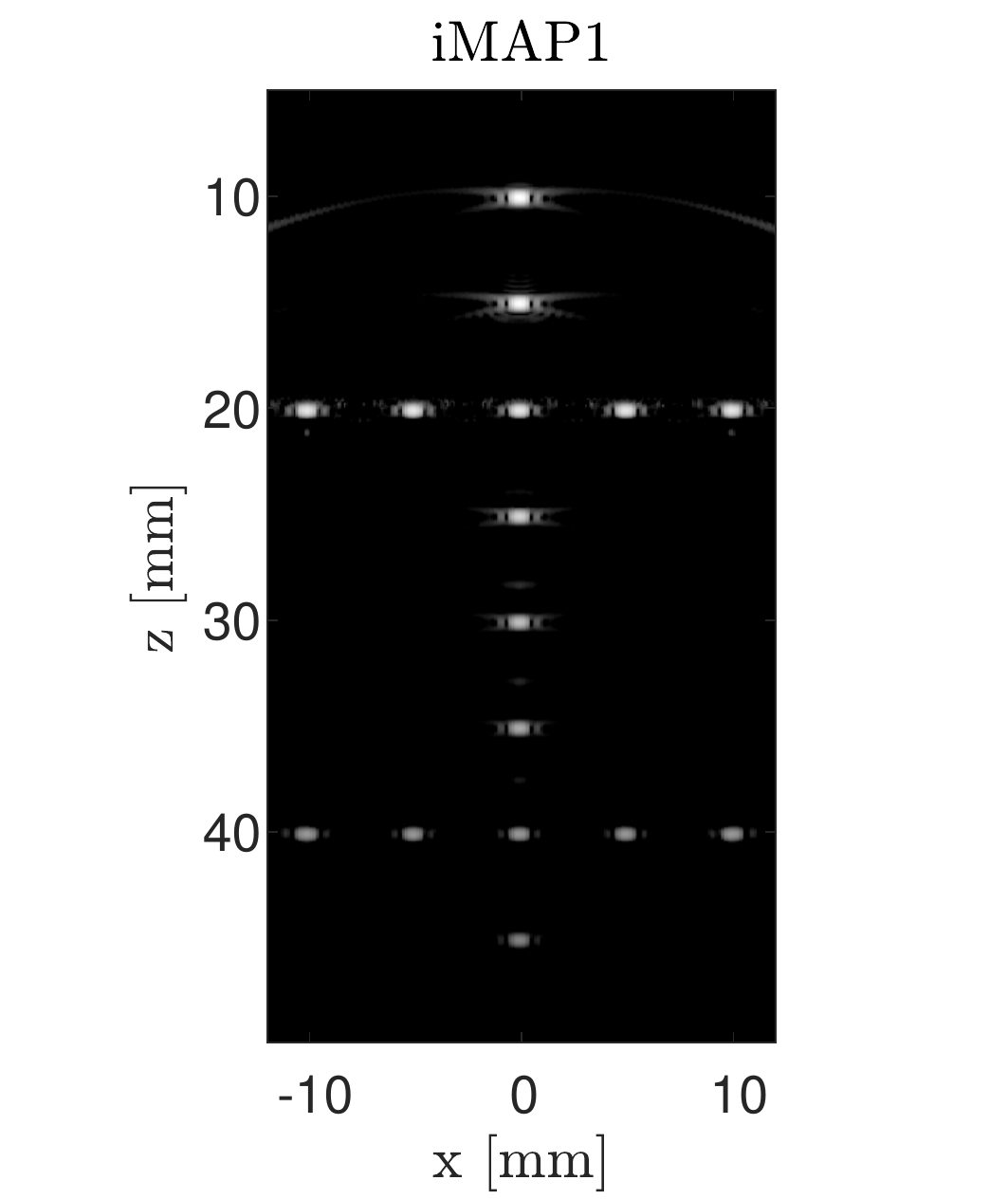}}%
  \centerline{(c)}\medskip
\end{minipage}
\begin{minipage}[b]{0.16\linewidth}
  \centering
  \centerline{\includegraphics[width=3.5cm]{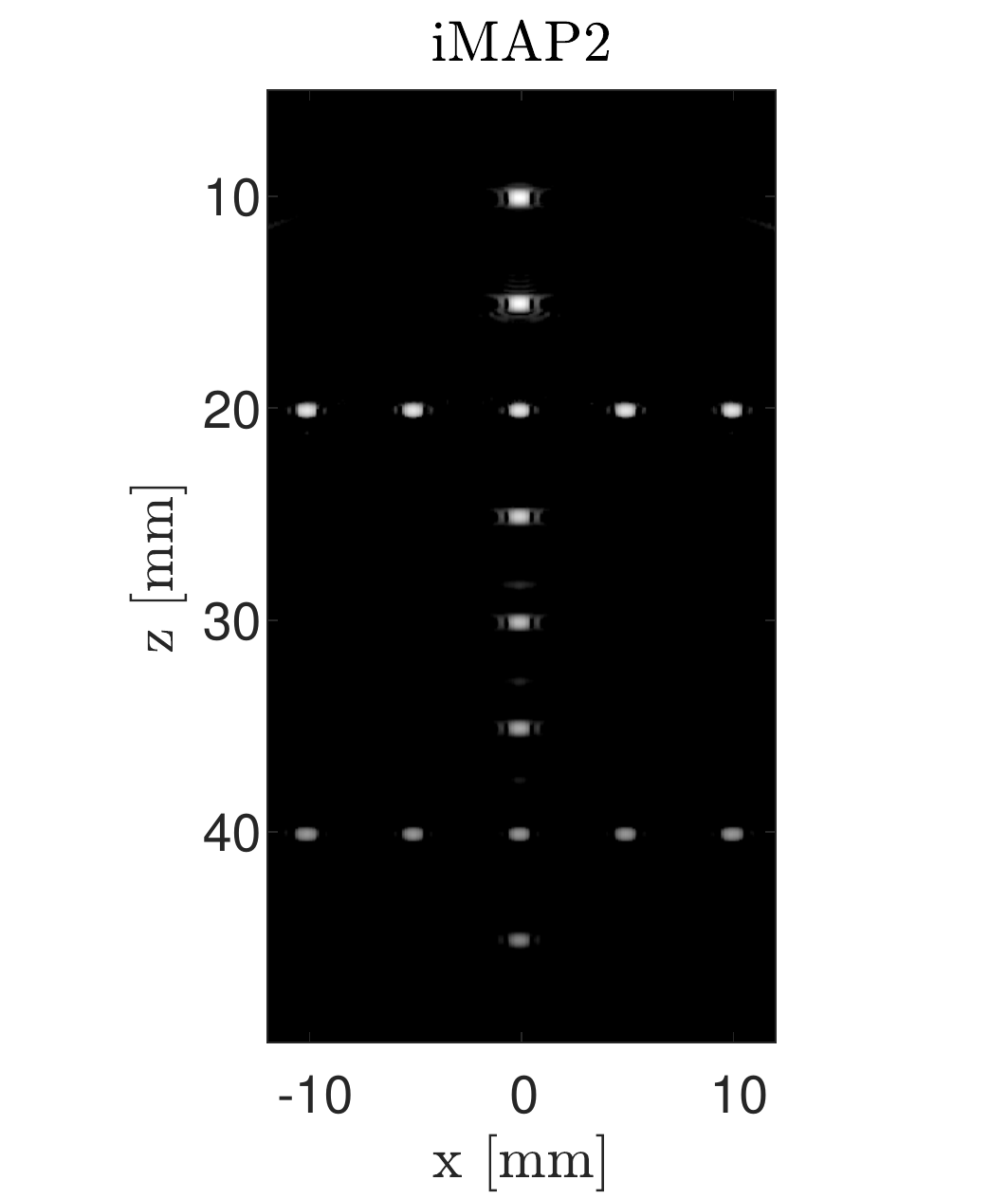}}%
  \centerline{(d)}\medskip
\end{minipage}
\begin{minipage}[b]{0.16\linewidth}
  \centering
  \centerline{\includegraphics[width=3.5cm]{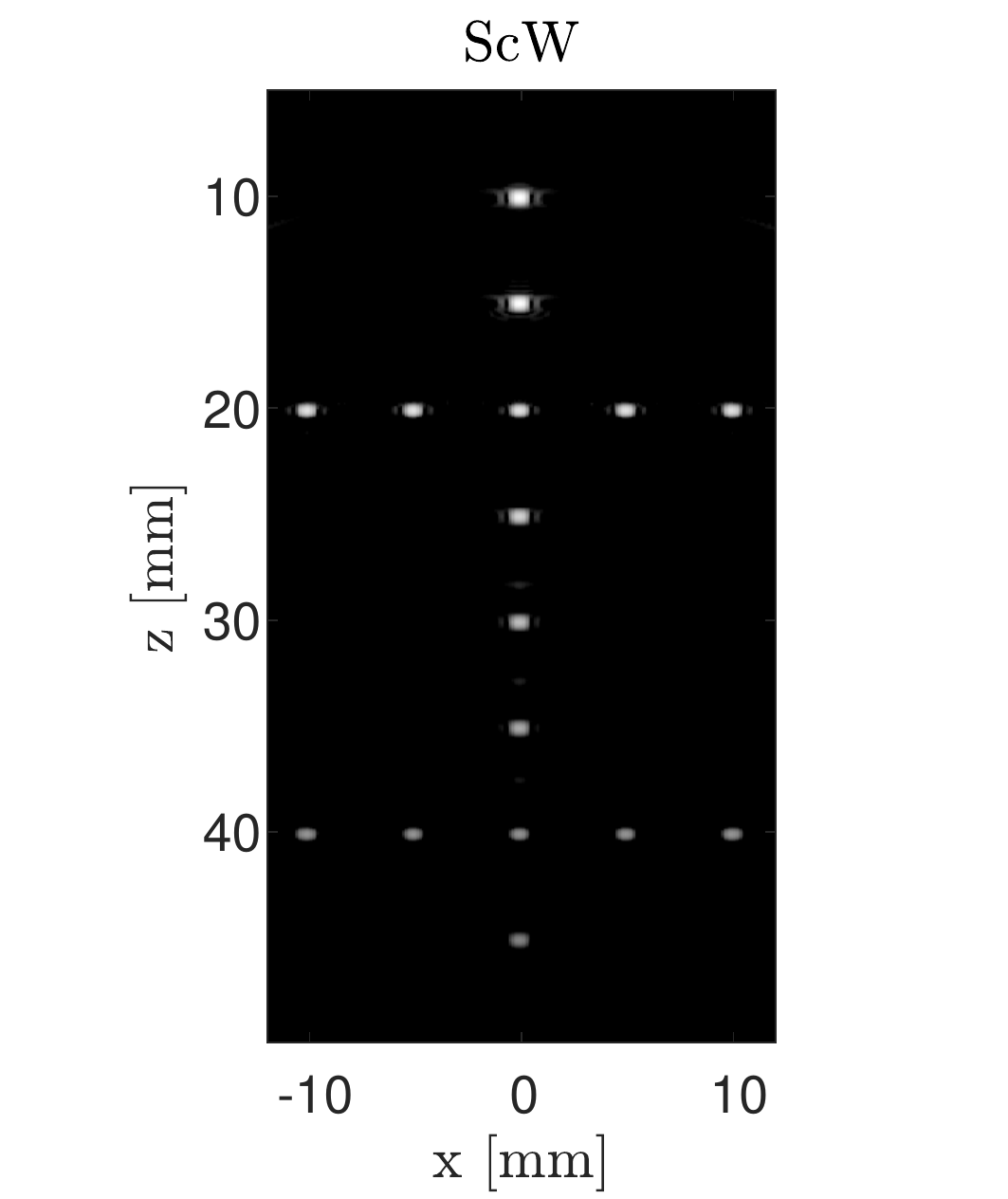}}%
  \centerline{(e)}\medskip
\end{minipage}
\begin{minipage}[b]{0.16\linewidth}
  \centering
  \centerline{\includegraphics[width=3.5cm]{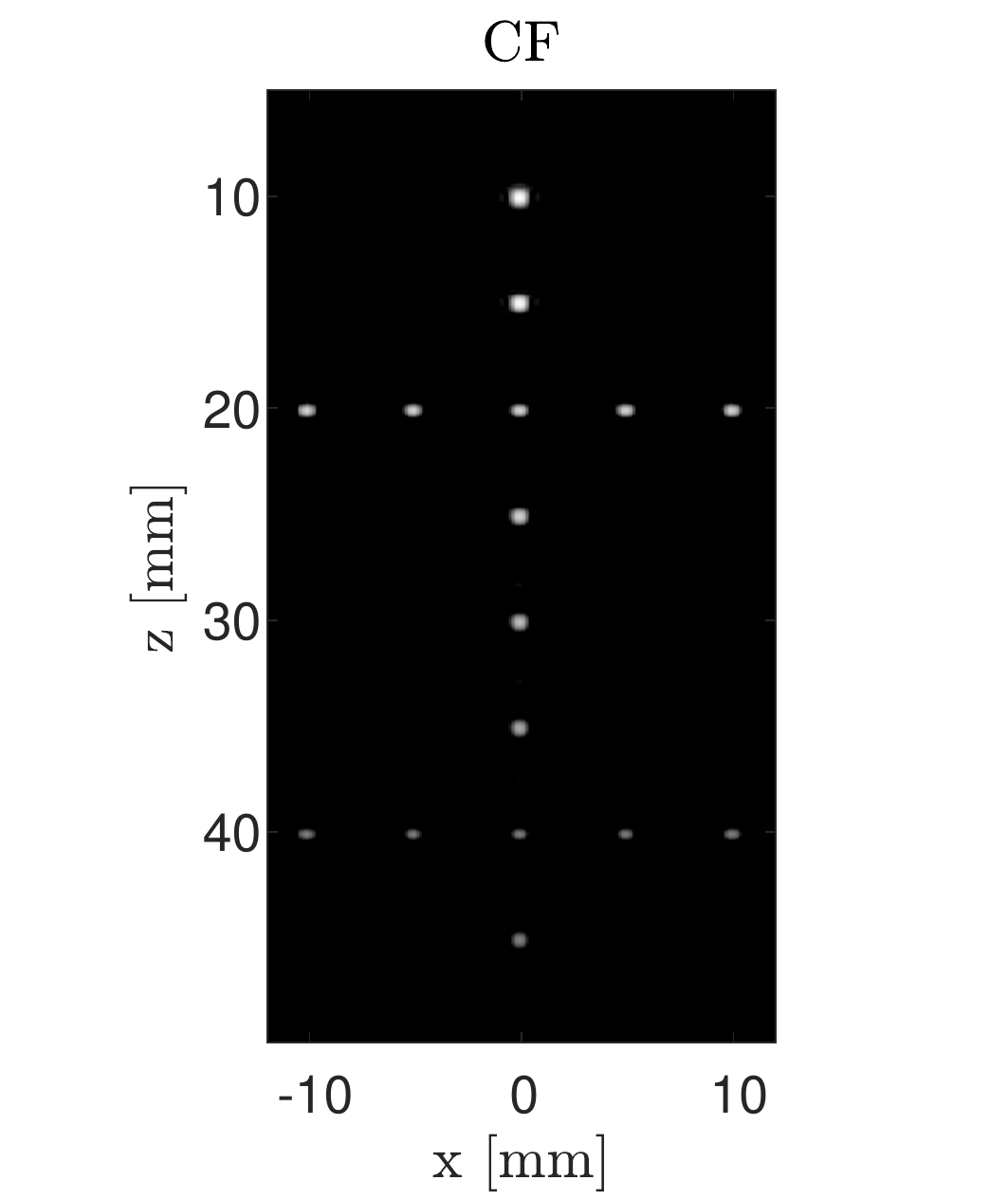}}%
  \centerline{(f)}\medskip
\end{minipage}
\caption{Images of  simulated point-reflector phantom obtained by (a) DAS, (b) Wiener postfilter, (c) iMAP1, (d) iMAP2, (e) ScW, (f) CF. The dynamic range of all the images is 50 dB.}
\label{fig:simResolution_1PW}
\end{figure*}
 \begin{figure*}[h]
\begin{minipage}[b]{0.48\linewidth}
  \centering
  \centerline{\includegraphics[width=9cm]{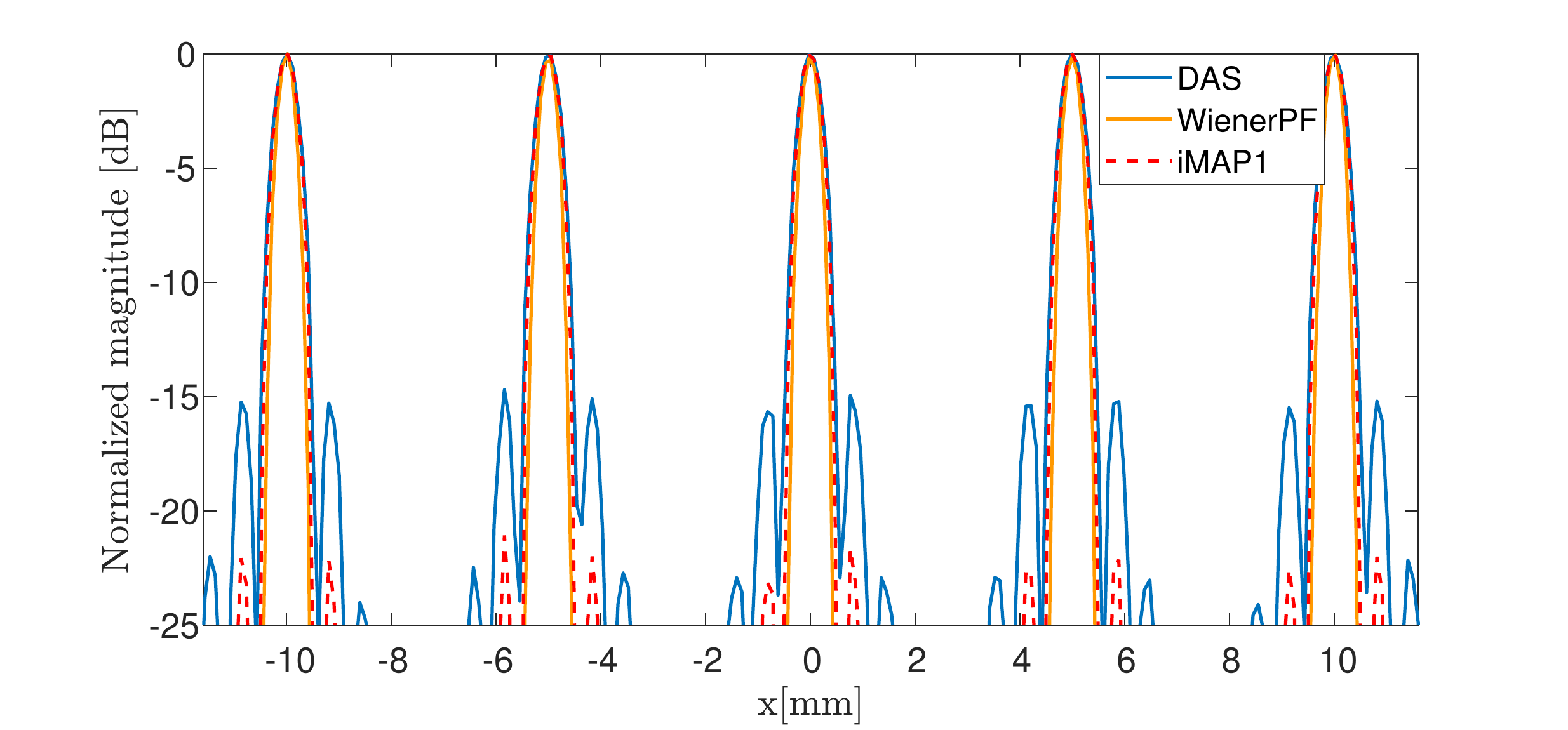}}%
  \centerline{(a)}\medskip
\end{minipage}
\begin{minipage}[b]{0.48\linewidth}
  \centering
  \centerline{\includegraphics[width=9cm]{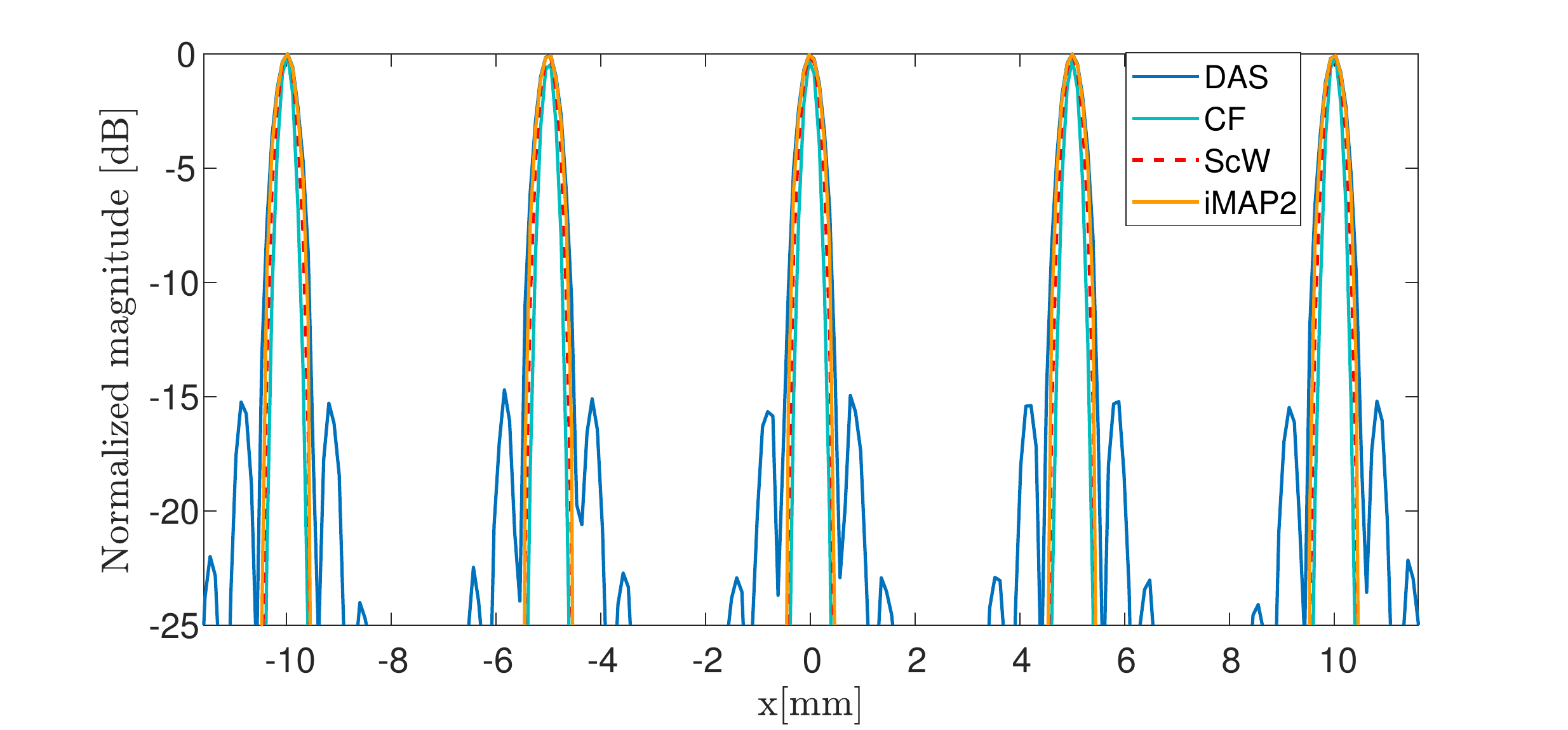}}%
  \centerline{(b)}\medskip
\end{minipage}
\caption{Lateral cross-section of point reflectors located at a depth of $20$ mm from Fig. \ref{fig:simResolution_1PW}, (a) DAS, Wiener postfilter, iMAP1, (b) DAS, CF, ScW, iMAP2.}
\label{fig:simRes_lat_PW1}
\end{figure*}
%


To compare the effect of iMAP and other interference suppression techniques on speckle, we evaluated the speckle area within the images obtained by different methods.
For each image we performed a Kolmogorov-Smirnov (K-S) test to find the regions obeying a Rayleigh probability density function (pdf) \cite{burckhardt1978speckle, wagner1983statistics}.
The K-S test is a common statistical hypothesis test that
verifies whether there is enough evidence in data to deduce that the hypothesis under consideration is correct. In our case, the hypothesis is that the investigated pattern obeys a Rayleigh pdf. 
To this end, the envelope data of each image is divided into overlapping patches of $20\times15$  pixels and the K-S test is applied to each patch. The patches that pass the K-S test with significance level $\alpha=0.05$ are included into the speckle region of each image.  The patches that did not pass the test are zeroed out. To quantify the similarity, we define the speckle region of the DAS image as a reference and compute what percentage of it is defined as speckle in images obtained by other techniques.
%
\begin{figure*}[h!]
\begin{minipage}[b]{0.32\linewidth}
  \centering
  \centerline{\includegraphics[width=5.09cm]{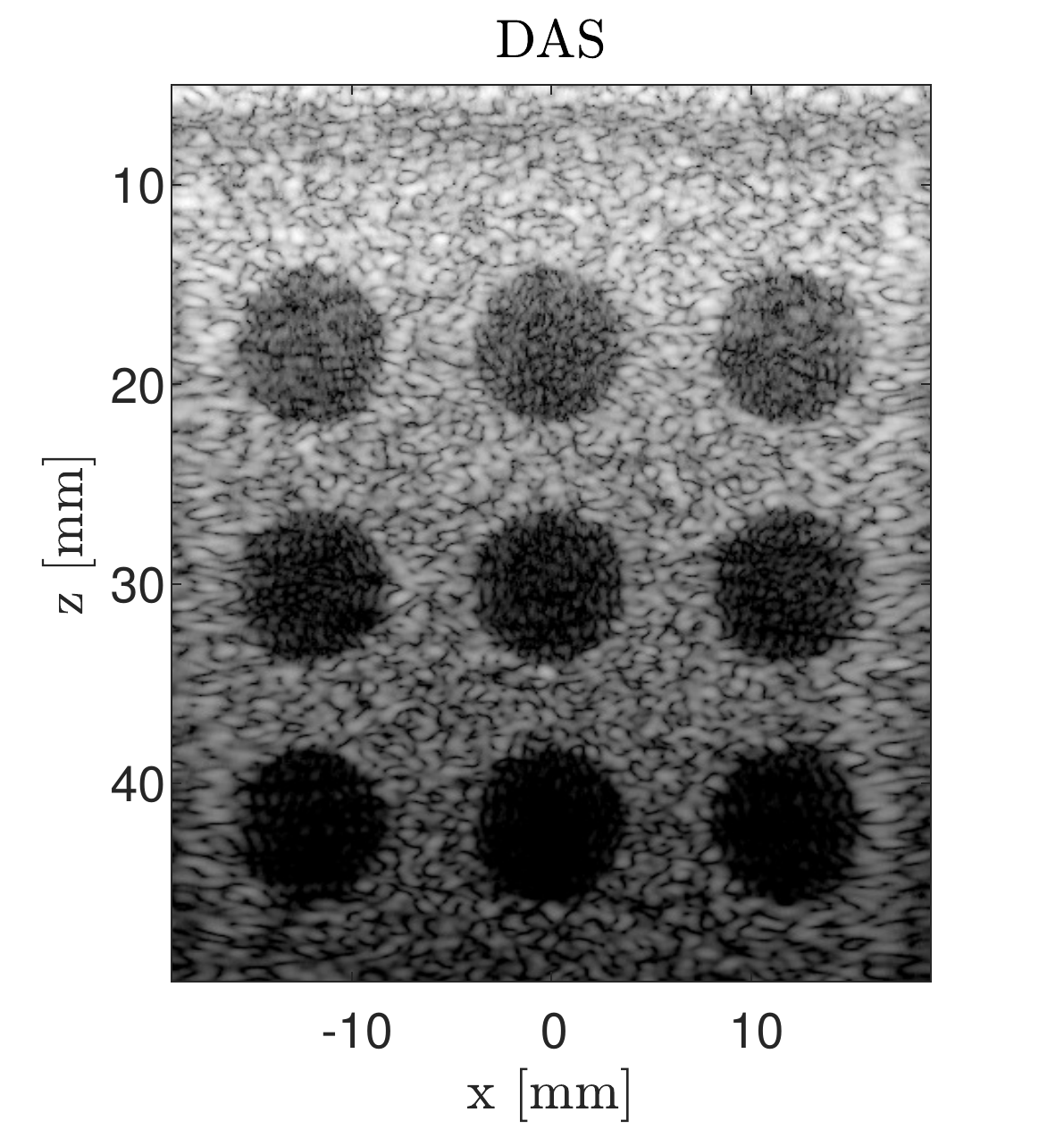}}%
  \centerline{(a)}\medskip
\end{minipage}
\begin{minipage}[b]{0.32\linewidth}
  \centering
  \centerline{\includegraphics[width=5cm]{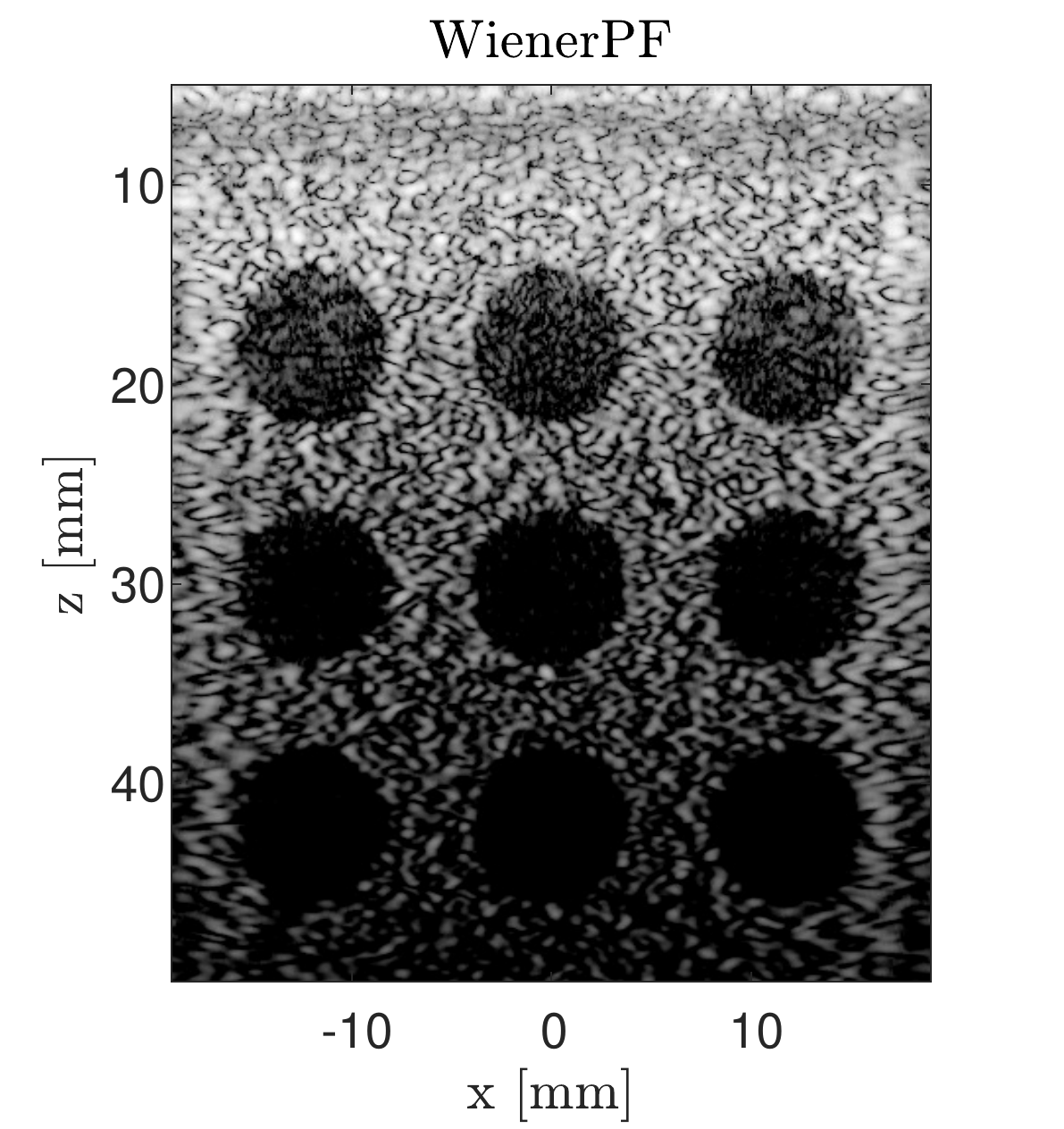}}%
  \centerline{(b)}\medskip
\end{minipage}
\begin{minipage}[b]{0.32\linewidth}
  \centering
  \centerline{\includegraphics[width=5cm]{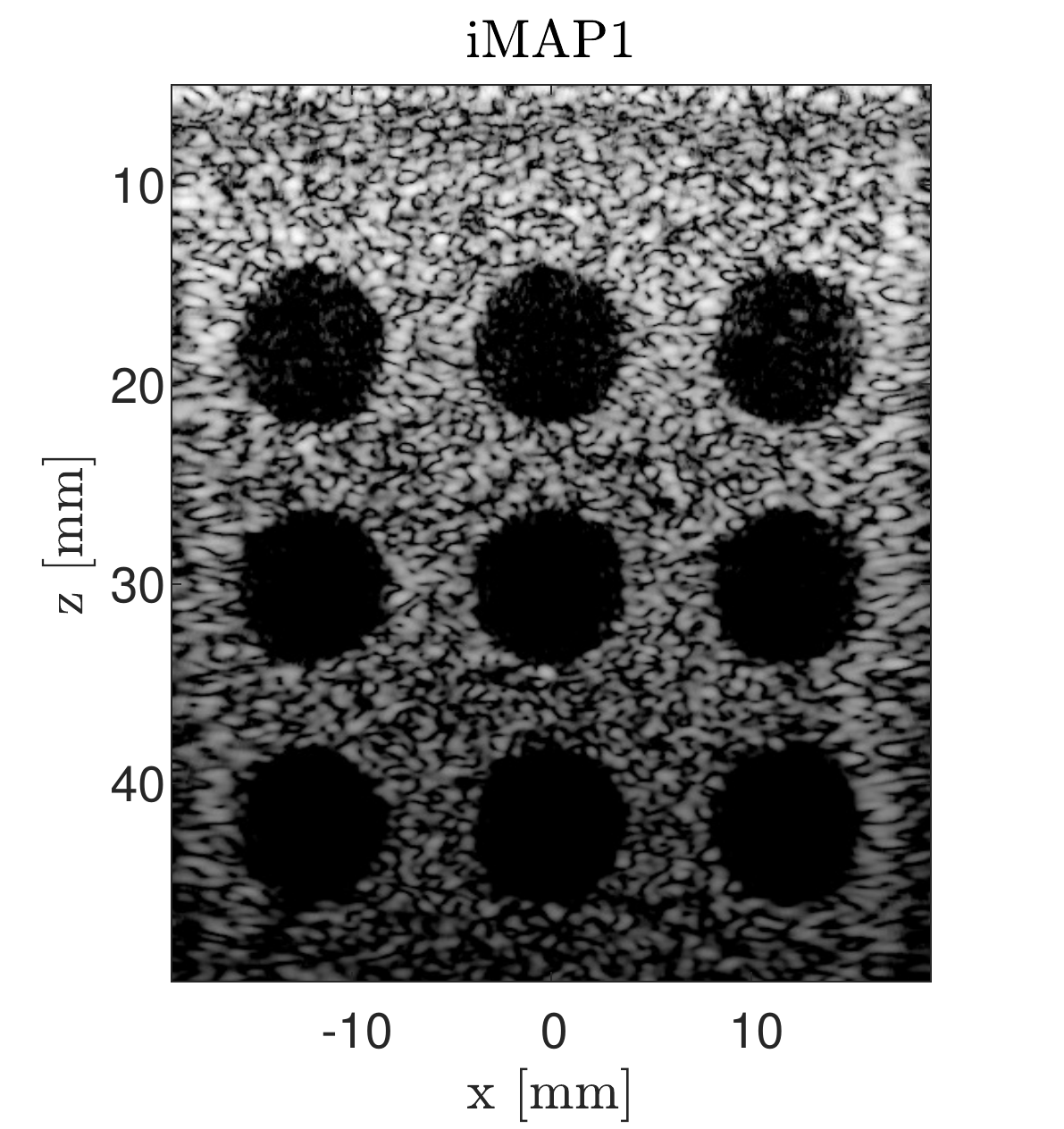}}%
  \centerline{(c)}\medskip
\end{minipage} \\
\begin{minipage}[b]{0.32\linewidth}
  \centering
  \centerline{\includegraphics[width=5cm]{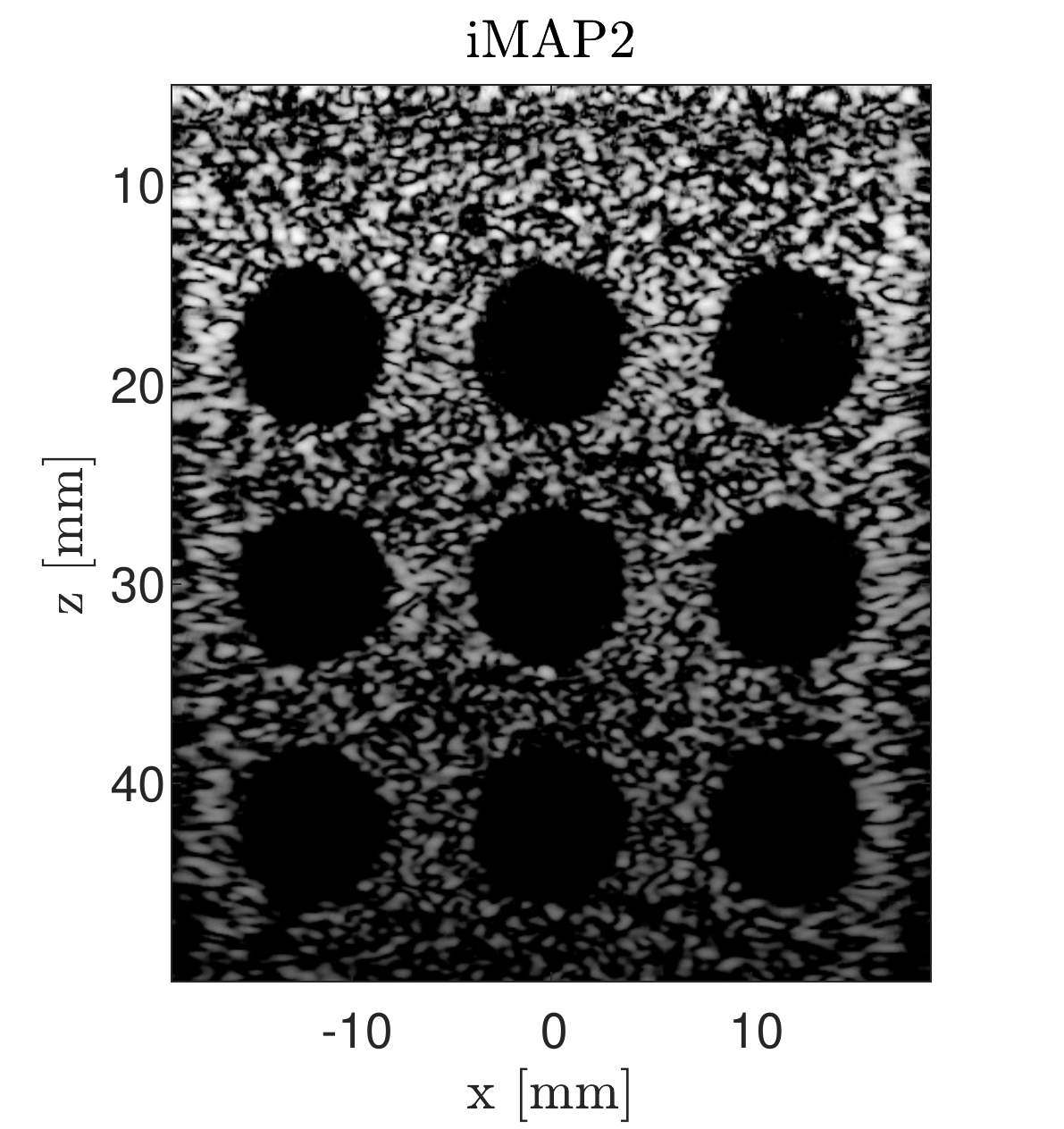}}%
  \centerline{(d)}\medskip
\end{minipage}
\begin{minipage}[b]{0.32\linewidth}
  \centering
  \centerline{\includegraphics[width=5cm]{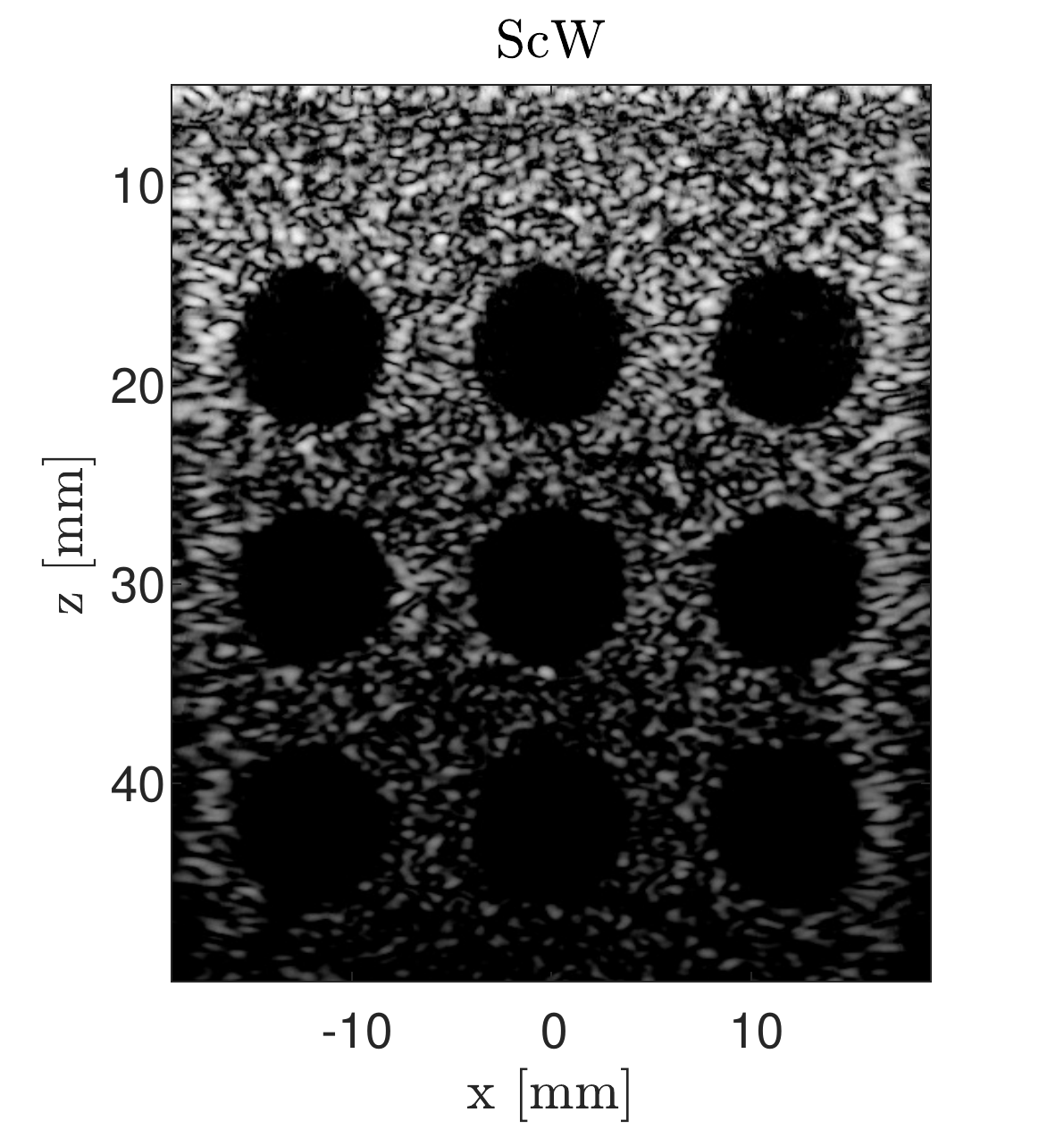}}%
  \centerline{(e)}\medskip
\end{minipage}
\begin{minipage}[b]{0.32\linewidth}
  \centering
  \centerline{\includegraphics[width=5cm]{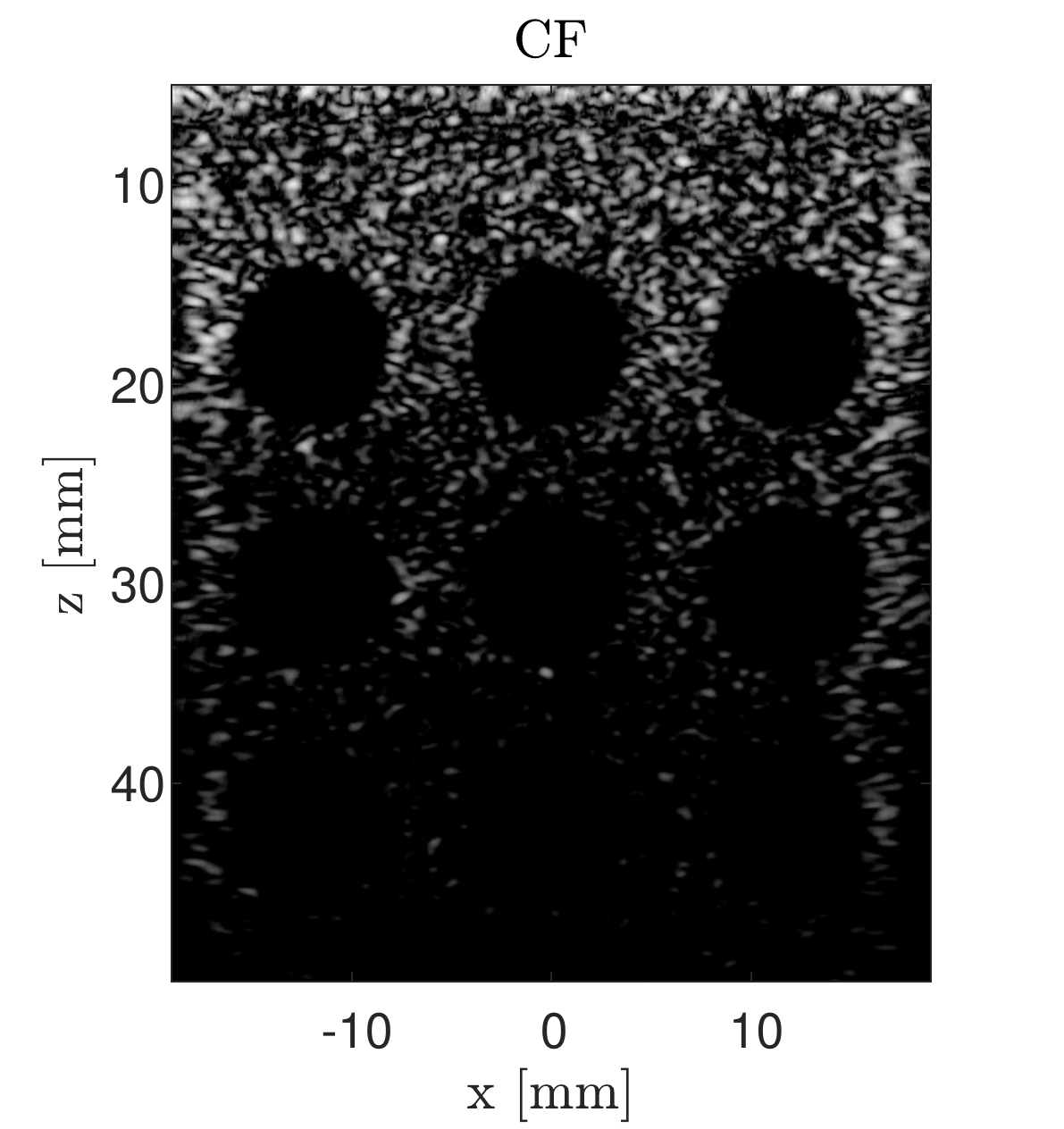}}%
  \centerline{(f)}\medskip
\end{minipage}
\caption{Images of  simulated anechoic cyst phantom obtained by (a) DAS, (b) Wiener postfilter, (c) iMAP1, (d) iMAP2, (e) ScW, (f) CF. The dynamic range for all the images is 60 dB.}
\label{fig:simContrast_1PW}
\end{figure*}
\label{sssec:contrast}
\begin{figure*}[h!]
\begin{minipage}[b]{0.48\linewidth}
  \centering
  \centerline{\includegraphics[width=8cm]{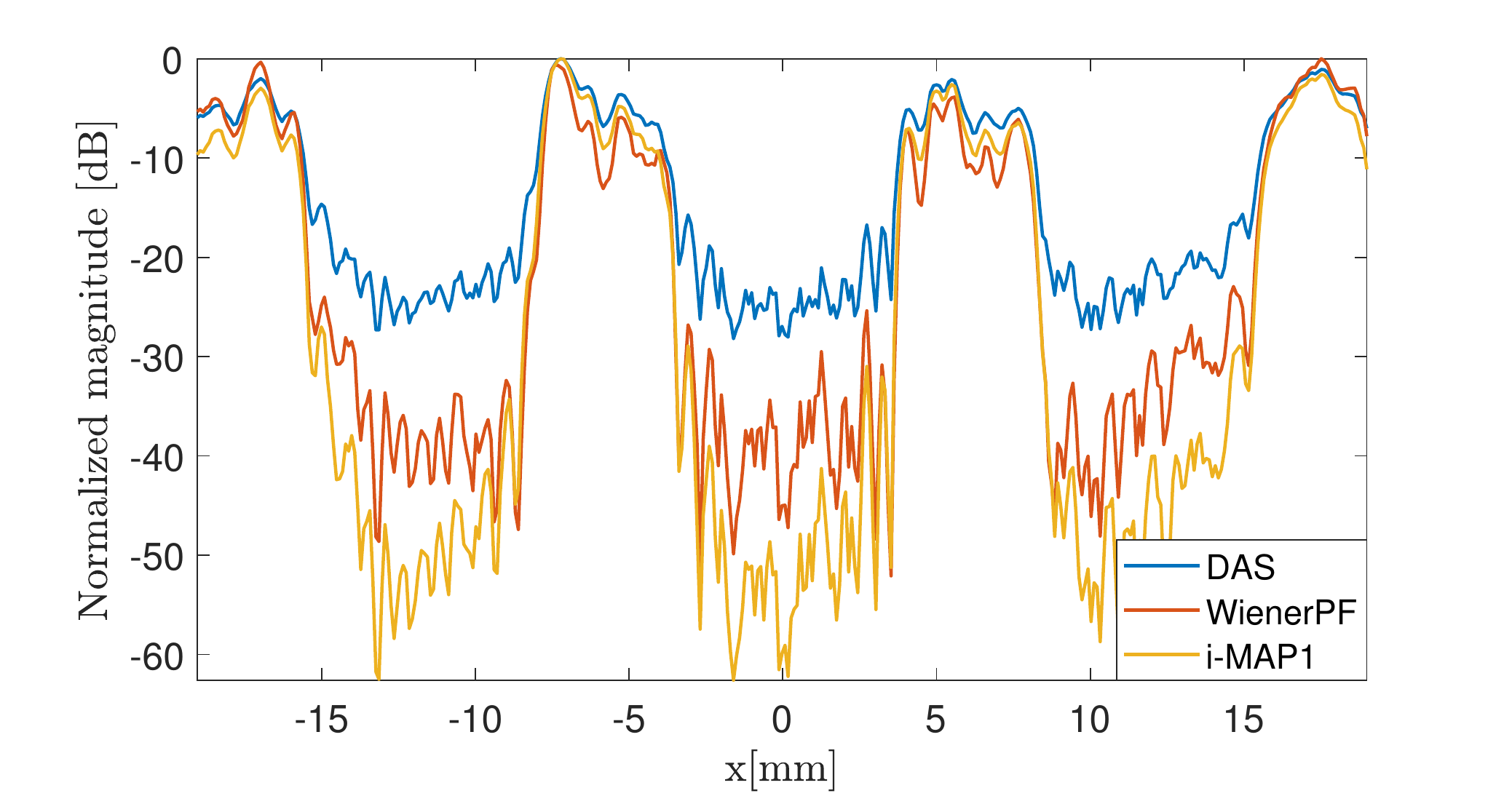}}%
  \centerline{(a)}\medskip
\end{minipage}
\begin{minipage}[b]{0.48\linewidth}
  \centering
  \centerline{\includegraphics[width=8cm]{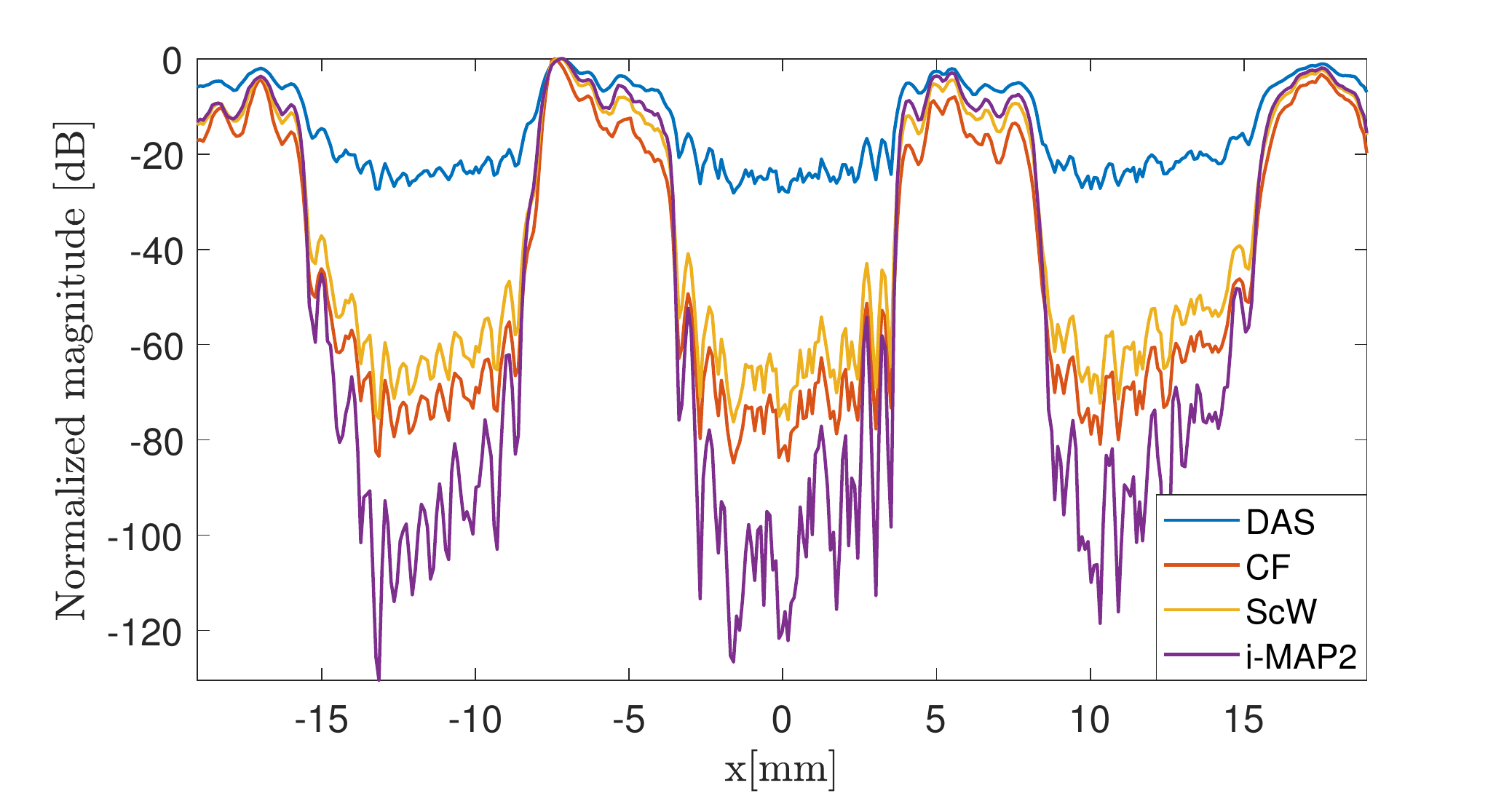}}%
  \centerline{(b)}\medskip
\end{minipage}
\caption{Lateral cross-section of three cysts located at a depth of $30$ mm from Fig. \ref{fig:simContrast_1PW}, (a) DAS, Wiener postfilter, iMAP1, (b) DAS, CF, ScW, iMAP2.}
\label{fig:simLatScanline_PW1_iMAP}
\end{figure*}
%
%
\section{Results}
\label{sec:results}
\subsection{Simulations}
\label{ssec:simulations}
 \subsubsection{Resolution}
\label{sssect:resolution}
We start with an image of a point-reflector phantom obtained by simulation of a single plane-wave transmission. Figure \ref{fig:simResolution_1PW} presents the results of DAS, Wiener postfilter, \mbox{ iMAP1 }, \mbox{iMAP2}, ScW and CF.
The axial and lateral resolution is computed as an average over all point reflectors within the image. The resulting values are summarized in \mbox{Table \ref{table:measured axial and lat res}}. 
The lateral cross-section of point reflectors located at a depth of 20 mm is presented in Fig. \ref{fig:simContrast_1PW}.
It can be noted that CF provides the most prominent improvement in both axial and lateral resolution. All the other methods lead to results that are better or comparable to standard DAS processing. 
As can be seen in Figs. \ref{fig:simResolution_1PW} and \ref{fig:simRes_lat_PW1}, the main effect of all the methods is in the reduction of the lateral side-lobes, which, as will be shown in Section \ref{sssec:contrast}, is expressed in contrast enhancement.
 \begin{table} [h!]
\caption{Measured axial and lateral resolution and CNR}
\label{table:measured axial and lat res}
\begin{center}
\begin{tabular}{| c | c | c |c|}
\hline
\textbf{Method} & \textbf{Axial res. [mm]} & \textbf{Lateral res. [mm]} & \textbf{CNR [dB]}\\
\hline
{DAS}                            &{0.4}       & {0.74}    &{2.57} \\ \hline
{Wiener Postfilter}  &{0.4}       & {0.53}   &{1.83} \\ \hline
{iMAP1}                      &{0.4}       & {0.7}     &{3.01}\\ \hline
{iMAP2}                      &{0.4}     & {0.69}     &{3.09}\\ \hline
{ScW}                            &{0.37}     & {0.6}     &{2.78}  \\ \hline
{CF}                               &{0.33}     &{0.44}   &{2.66}\\ \hline
\end{tabular}
\end{center}
\end{table}
\subsubsection{Contrast}
To evaluate the contrast improvement obtained by iMAP compared to other techniques, we use  an anechoic cyst phantom. Figure~\ref{fig:simContrast_1PW} presents the images  obtained with DAS, Wiener postfilter, \mbox{ iMAP1 }, \mbox{iMAP2}, ScW and CF and gives a qualitative impression of the contrast resulting from each approach.
The lateral cross section of three cysts located at a depth of 30 mm is shown in Fig. \ref{fig:simLatScanline_PW1_iMAP}.
The acquisition is performed with  a single plane-wave transmission. 
To evaluate the contrast quantitatively we computed the CNR of an image as an average CNR over all nine cysts of the simulated phantom. The values of CNR are presented in \mbox{Table \ref{table:measured axial and lat res}}. 

\begin{figure*}[t]
\begin{minipage}[b]{0.48\linewidth}
  \centering
  \centerline{\includegraphics[width=8cm]{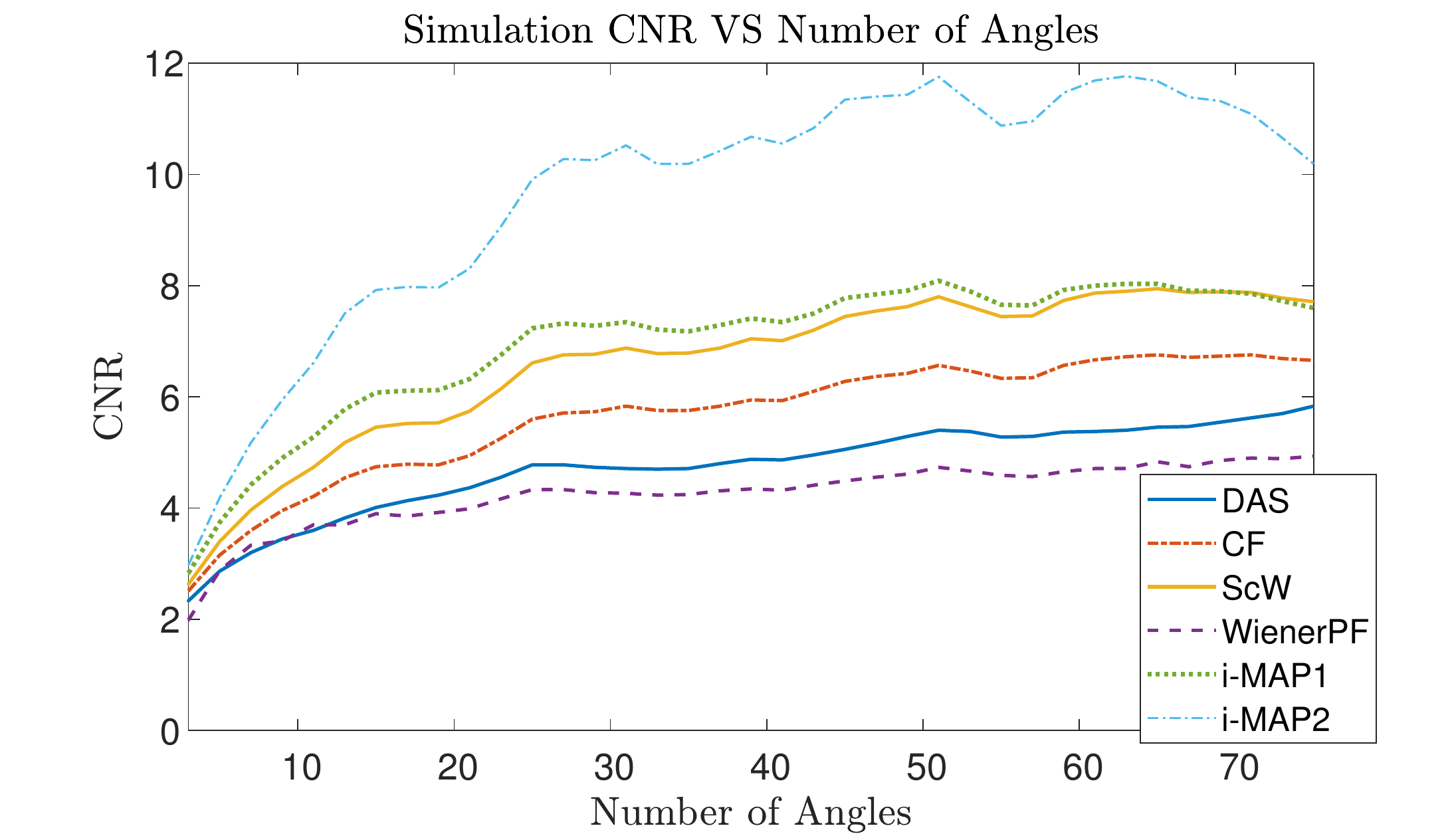}}%
  \centerline{(a)}\medskip
\end{minipage}
\begin{minipage}[b]{0.48\linewidth}
  \centering
  \centerline{\includegraphics[width=8cm]{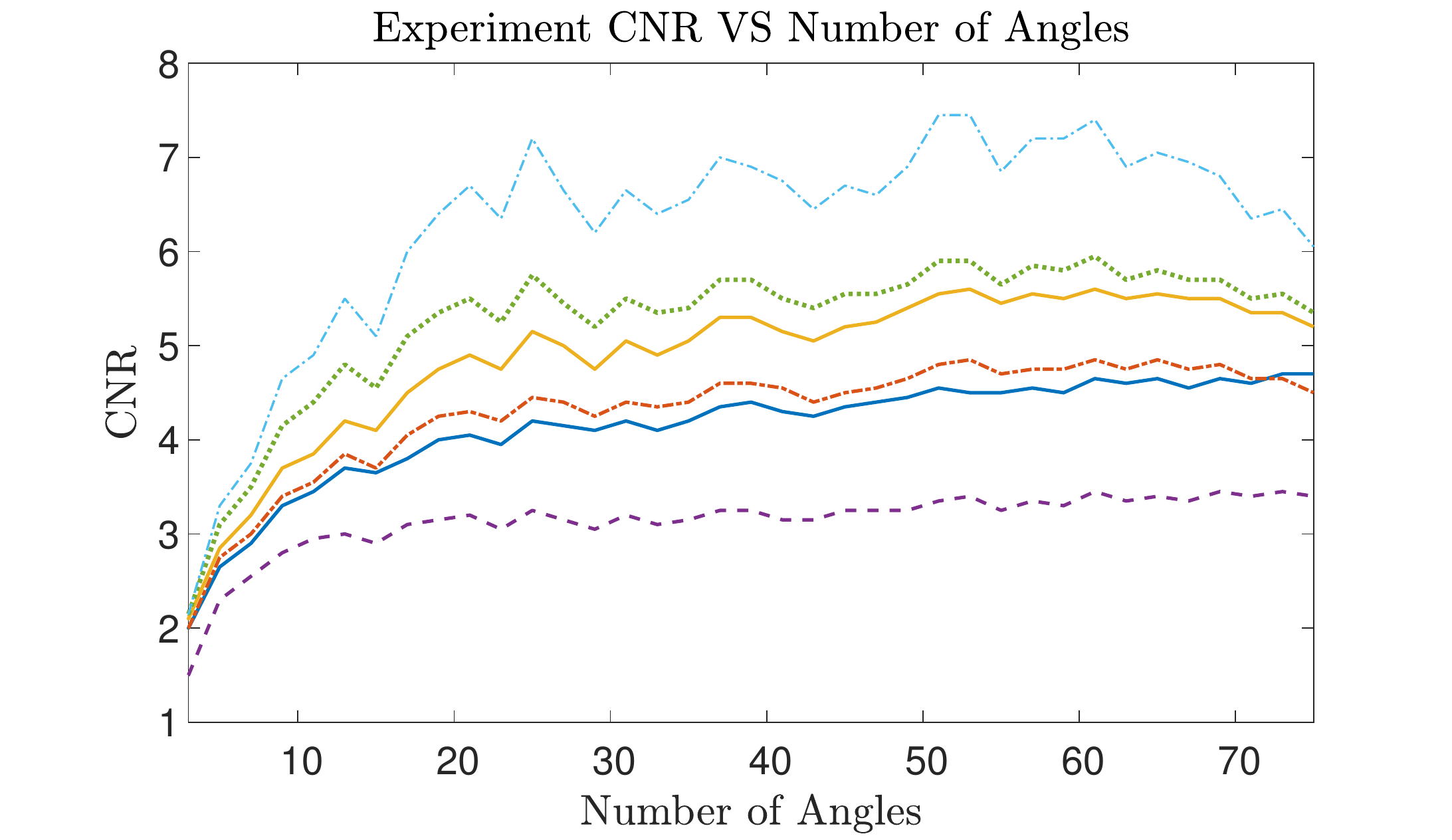}}%
  \centerline{(b)}\medskip
\end{minipage}
\caption{CNR as a function of the number of transmitted plane-waves, (a) simulated image, (b) experimental scan. The legend is the same for both figures.}
\label{fig:CNR}
\end{figure*}
%
%
\begin{figure*}[h!]
\begin{minipage}[b]{0.24\linewidth}
  \centering
  \centerline{\includegraphics[width=5cm]{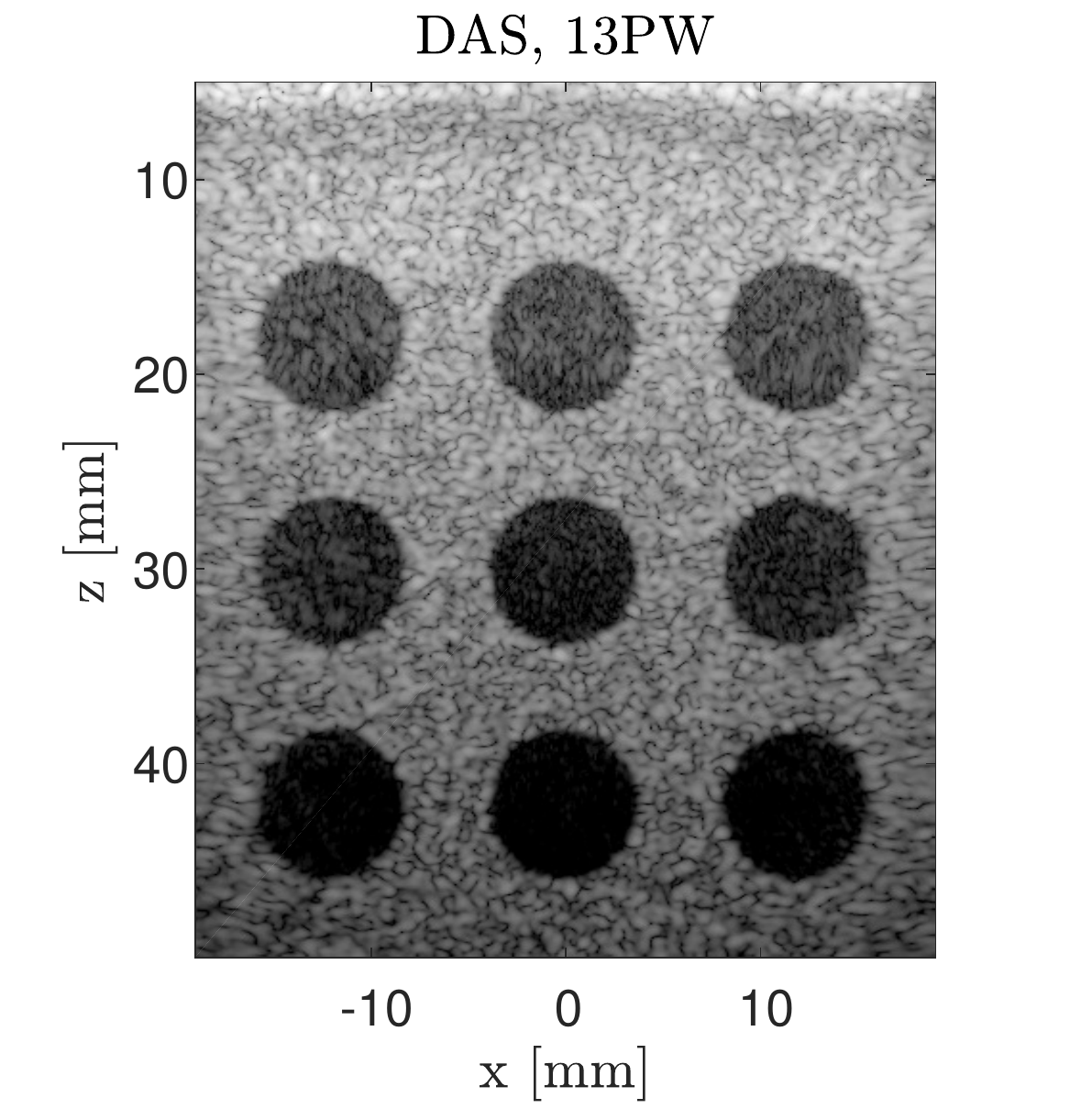}}%
  \centerline{(a)}\medskip
\end{minipage}
\begin{minipage}[b]{0.24\linewidth}
  \centering
  \centerline{\includegraphics[width=5cm]{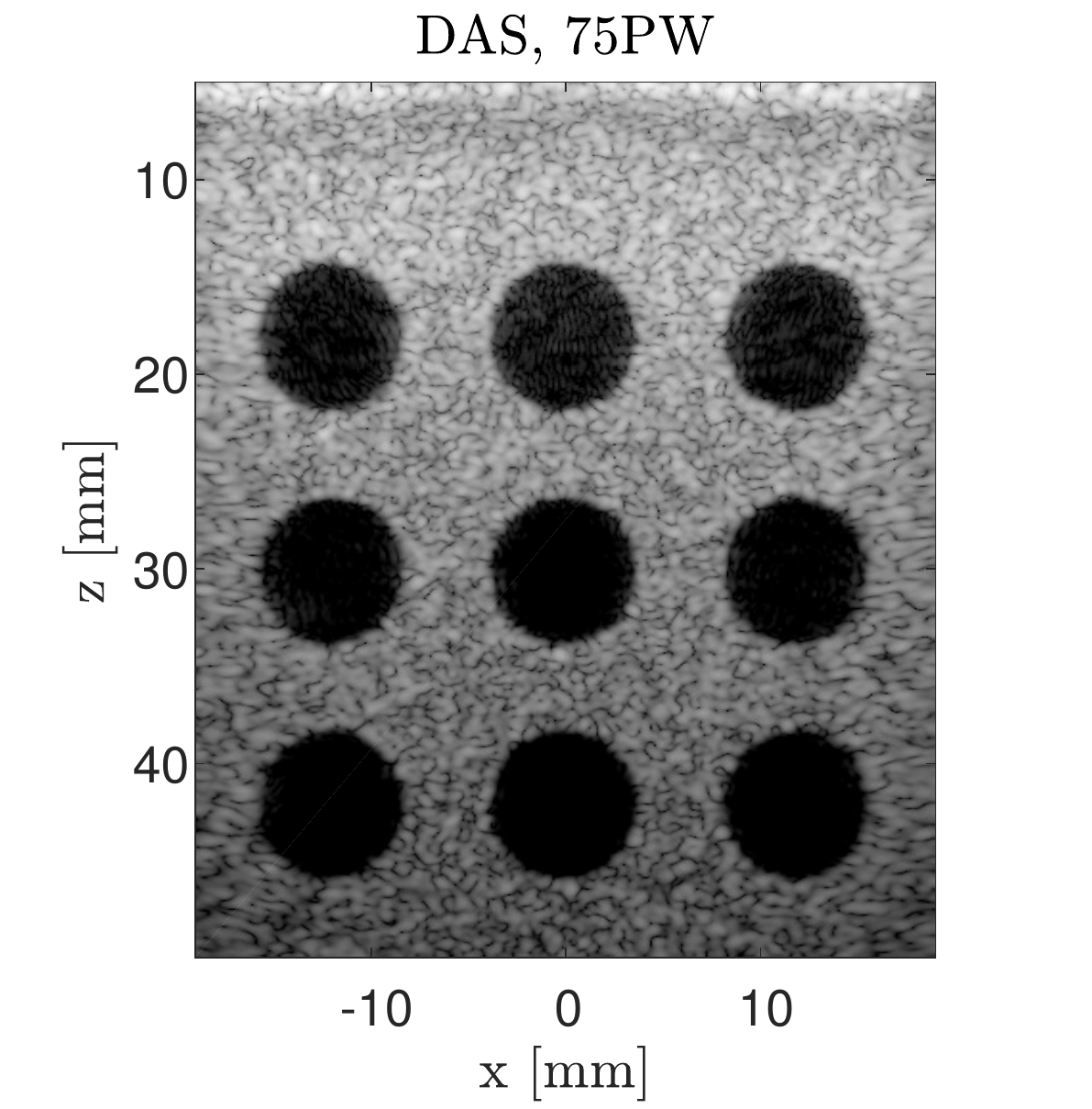}}%
  \centerline{(b)}\medskip
\end{minipage}
\begin{minipage}[b]{0.24\linewidth}
  \centering
  \centerline{\includegraphics[width=5cm]{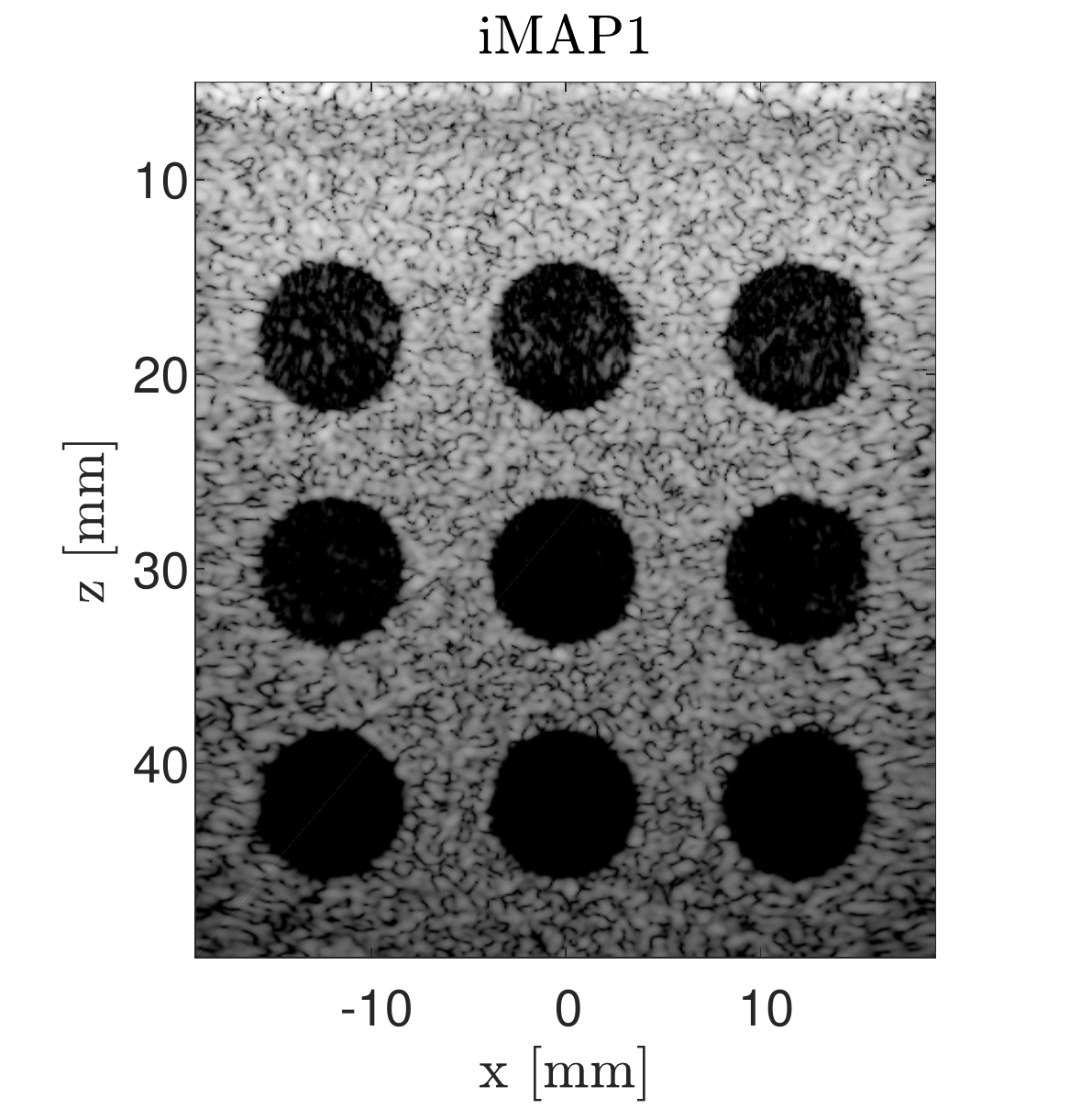}}%
  \centerline{(c)}\medskip
\end{minipage}
\begin{minipage}[b]{0.24\linewidth}
  \centering
  \centerline{\includegraphics[width=5cm]{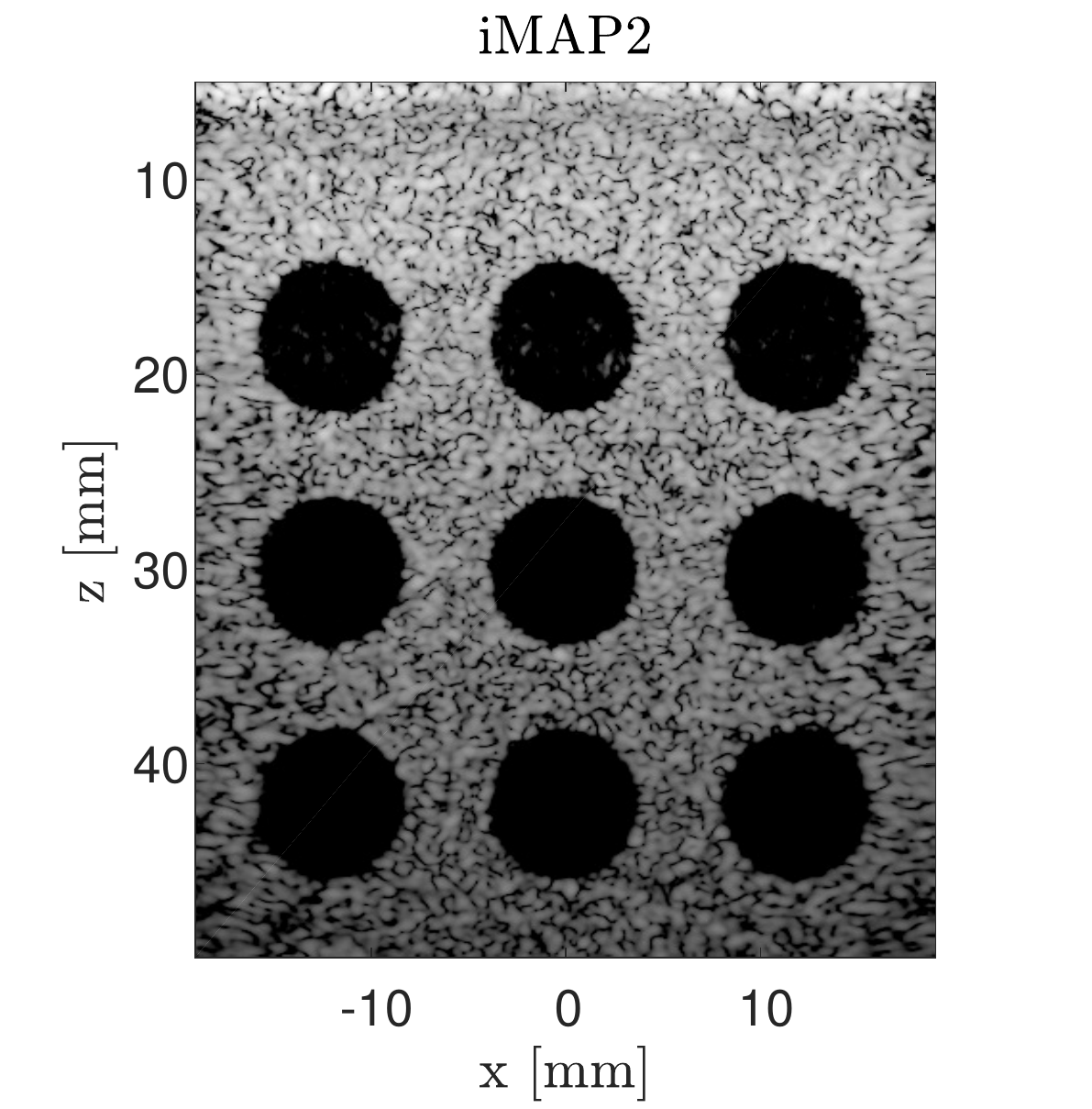}}%
  \centerline{(d)}\medskip
\end{minipage}\\
\begin{minipage}[b]{0.33\linewidth}
\centerline{\includegraphics[width=5cm]{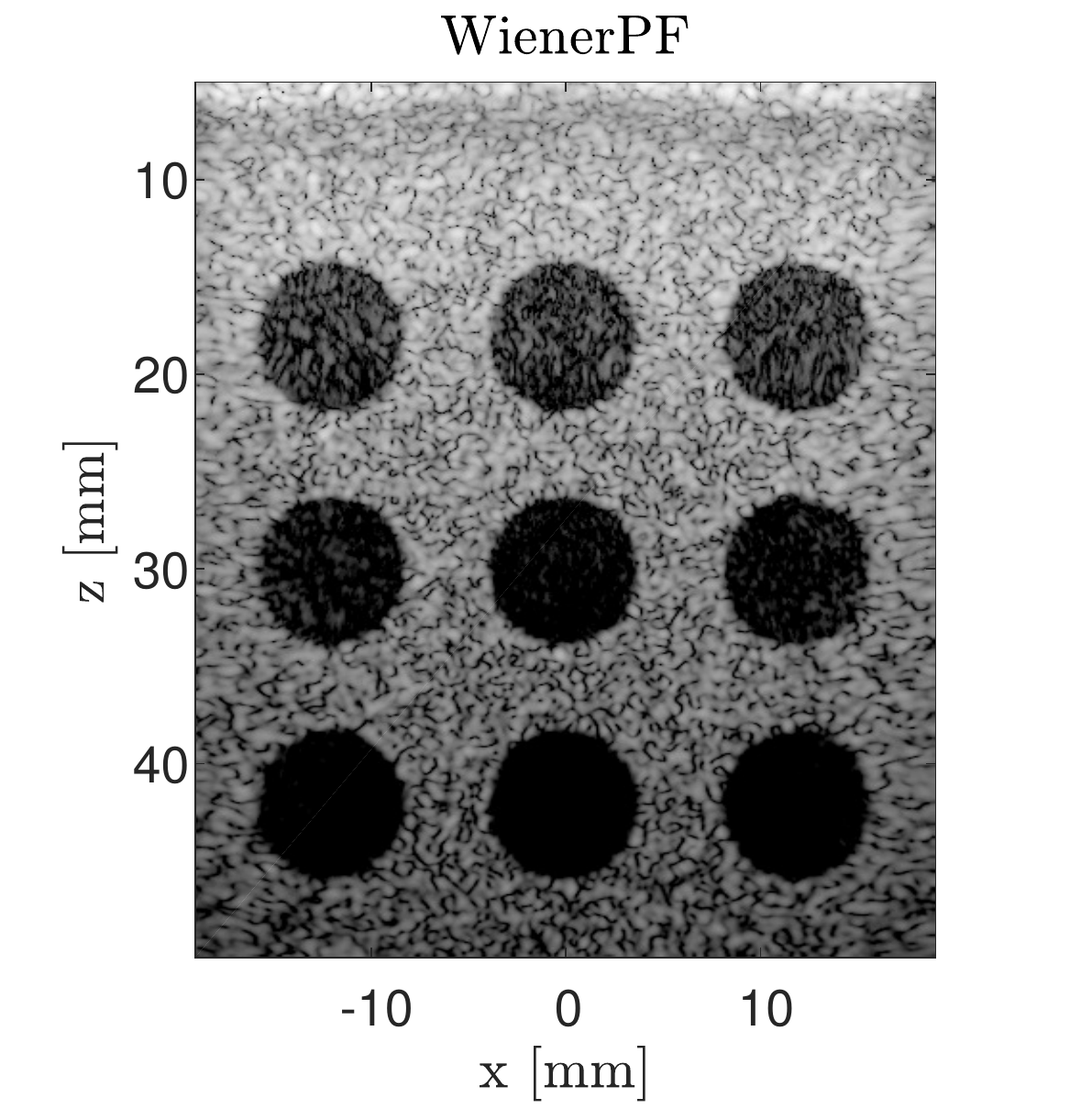}}%
  \centerline{(e)}\medskip
\end{minipage}
\begin{minipage}[b]{0.33\linewidth}
  \centerline{\includegraphics[width=5cm]{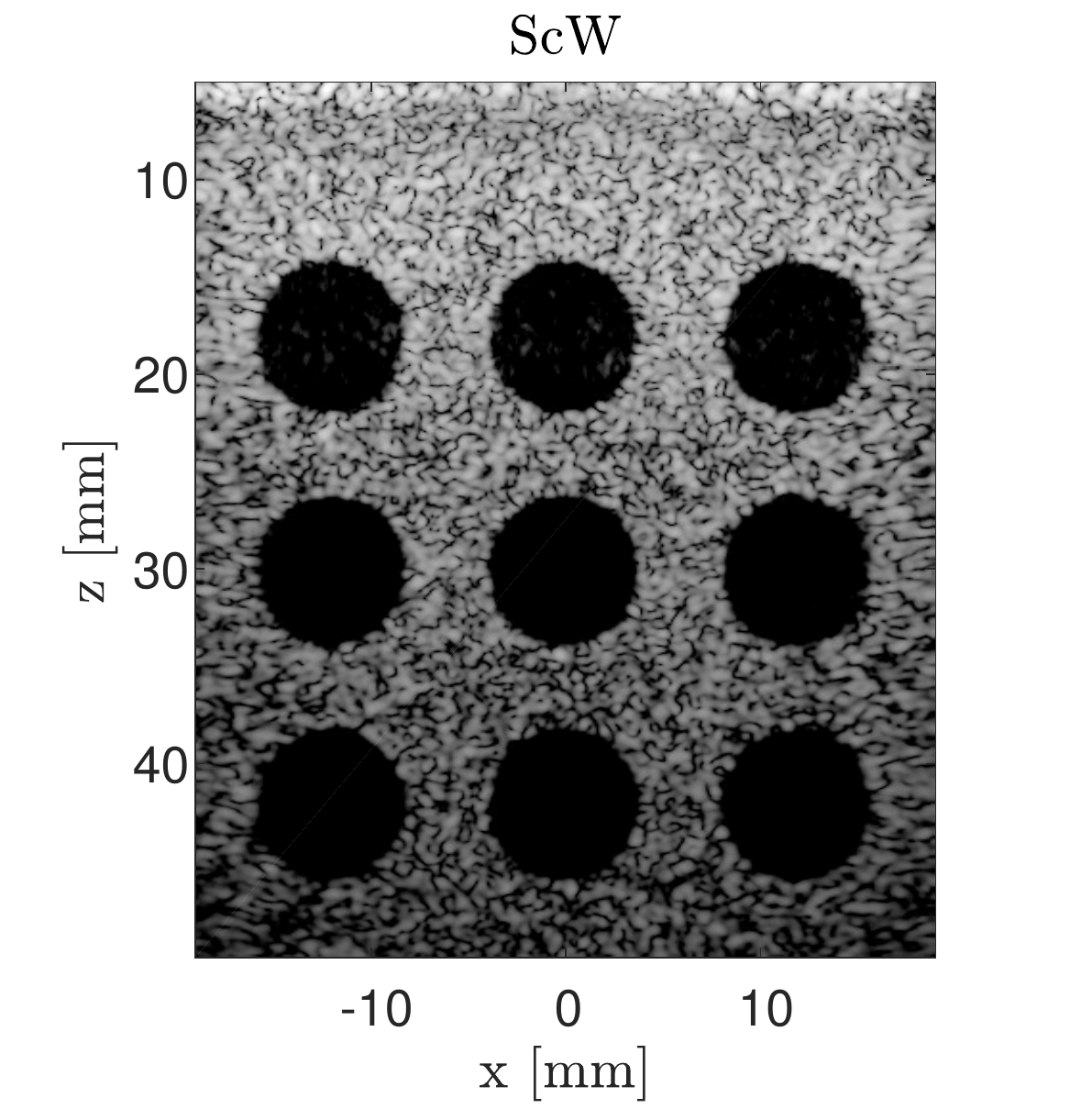}}%
  \centerline{(f)}\medskip
\end{minipage}
\begin{minipage}[b]{0.33\linewidth}
  \centerline{\includegraphics[width=5cm]{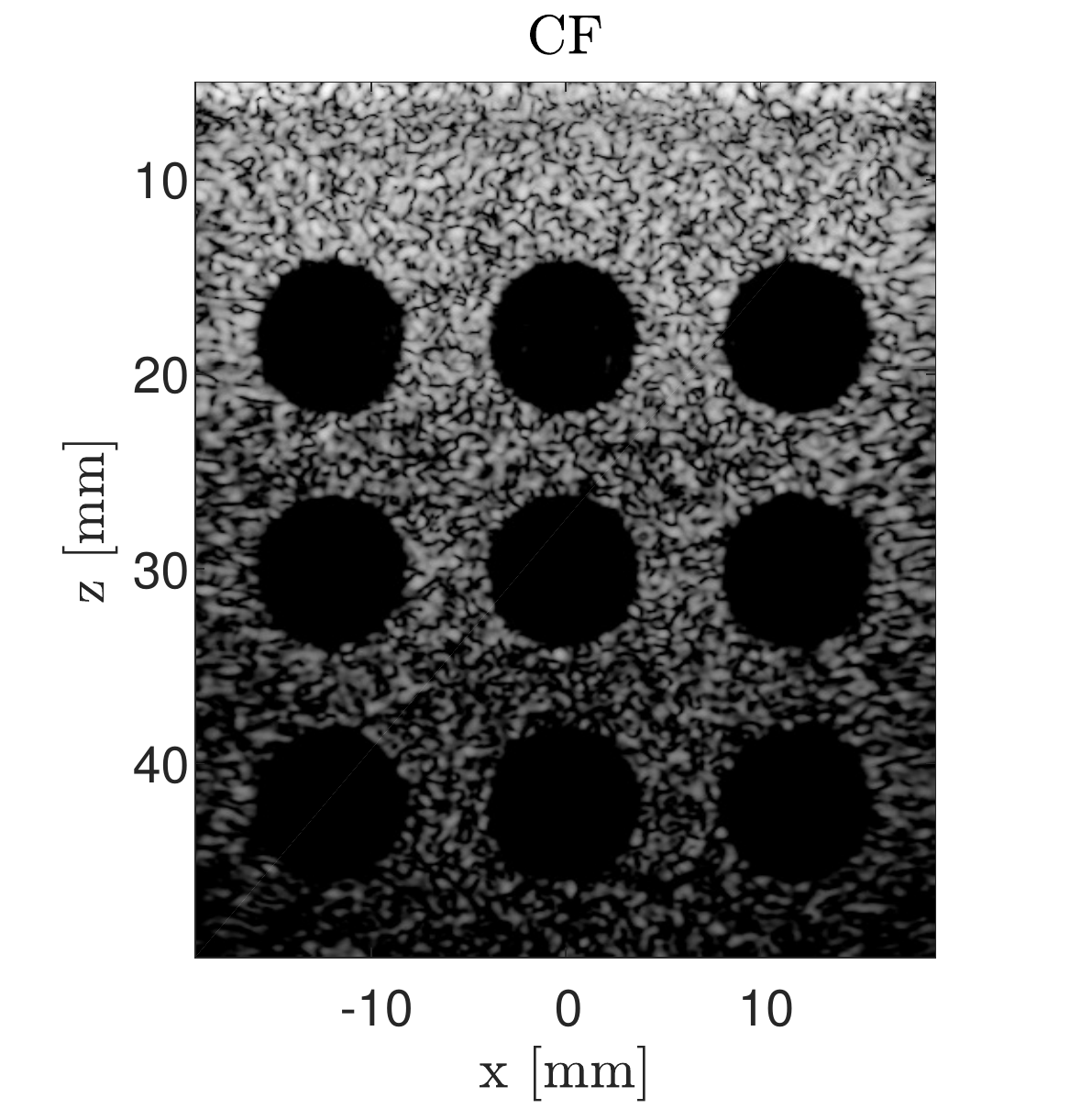}}%
  \centerline{(g)}\medskip
\end{minipage}
\caption{Images of  simulated anechoic cyst phantom obtained by (a) DAS with 13 plane-waves, (b) DAS with 75 plane-waves, (c) iMAP1 with 13 plane-waves, (d) iMAP2 with 13 plane-waves, (e) Wiener postfilter with 13 plane-waves, (f) ScW with 13 plane-waves, (g) CF with 13 plane-waves. All the images are presented with a dynamic range of 70 dB.}
\label{fig:simContrast_multiplePW}
\end{figure*}
The CNR obtained with Wiener postfilter is reduced compared to DAS, while iMAP1  provides $17.12\%$ improvement. In addition, it can be seen in Fig. \ref{fig:simLatScanline_PW1_iMAP}(a) that iMAP1 provides better interference suppression within the cyst while being closer to DAS result in the speckle regions in between the cysts.
The superior performance of iMAP1 compared to Wiener postfilter may be explained by poor estimation of the noise covariance matrix. The estimator of $\mathbf{R}_{n}$ in \eqref{eq:estimators for Wiener postfilter} does not consider an assumption of spatially white noise. This assumption is used by iMAP and leads to improved performance. This provides empirical justification for the spatially white interference model used in this work. 

We next compare the results of CF, ScW and iMAP2. All these techniques are characterized by a very sharp magnitude drop-off  at the transition from speckle to cyst regions, noted in Fig. \ref{fig:simLatScanline_PW1_iMAP}(b).  As a result,  cyst boundaries in Fig. \ref{fig:simContrast_1PW} are very clear even for an image obtained with a single plane-wave transmission.
As expected, an image obtained by CF suffers from severely reduced brightness. Even though noise  within the cyst is highly suppressed, the improvement in CNR obtained by CF is moderate due to reduced brightness of the surrounding area. The performance of ScW is improved compared to CF. Visually, the resulting image resembles the one obtained by iMAP2, however, the value of CNR for \mbox{iMAP2} is higher. Figure \ref{fig:simLatScanline_PW1_iMAP}(b) shows that iMAP2 outperforms both CF and ScW in terms of noise suppression within the cyst. Despite the superior noise suppression, the effect of \mbox{iMAP2} within the speckle regions is less prominent compared to CF and ScW.
This shows that iMAP2 does not suppress the magnitude uniformly over all image regions and has better adaptive capabilities compared to CF and ScW.

To evaluate the performance of different methods for transmission of multiple plane-waves with different inclinations we present the CNR as a function of the number of transmitted plane-waves,  varying from 3 to 75.
The angles are spaced uniformly between $-16^{\circ}$ and $16^{\circ}$ to keep the lateral resolution comparable for different number of plane-waves.
As can be seen in Fig.~\ref{fig:CNR}~(a), the CNR increases with the number of transmitted plane-waves for all approaches. In terms of CNR, iMAP for both one and two iterations outperforms other methods being studied.  
\mbox{iMAP1} provides on average 2.17 dB improvement compared to DAS. For \mbox{iMAP2}, the average improvement is 4.82 dB. For CF and ScW the average improvement is 0.91 and 1.84 dB, respectively.
  
We note that only 13 plane-waves are required for \mbox{iMAP1} to yield the contrast compared to that of DAS with 75 transmissions. 
Figures \ref{fig:simContrast_multiplePW}(a) and (b) present images obtained by DAS from 13 and 75 transmissions. Images formed from 13 plane-waves using iMAP1, iMAP2, Wiener postfilter, ScW and CF  are shown in Figs. \ref{fig:simContrast_multiplePW}(c), (d), (e), (f) and (g), respectively.

%
\begin{figure*}[t]
\begin{minipage}[b]{0.24\linewidth}
  \centering
  \centerline{\includegraphics[width=5cm]{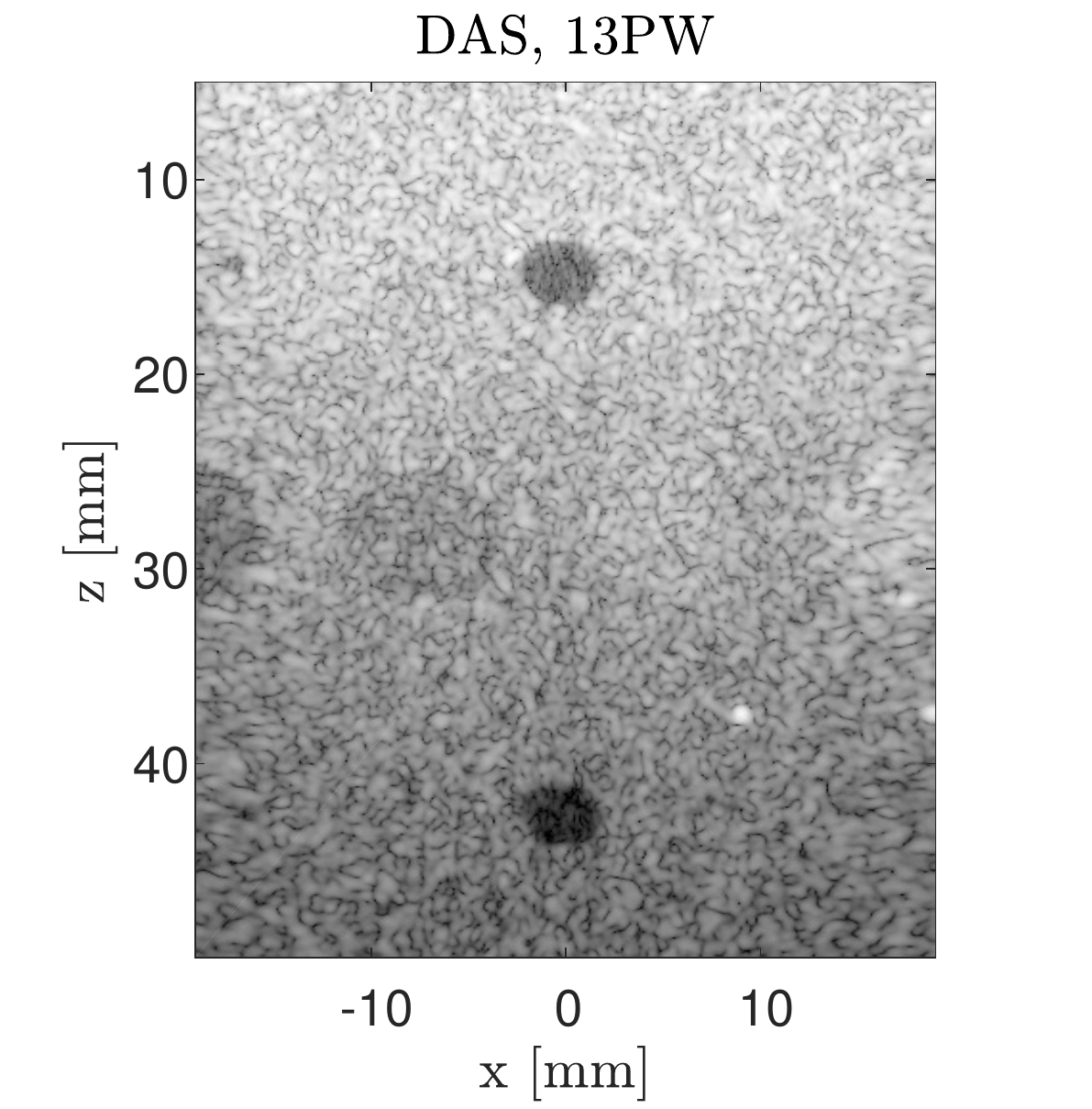}}%
  \centerline{(a)}\medskip
\end{minipage}
\begin{minipage}[b]{0.24\linewidth}
  \centering
  \centerline{\includegraphics[width=5cm]{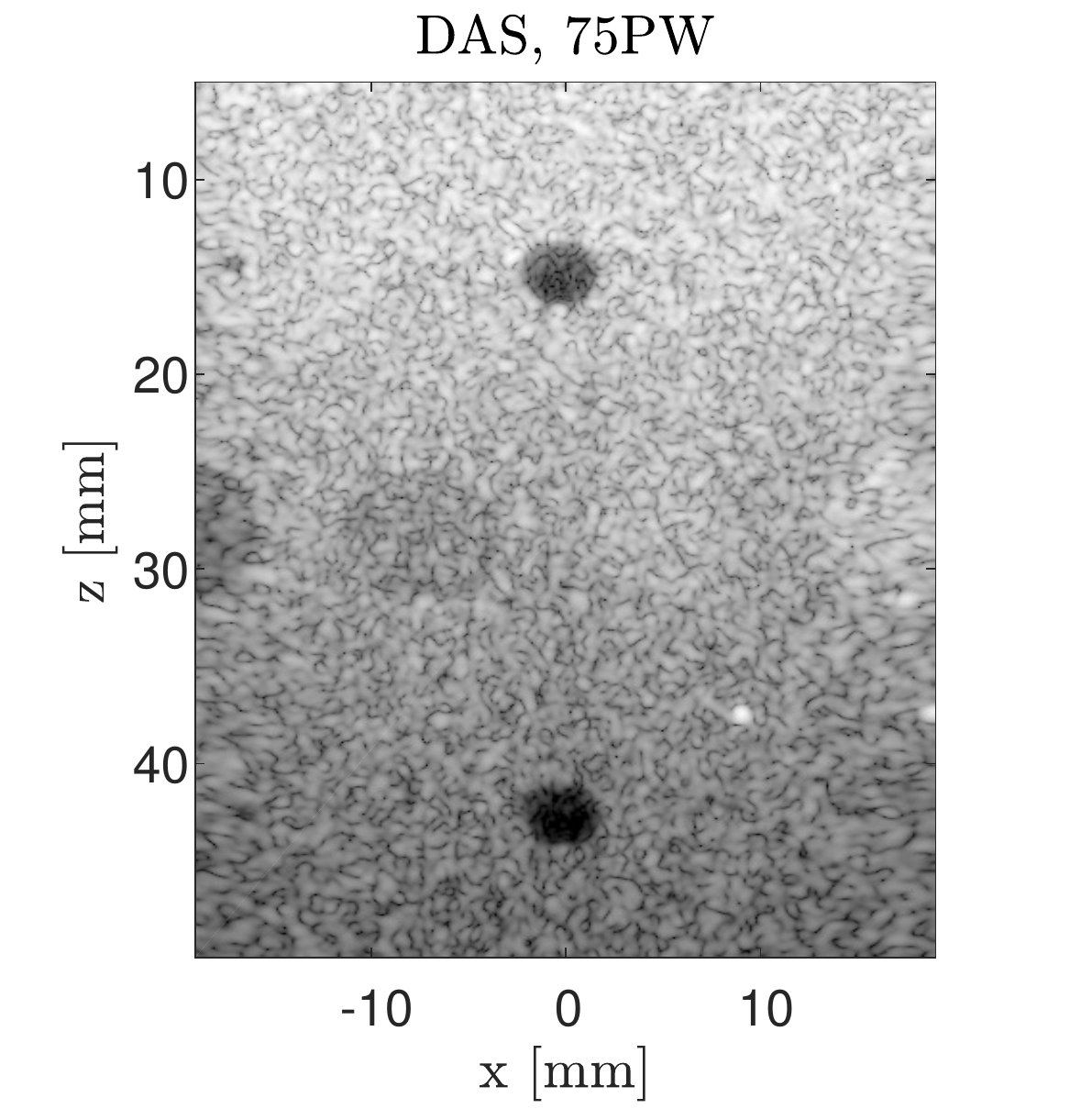}}%
  \centerline{(b)}\medskip
\end{minipage}
\begin{minipage}[b]{0.24\linewidth}
  \centering
  \centerline{\includegraphics[width=5cm]{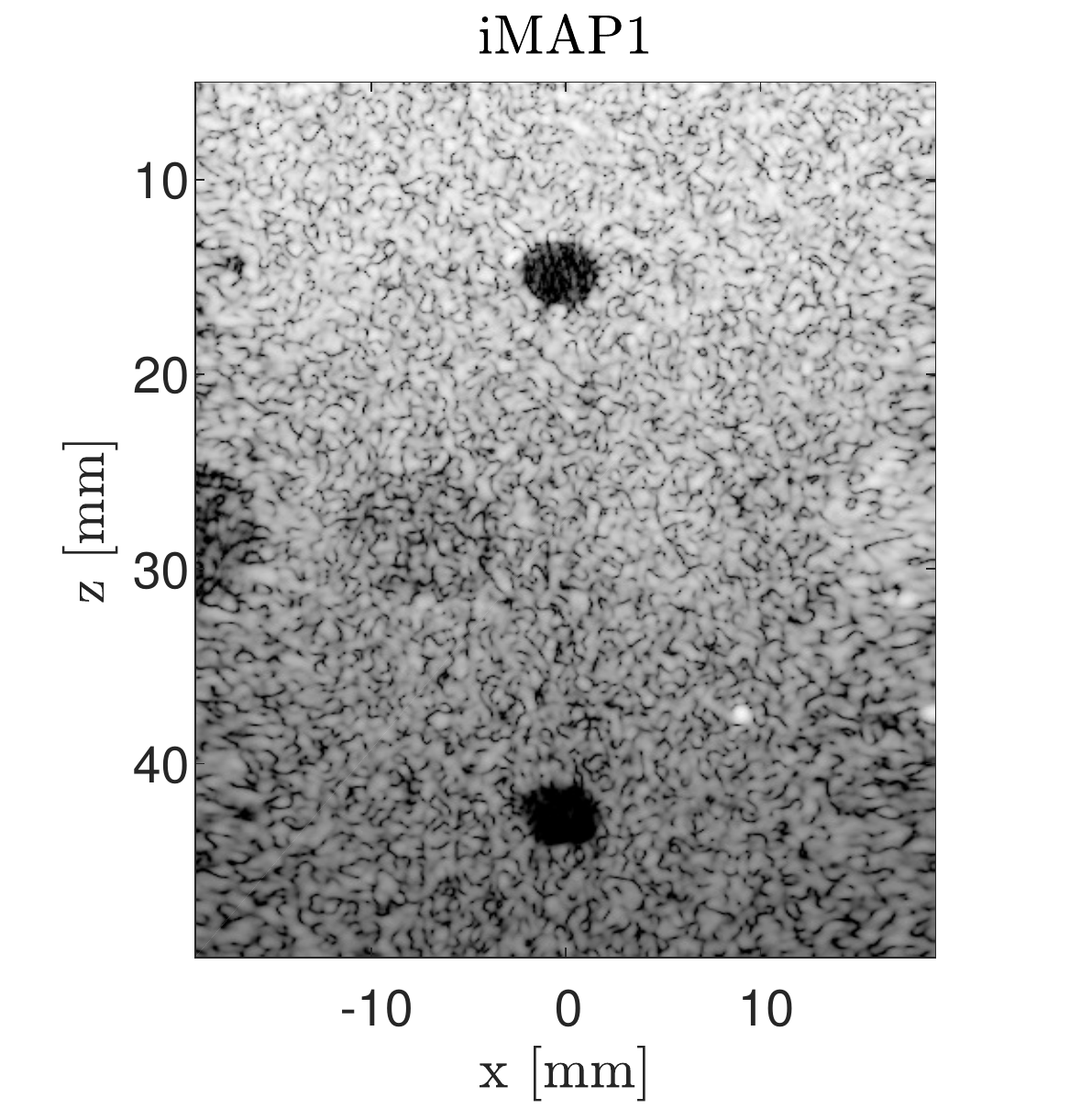}}%
  \centerline{(c)}\medskip
\end{minipage}
\begin{minipage}[b]{0.24\linewidth}
  \centering
  \centerline{\includegraphics[width=5cm]{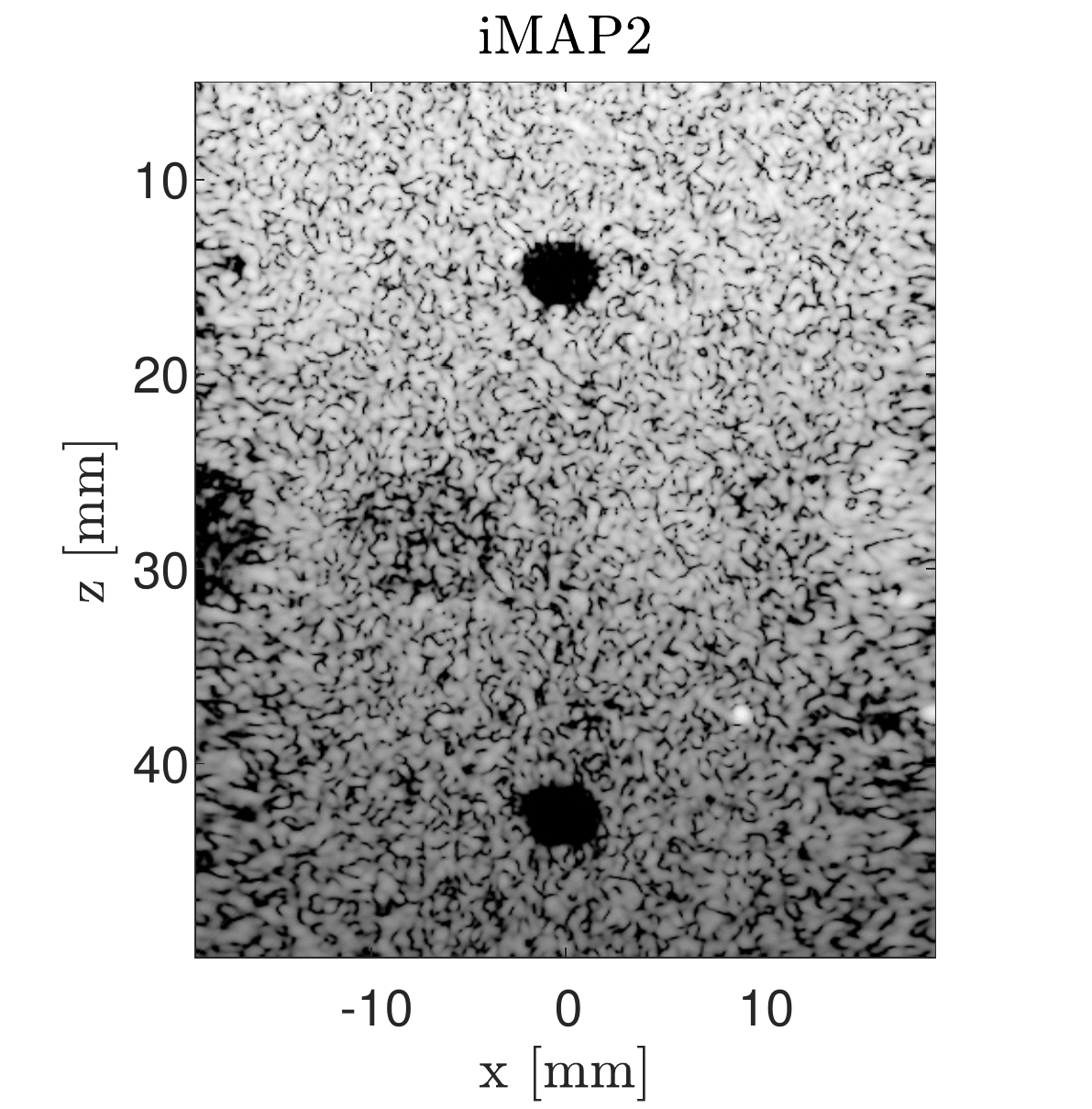}}%
  \centerline{(d)}\medskip
\end{minipage}\\
\begin{minipage}[b]{0.33\linewidth}
  \centerline{\includegraphics[width=5cm]{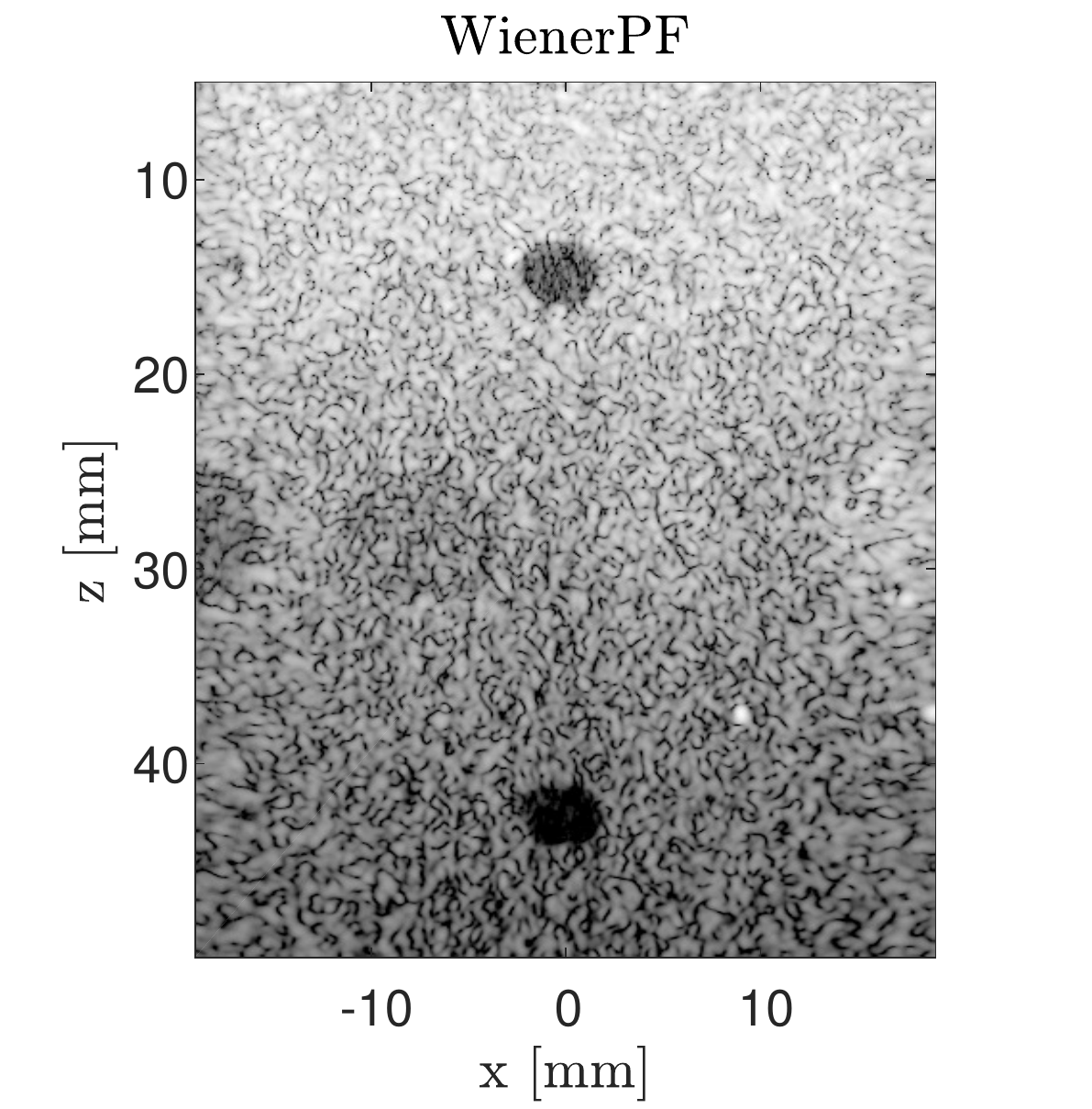}}%
  \centerline{(e)}\medskip
\end{minipage}
\begin{minipage}[b]{0.33\linewidth}
  \centerline{\includegraphics[width=5cm]{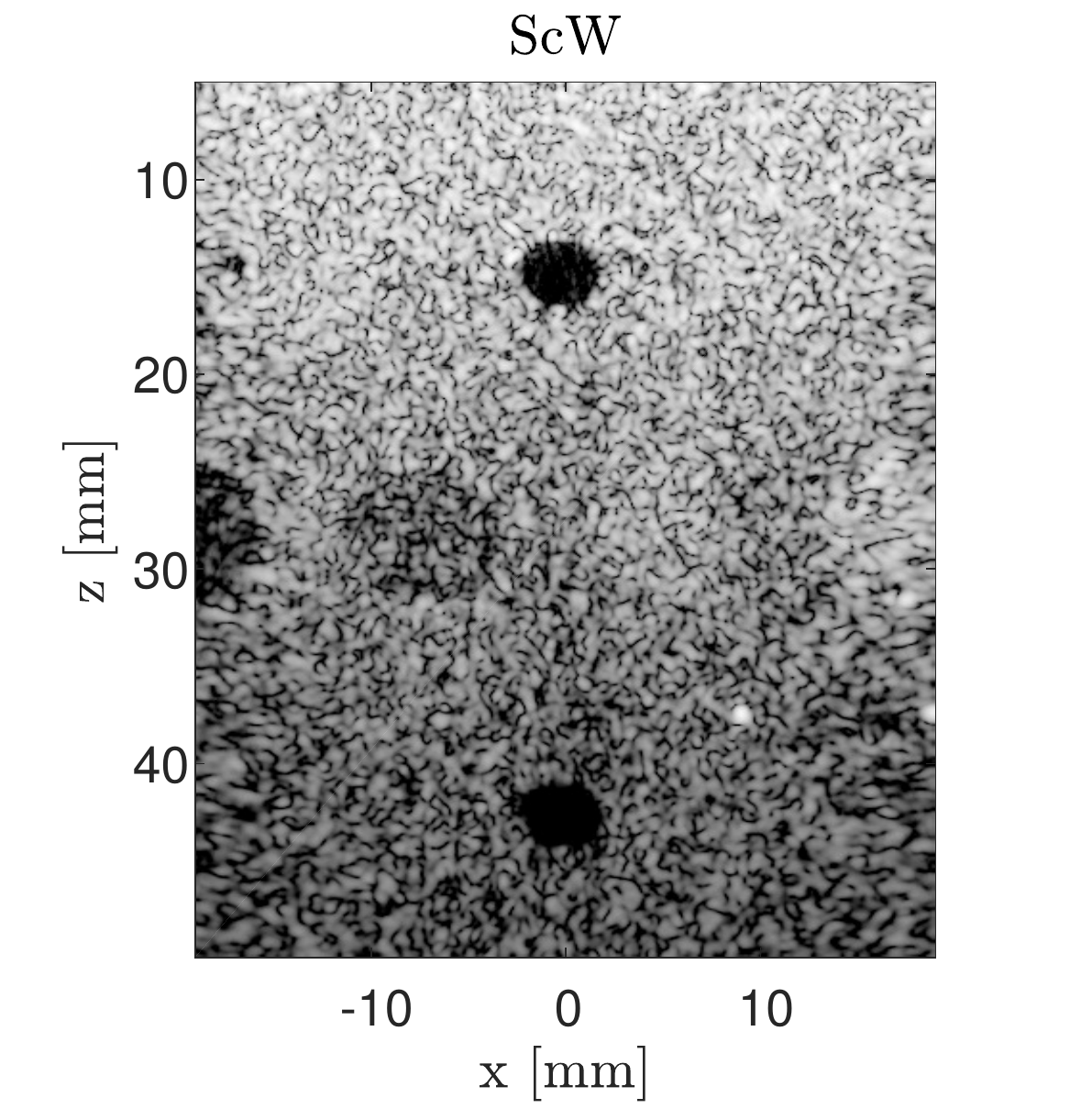}}%
  \centerline{(f)}\medskip
  \end{minipage}
\begin{minipage}[b]{0.33\linewidth}
  \centerline{\includegraphics[width=5cm]{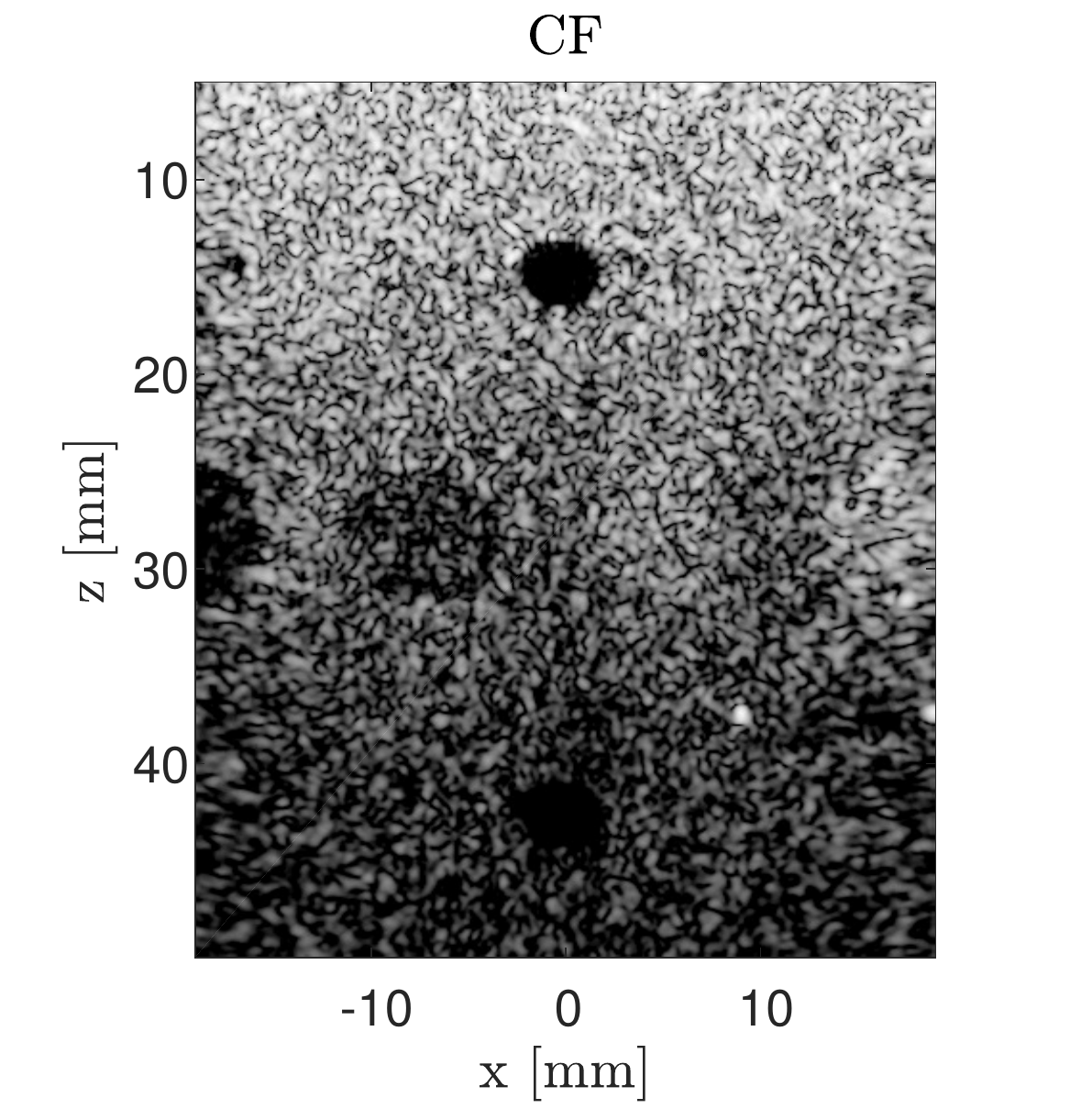}}%
  \centerline{(g)}\medskip
\end{minipage}
\caption{Images of  anechoic cyst phantom obtained by (a) DAS with 13 plane-waves, (b) DAS with 75 plane-waves, (c) iMAP1 with 13 plane-waves, (d) iMAP2 with 13 plane-waves, (e) Wiener postfilter with 13 plane-waves, (f) ScW with 13 plane-waves, (g) CF with 13 plane-waves. All the images are presented with a dynamic range of 70 dB.}
\label{fig:expContrast_multiplePW}
\end{figure*}
%
\begin{figure*}[h!]
\begin{minipage}[b]{0.48\linewidth}
  \centering
  \centerline{\includegraphics[width=9cm]{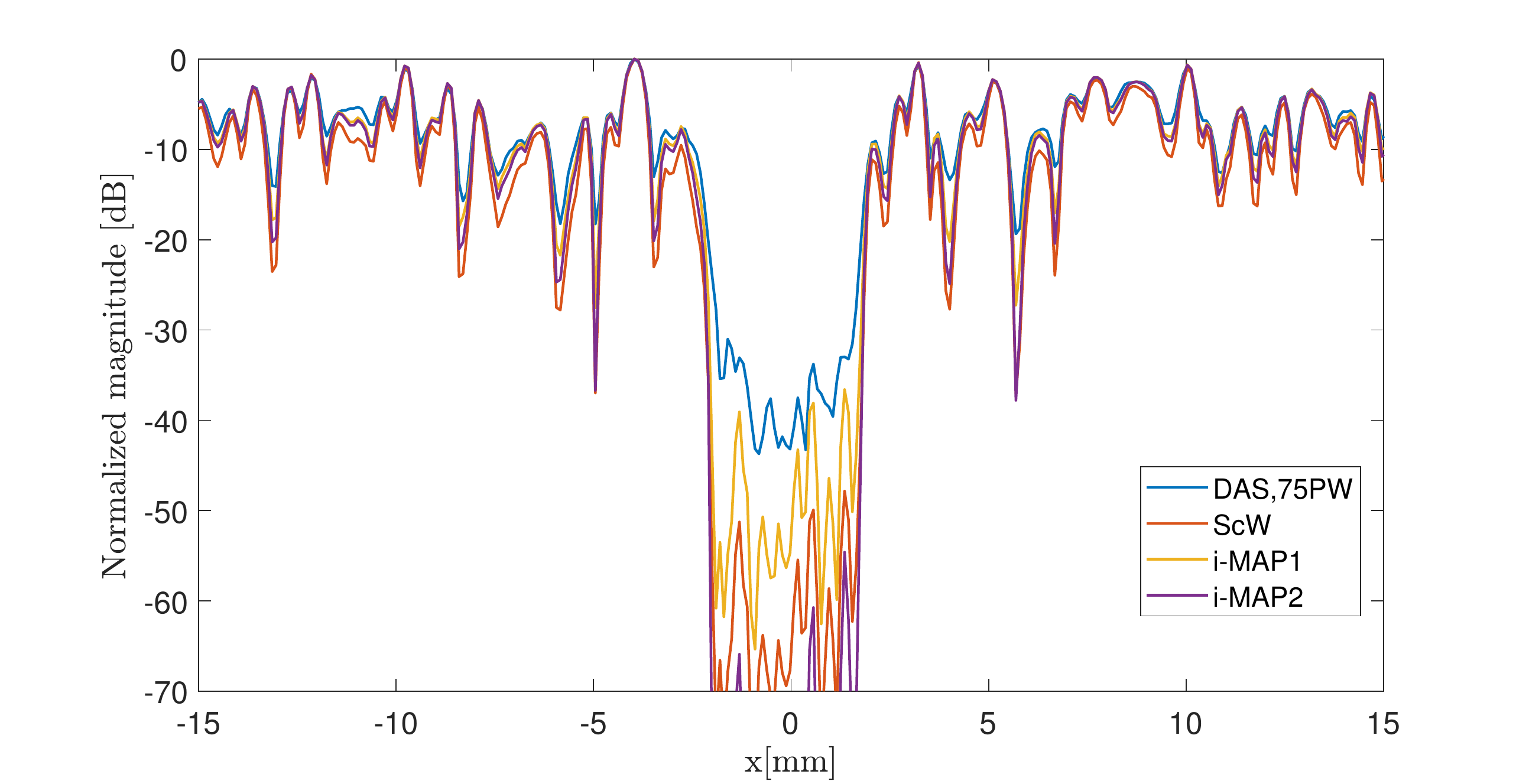}}%
  \centerline{(a)}\medskip
\end{minipage}
\begin{minipage}[b]{0.48\linewidth}
  \centering
  \centerline{\includegraphics[width=9cm]{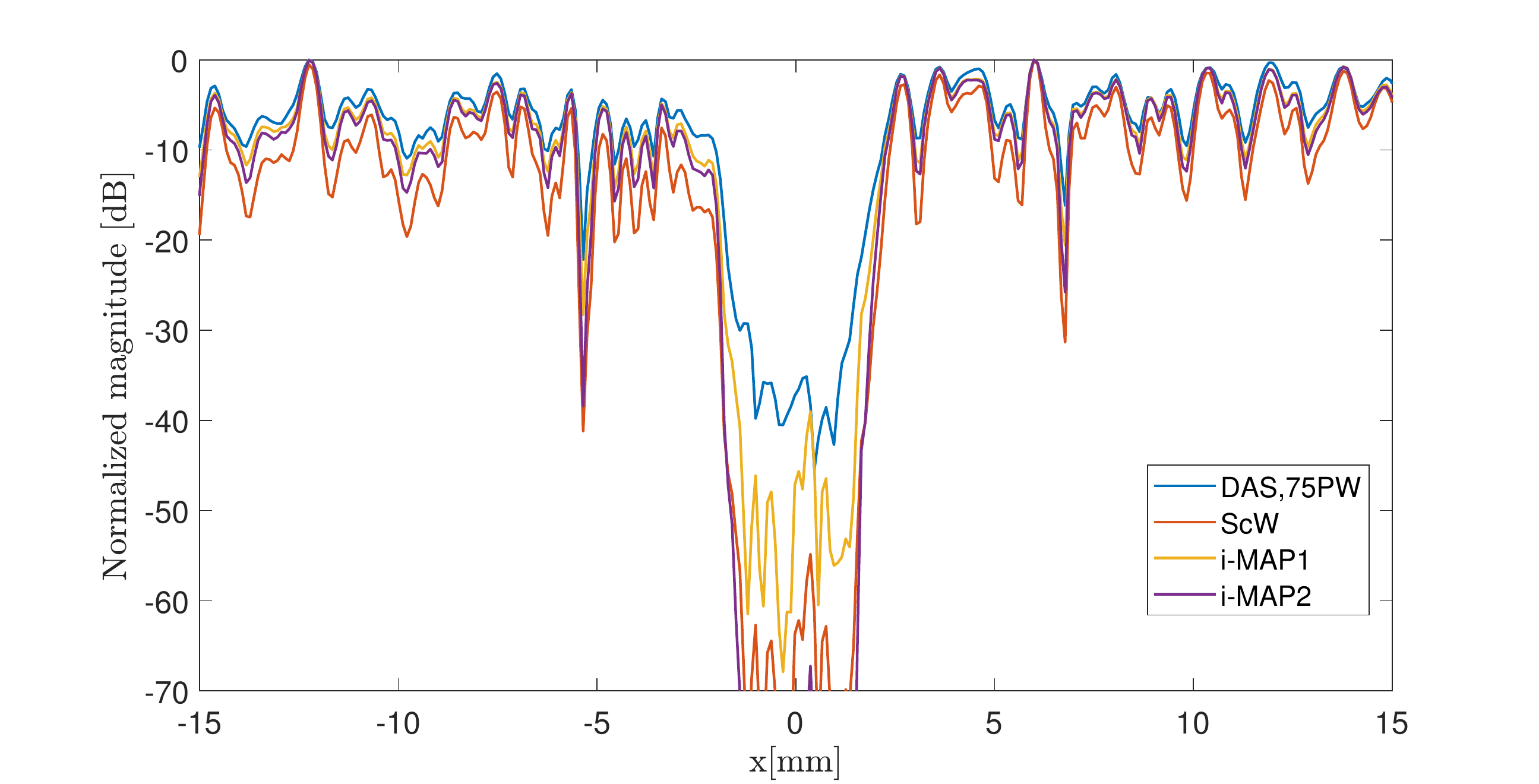}}%
  \centerline{(b)}\medskip
\end{minipage}
\caption{Lateral cross-section of the two cysts from Fig. \ref{fig:expContrast_multiplePW}, (a) upper cyst (b) lower cyst. iMAP1, iMAP2 and ScW are using 13 transmitted plane-waves. DAS is performed with 75 transmissions.}
\label{fig:expLatScanline_PW1_iMAP}
\end{figure*}
%
  %
\begin{figure*}[h!]
\begin{minipage}[b]{0.24\linewidth}
  \centering
  \centerline{\includegraphics[width=5cm]{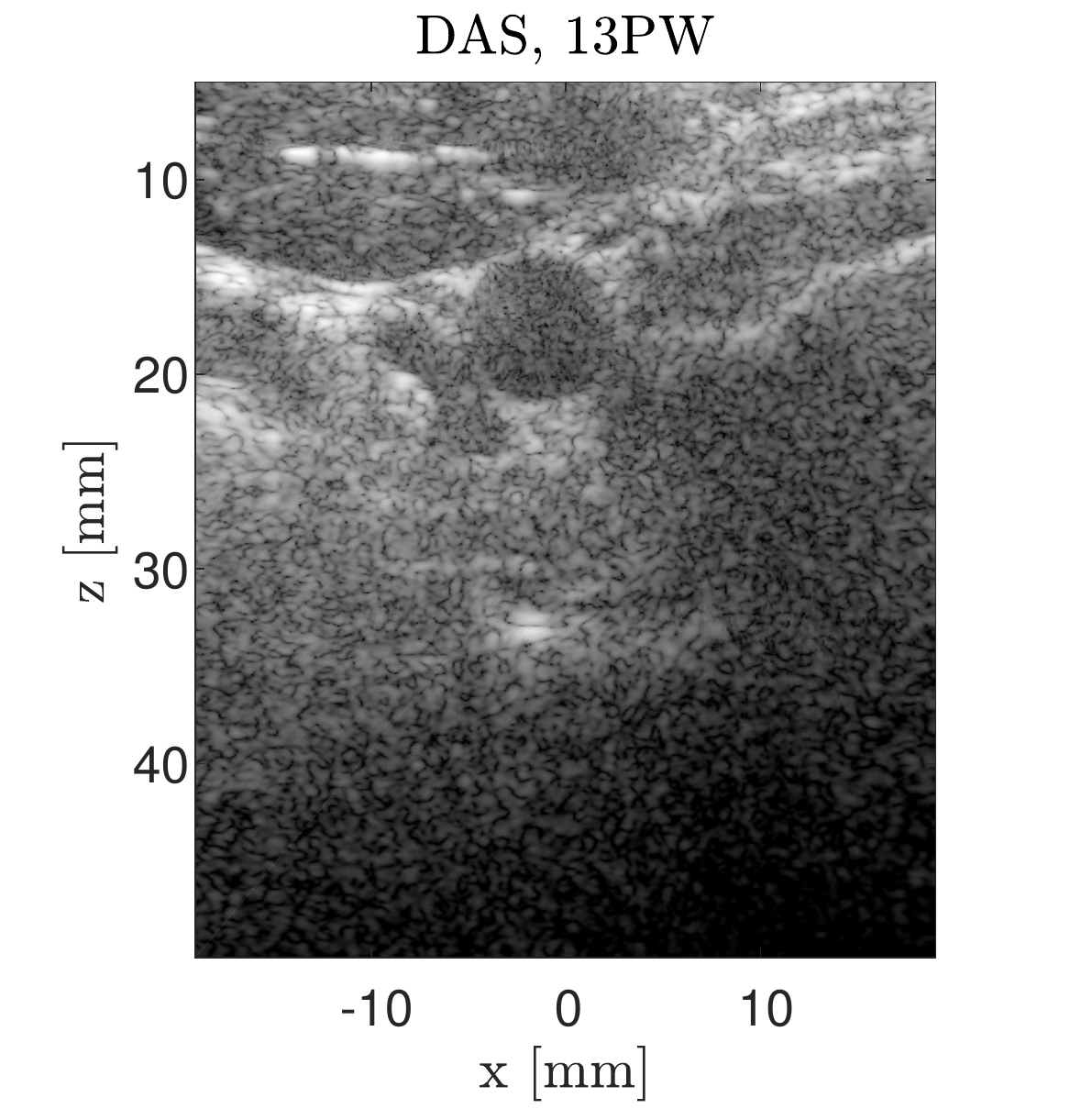}}%
  \centerline{(a)}\medskip
\end{minipage}
\begin{minipage}[b]{0.24\linewidth}
  \centering
  \centerline{\includegraphics[width=5cm]{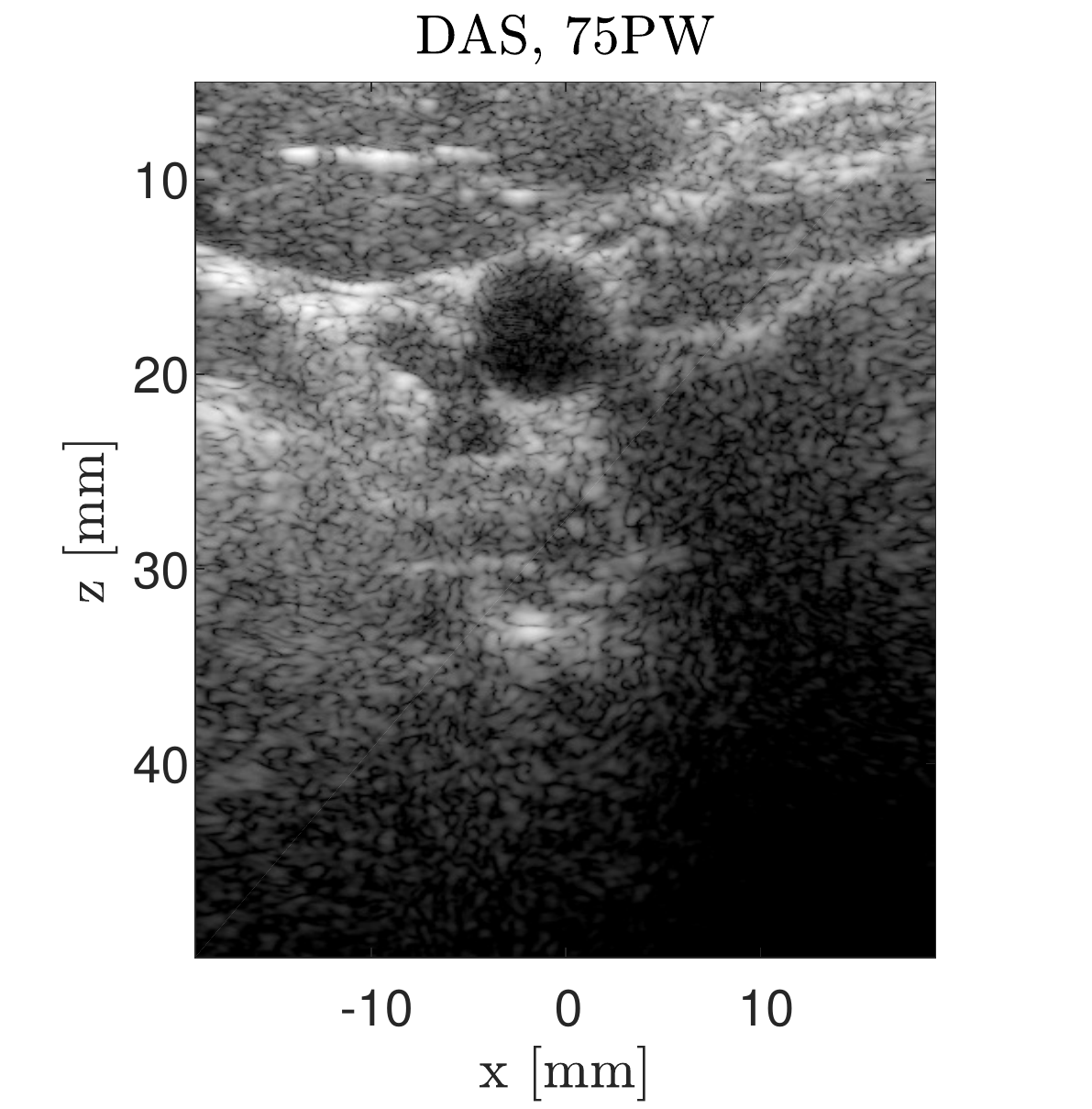}}%
  \centerline{(b)}\medskip
\end{minipage}
\begin{minipage}[b]{0.24\linewidth}
  \centering
  \centerline{\includegraphics[width=5.2cm]{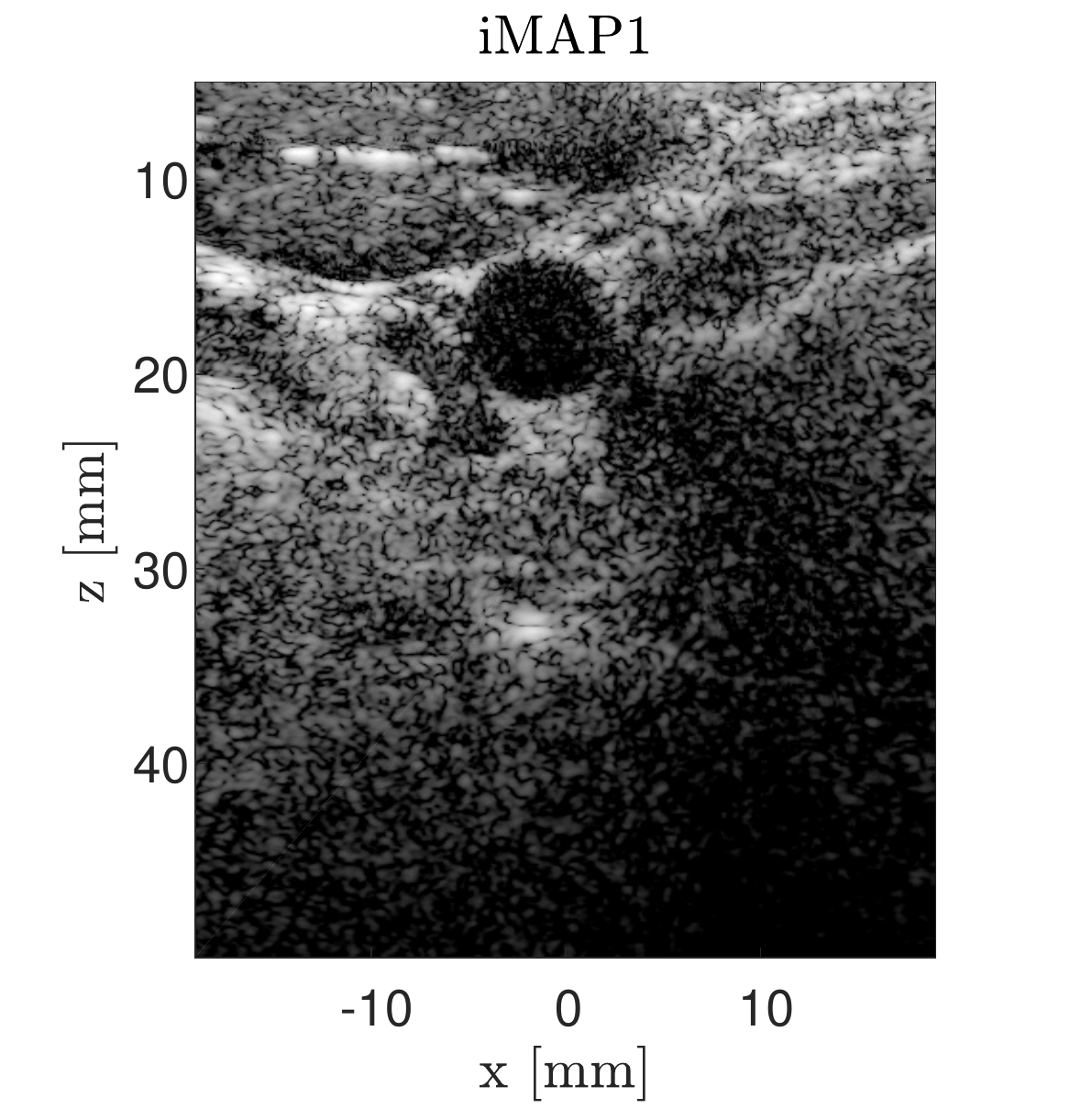}}%
  \centerline{(c)}\medskip
\end{minipage}
\begin{minipage}[b]{0.24\linewidth}
  \centering
  \centerline{\includegraphics[width=5.2cm]{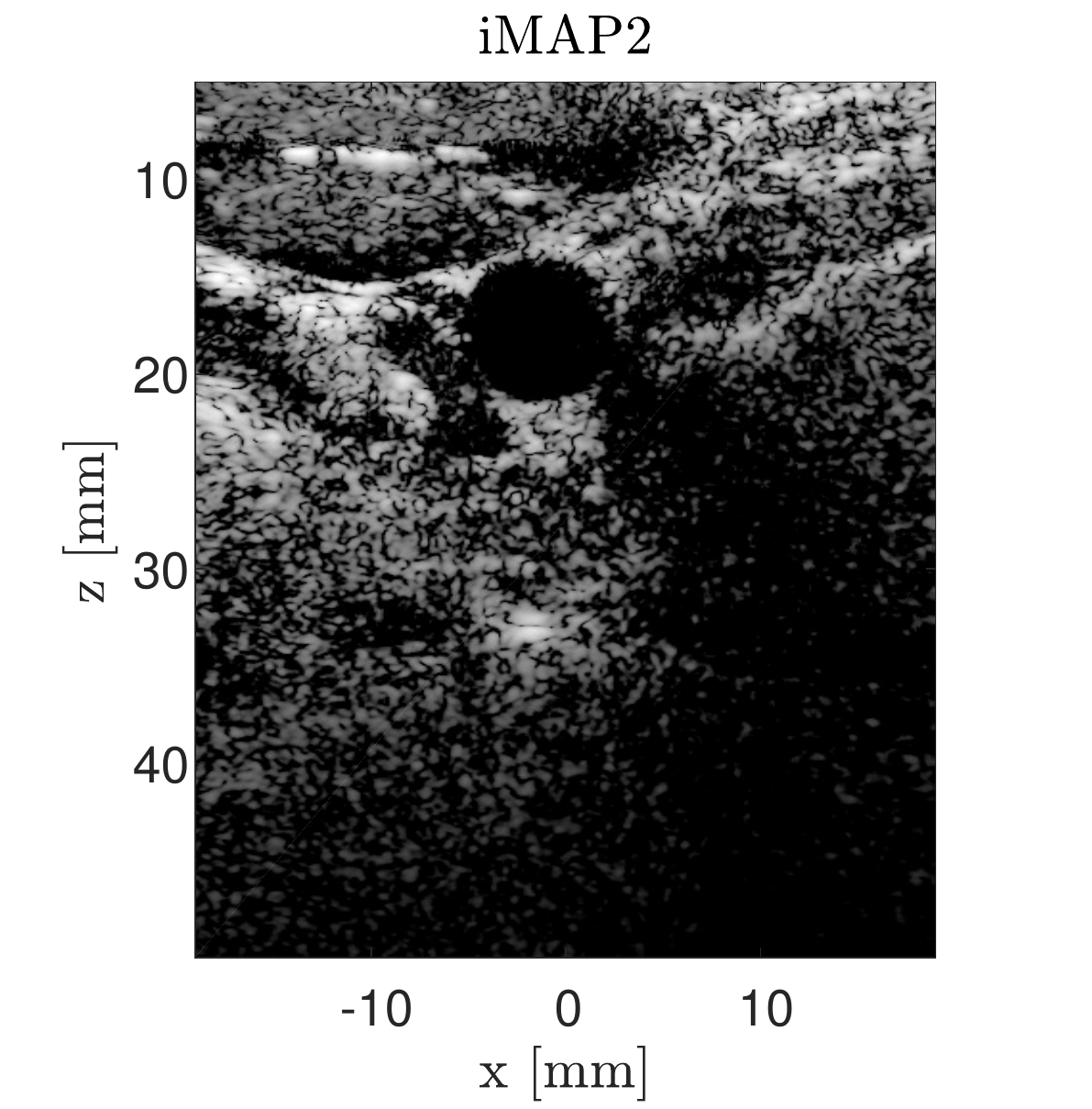}}%
  \centerline{(d)}\medskip
\end{minipage}\\
\begin{minipage}[b]{0.33\linewidth}
  \centerline{\includegraphics[width=5cm]{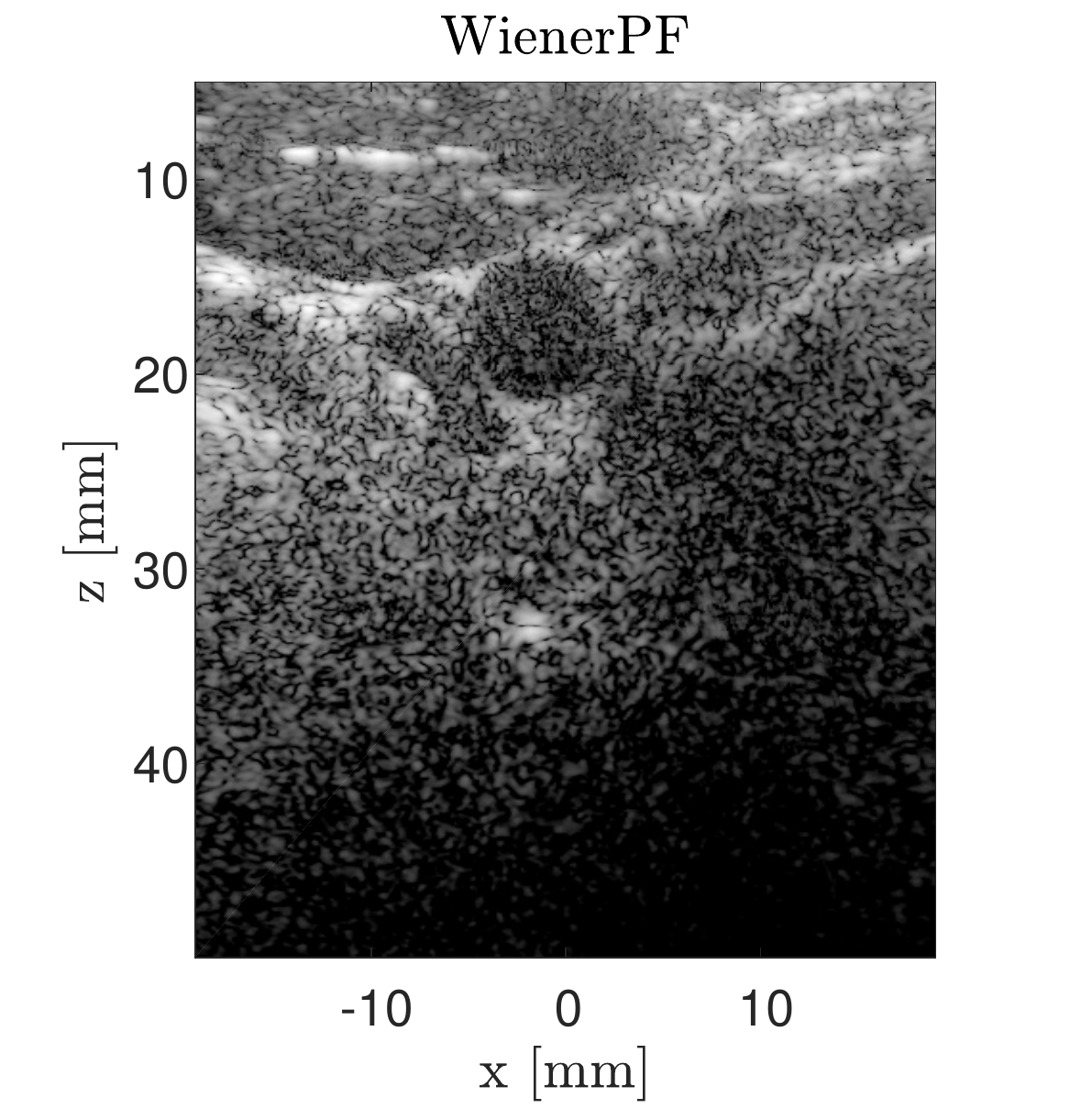}}%
  \centerline{(e)}\medskip
\end{minipage}
\begin{minipage}[b]{0.33\linewidth}
  \centerline{\includegraphics[width=5cm]{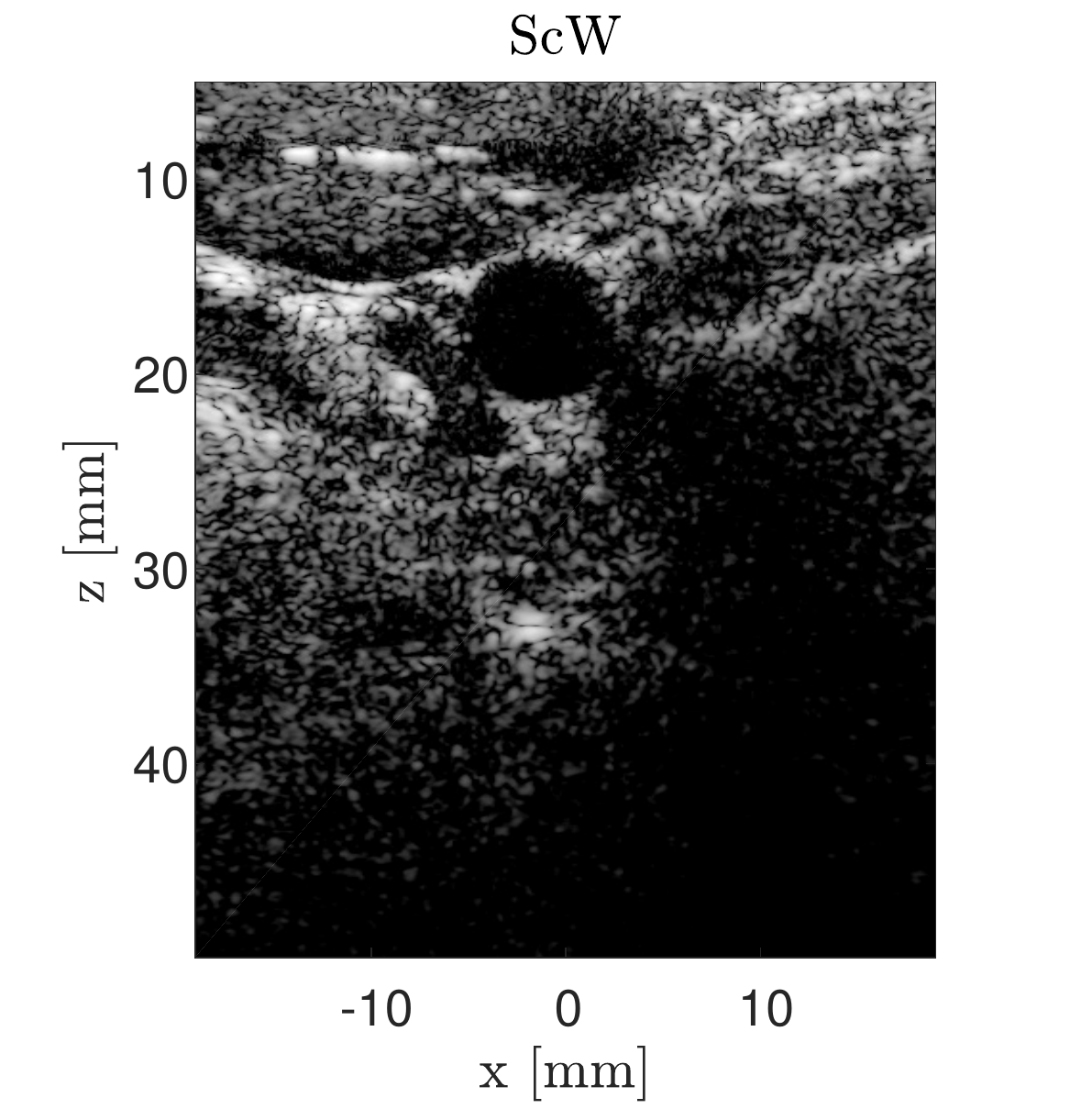}}%
  \centerline{(f)}\medskip
\end{minipage}
\begin{minipage}[b]{0.33\linewidth}
  \centerline{\includegraphics[width=5cm]{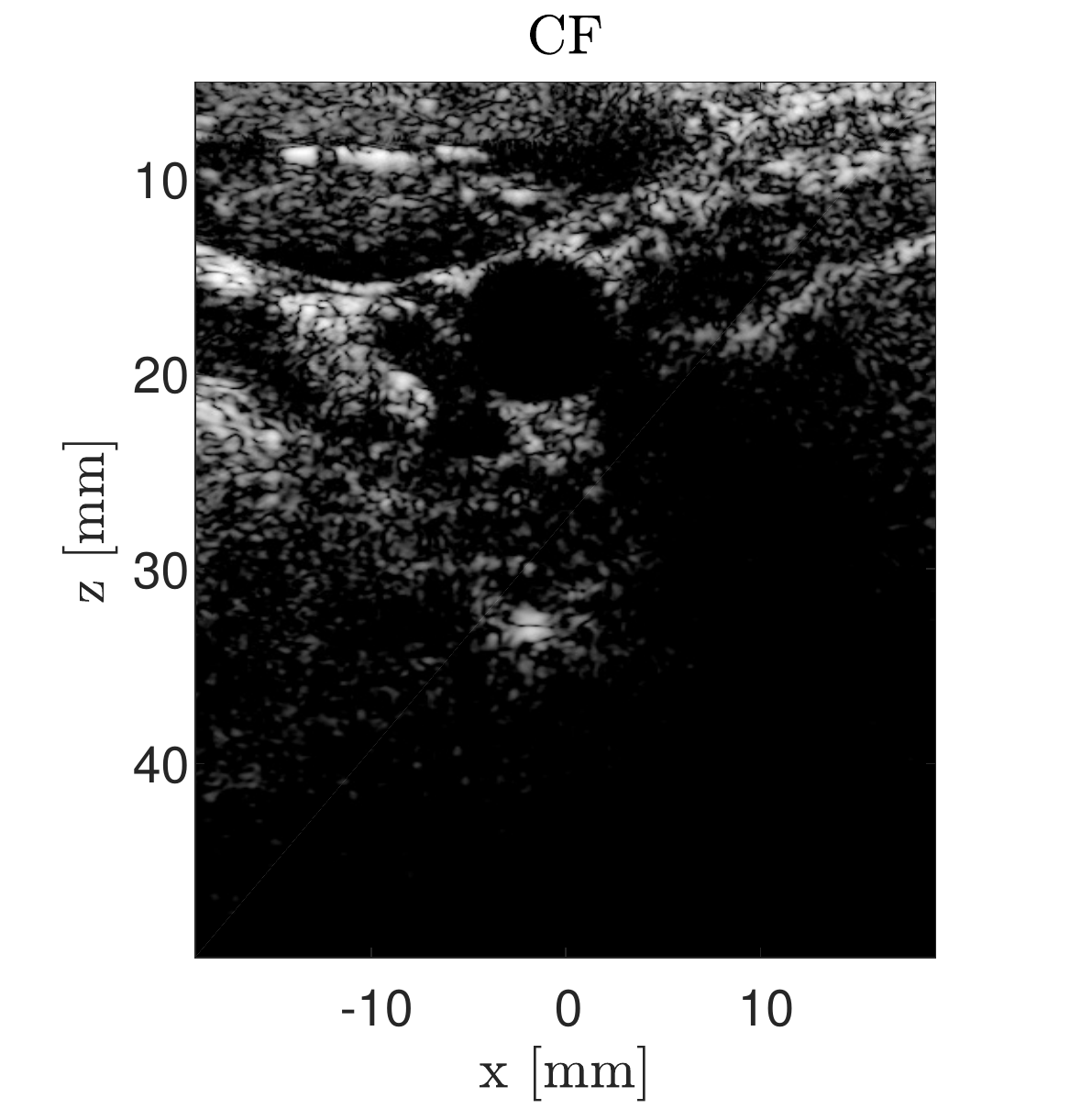}}%
  \centerline{(g)}\medskip
\end{minipage}
\caption{Images of  a cross section of a carotid artery obtained by (a) DAS with 13 plane-waves, (b) DAS with 75 plane-waves, (c) iMAP1 with 13 plane-waves, (d) iMAP2 with 13 plane-waves, (e) Wiener postfilter with 13 plane-waves, (f) ScW with 13 plane-waves, (g) CF with 13 plane-waves. All the images are presented with a dynamic range of 70 dB.}
\label{fig:carCross}
\end{figure*}

The similarity of speckle regions for images in Fig. \ref{fig:simContrast_multiplePW} is evaluated using the K-S test as elaborated in Section \ref{ssec:exp setup}. We use the DAS image with 75 transmissions as reference. The similarity values are presented in Table \ref{table:speckle similarity}. Evidently, the similarity of CF is extremely low, explicitly, less than one percent of the resulting pattern is defined as speckle by K-S test. The first iteration of iMAP yields the highest similarity of 74.12\%, while iMAP2 yields a similarity of 62.45\% and is comparable to Wiener postfilter. These results are in good agreement with the lateral cross-section of the cyst phantom obtained with a single transmission presented in Fig. \ref{fig:simLatScanline_PW1_iMAP}.

 \begin{table} [h!]
\caption{Similarity of speckle regions for different methods}
\label{table:speckle similarity}
\begin{center}
\hspace{-0.5 cm}
\begin{tabular}{| c | c | c |}
\hline
\textbf{Method} & \textbf{Simulation (Fig. \ref{fig:simContrast_multiplePW}), [\%]} & \textbf{Experiment (Fig. \ref{fig:expContrast_multiplePW}), [\%]} \\
\hline
{Wiener Postfilter}  &{63.36}       & {70.50}    \\ \hline
{iMAP1}                      &{74.12}       & {90.02} \\ \hline
{iMAP2}                      &{62.45}         &{61.26}\\ \hline
{ScW}                            &{25.10}         &{24.00}  \\ \hline
{CF}                               &{0.062}       &{1.52}\\ \hline
\end{tabular}
\end{center}
\end{table}

 \subsection{Experiments}
\label{ssec:experiments}
We next proceed to experimental data obtained by scanning a phantom containing both anechoic inclusions and a point reflector to allow for evaluation of both contrast and resolution. 

We start with the evaluation of contrast. A graph of  CNR as a function of the number of plane-waves is presented in Fig. \ref{fig:CNR}(b). Similar to simulated data, the contrast of iMAP1 using 13 plane-waves is comparable to that of DAS with 75 plane-waves. 
Figures \ref{fig:expContrast_multiplePW}(a) and (b) present images obtained by DAS from 13 and 75 transmissions. Images formed with 13 plane-waves using iMAP1, iMAP2, Wiener postfilter, ScW and CF  are shown in Figs. \ref{fig:expContrast_multiplePW}(c), (d), (e), (f) and (g), respectively. Similar to simulation results, iMAP with one and two iterations and ScW provide prominent improvement of contrast. To take a closer look at the performance of the above methods, lateral cross-sections of upper and lower cysts of Fig.~\ref{fig:expContrast_multiplePW} are presented in Fig. \ref{fig:expLatScanline_PW1_iMAP}(a) and (b), respectively.  We note that all the techniques with 13 transmitted plane-waves provide a sharper drop-off from speckle to cyst region compared to DAS with 75 transmissions. As in the simulated results, \mbox{iMAP2} outperforms ScW in terms of noise suppression within  the cyst, while maintaining a speckle pattern that is closer to DAS the output. This is especially prominent for lower depth. 

 The quantitative measurements of contrast and lateral resolution of images in Fig. \ref{fig:expContrast_multiplePW} are presented in Table \ref{table:measured cont and res}. The axial resolution is 0.58 mm for all the images.
 It can be seen that iMAP1 with 13 plane-waves is comparable to DAS based on 75 plane-waves in terms of both contrast and resolution. Similarity values for speckle regions are presented in Table \ref{table:speckle similarity}. The DAS image with 75 transmissions is used as reference.  

 \begin{table}[h!]
\caption{Measured lateral resolution and contrast}
\label{table:measured cont and res}
\begin{center}
\begin{tabular}{| c | c |c|}
\hline
\textbf{Method} &  \textbf{Lateral res. [mm]} & \textbf{CNR [dB]}\\
\hline
{DAS, 13 PW}                                 & {0.55}    &{3.7} \\ \hline
{DAS, 75 PW}                                 & {0.57}    &{4.7} \\ \hline
{Wiener Postfilter}       & { 0.41}   &{3} \\ \hline
{iMAP1}                          & {0.55}     &{4.8}\\ \hline
{iMAP2}                          & {0.55}     &{5.5}\\ \hline
{ScW}                               & { 0.51}     &{4.2}  \\ \hline
{CF}                                  &{0.39 }   &{3.7}\\ \hline
\end{tabular}
\end{center}
\end{table}

The experimental results are in close agreement with simulations. Both simulated and experimental acquisitions, therefore, are consistent with our expectation and verify that iMAP2 provides improved contrast compared to ScW, while better preserving the speckle pattern.

%
\begin{figure*}[h!]
\begin{minipage}[b]{0.24\linewidth}
  \centering
  \centerline{\includegraphics[width=5cm]{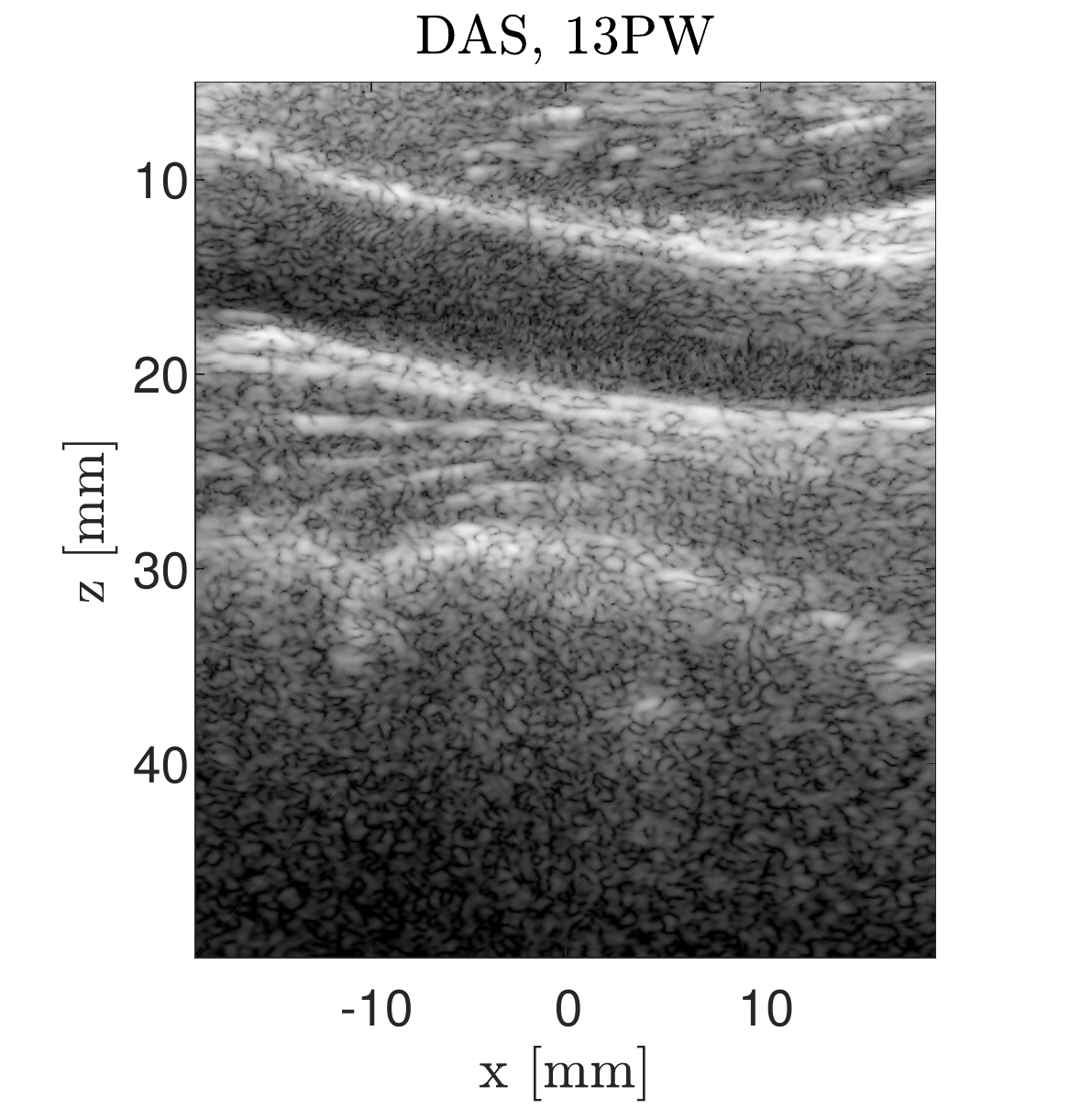}}%
  \centerline{(a)}\medskip
\end{minipage}
\begin{minipage}[b]{0.24\linewidth}
  \centering
  \centerline{\includegraphics[width=5cm]{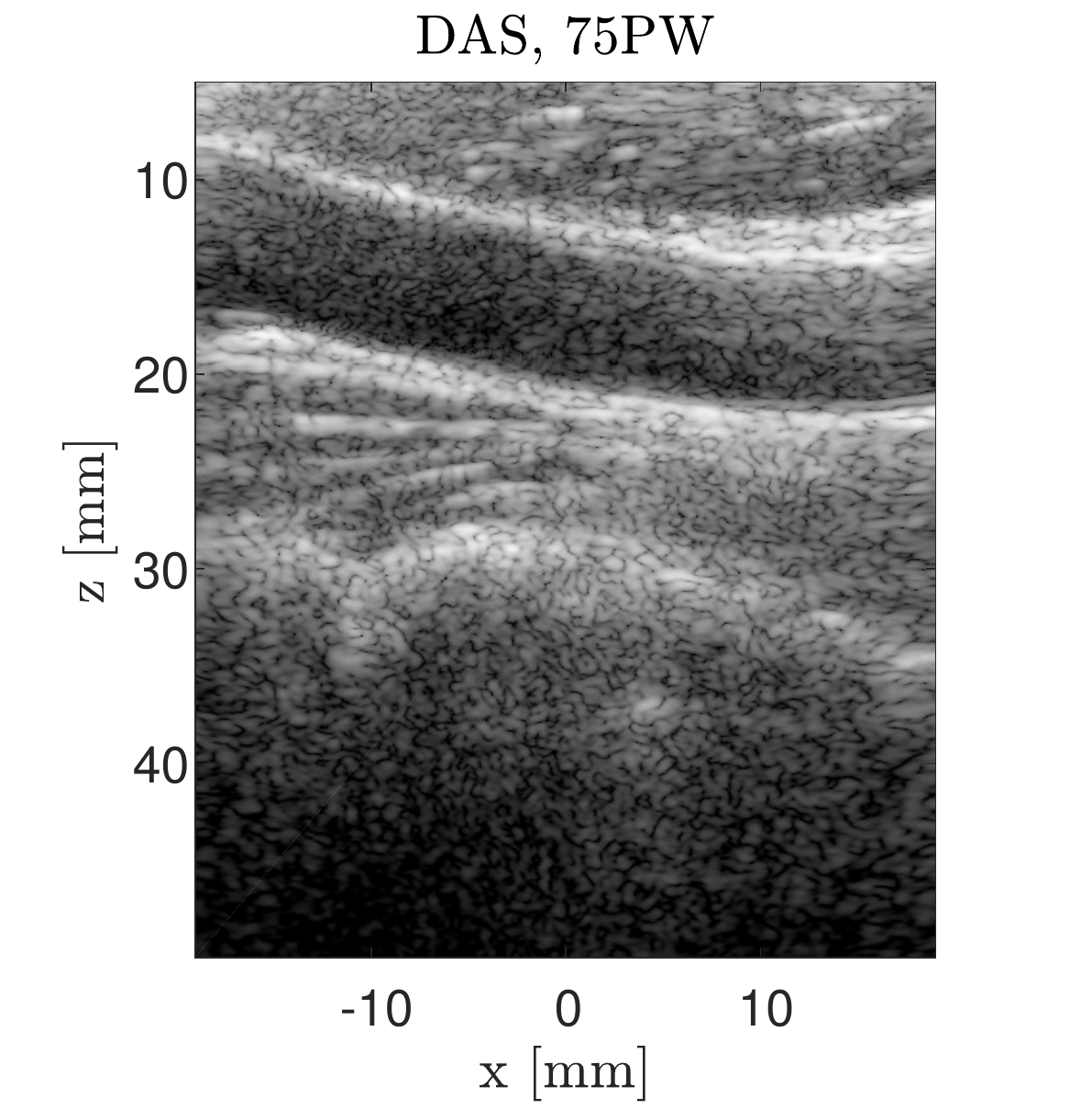}}%
  \centerline{(b)}\medskip
\end{minipage}
\begin{minipage}[b]{0.24\linewidth}
  \centering
  \centerline{\includegraphics[width=5cm]{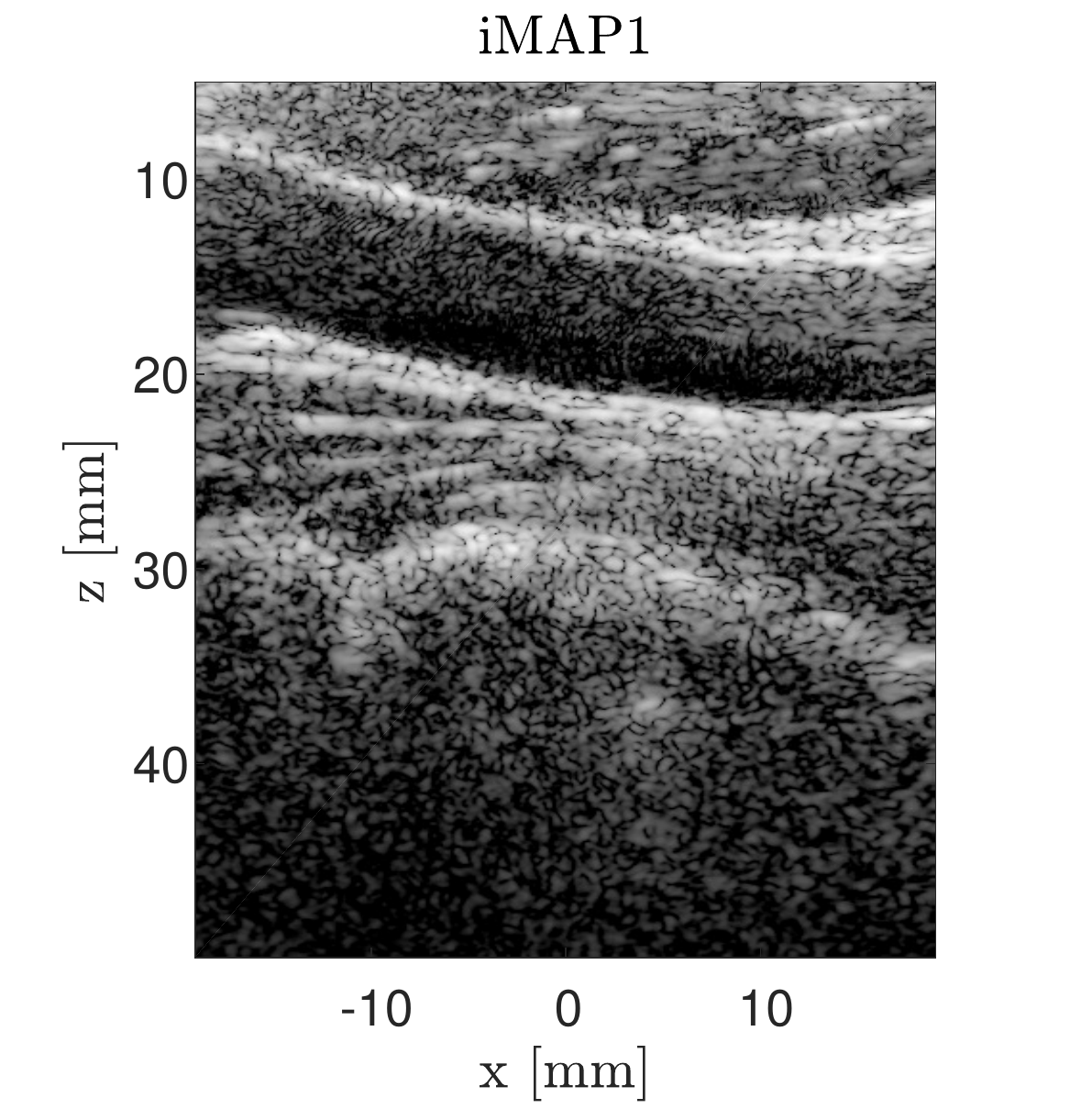}}%
  \centerline{(c)}\medskip
\end{minipage}
\begin{minipage}[b]{0.24\linewidth}
  \centering
  \centerline{\includegraphics[width=5cm]{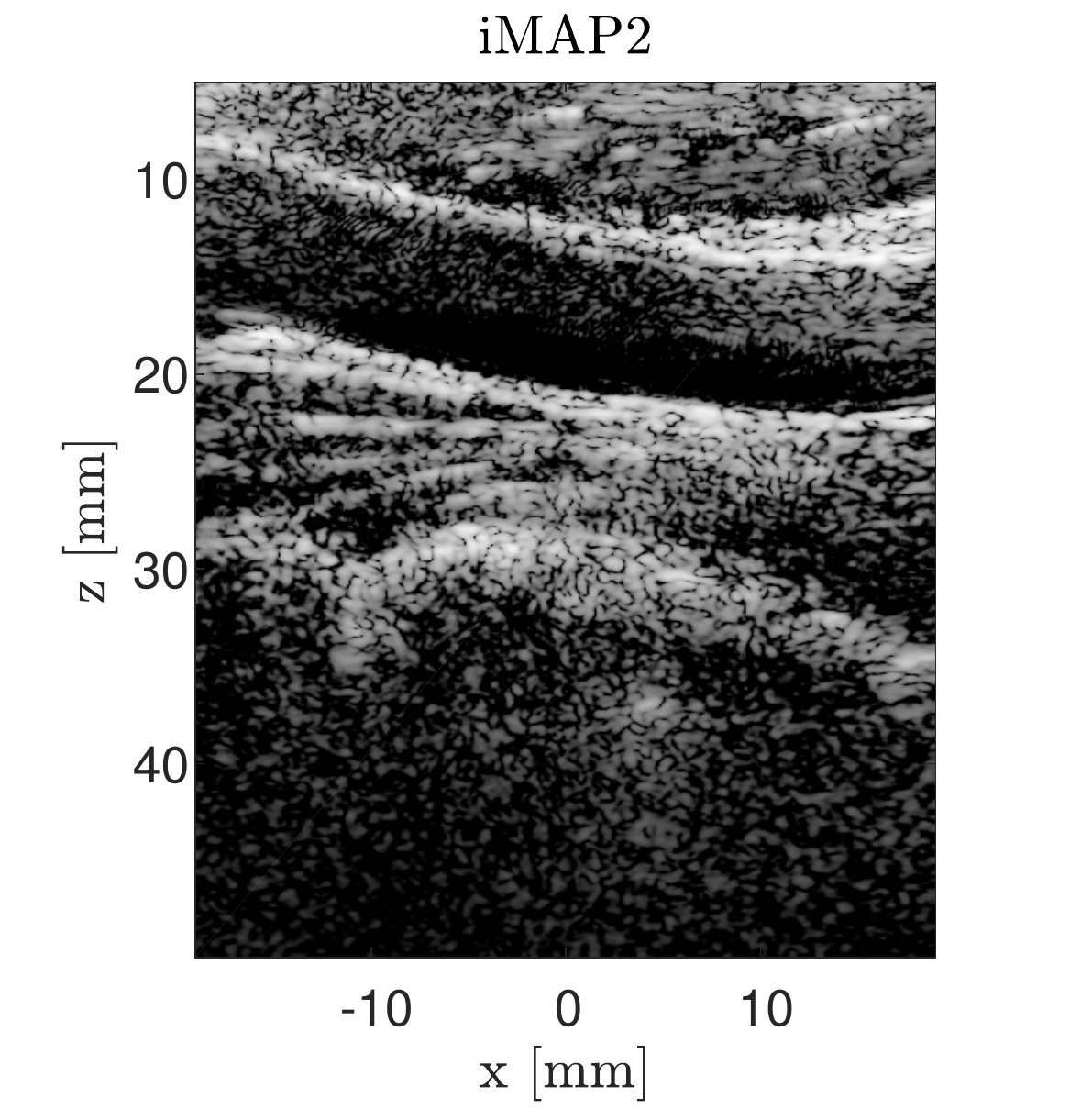}}%
  \centerline{(d)}\medskip
\end{minipage}\\
\begin{minipage}[b]{0.33\linewidth}
  \centering
  \centerline{\includegraphics[width=5cm]{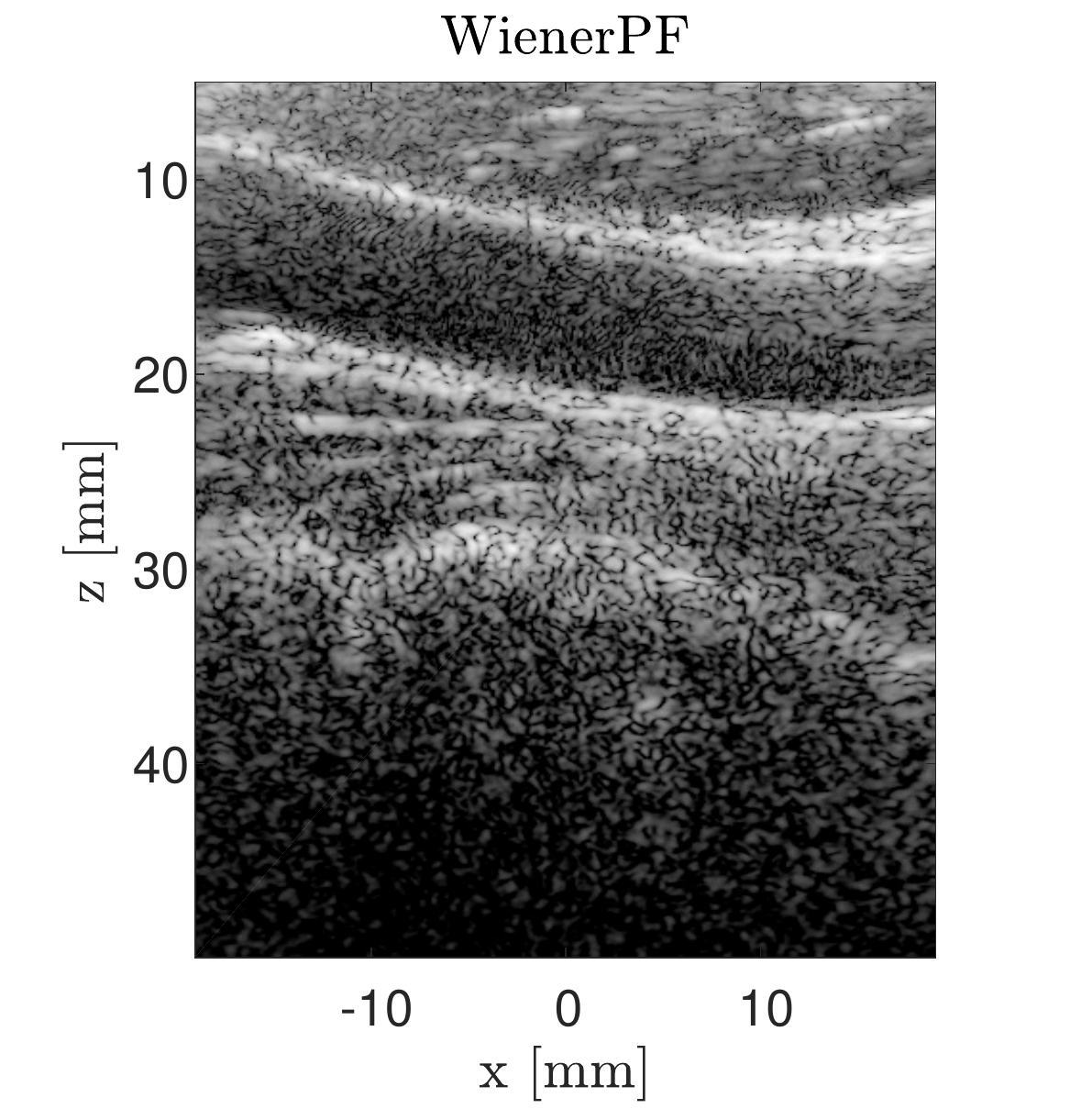}}%
  \centerline{(e)}\medskip
\end{minipage}
\begin{minipage}[b]{0.33\linewidth}
  \centering
  \centerline{\includegraphics[width=5cm]{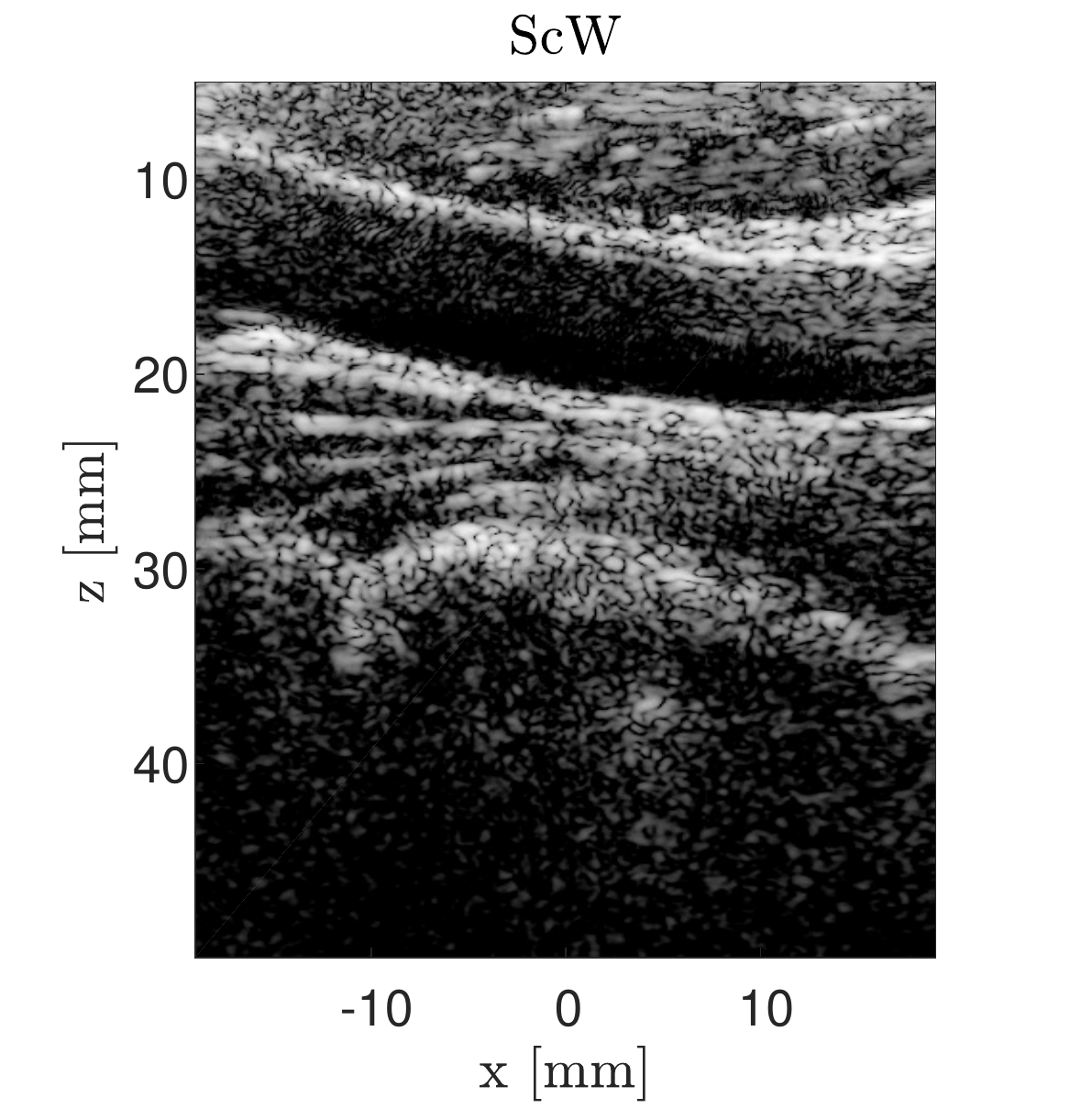}}%
  \centerline{(f)}\medskip
\end{minipage}
\begin{minipage}[b]{0.33\linewidth}
  \centering
  \centerline{\includegraphics[width=5cm]{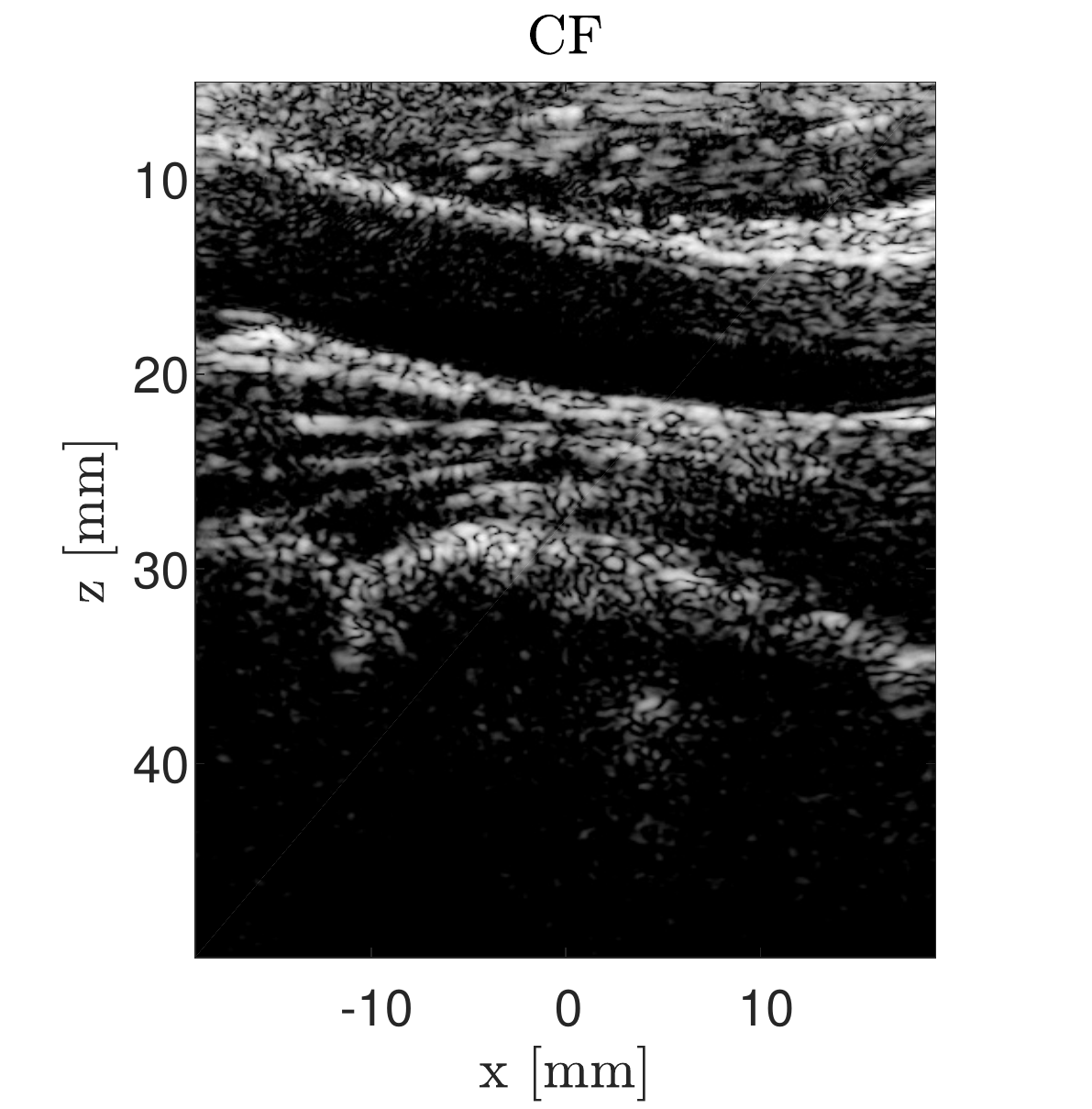}}%
  \centerline{(g)}\medskip
\end{minipage}
\caption{Images of  a longitudinal scan of a carotid artery obtained by (a) DAS with 13 plane-waves, (b) DAS with 75 plane-waves, (c) iMAP1 with 13 plane-waves, (d) iMAP2 with 13 plane-waves, (e) Wiener postfilter with 13 plane-waves, (f) ScW with 13 plane-waves, (g) CF with 13 plane-waves. All the images are presented with a dynamic range of 70 dB.}
\label{fig:carLong}
\end{figure*}

\begin{figure*}[h]
\begin{minipage}[b]{0.48\linewidth}
  \centering
  \centerline{\includegraphics[width=8cm]{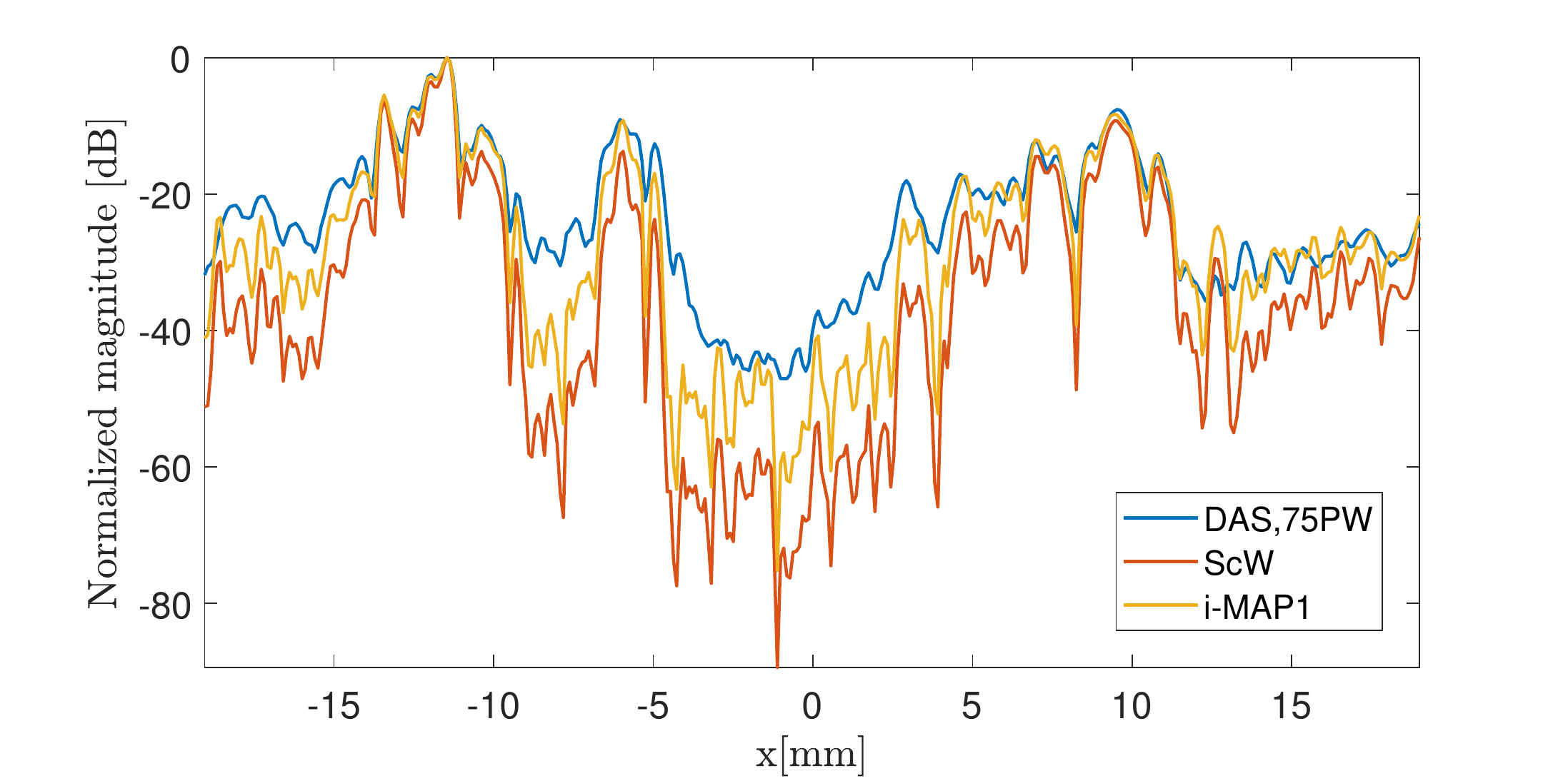}}%
  \centerline{(a)}\medskip
\end{minipage}
\begin{minipage}[b]{0.48\linewidth}
  \centering
  \centerline{\includegraphics[width=8cm]{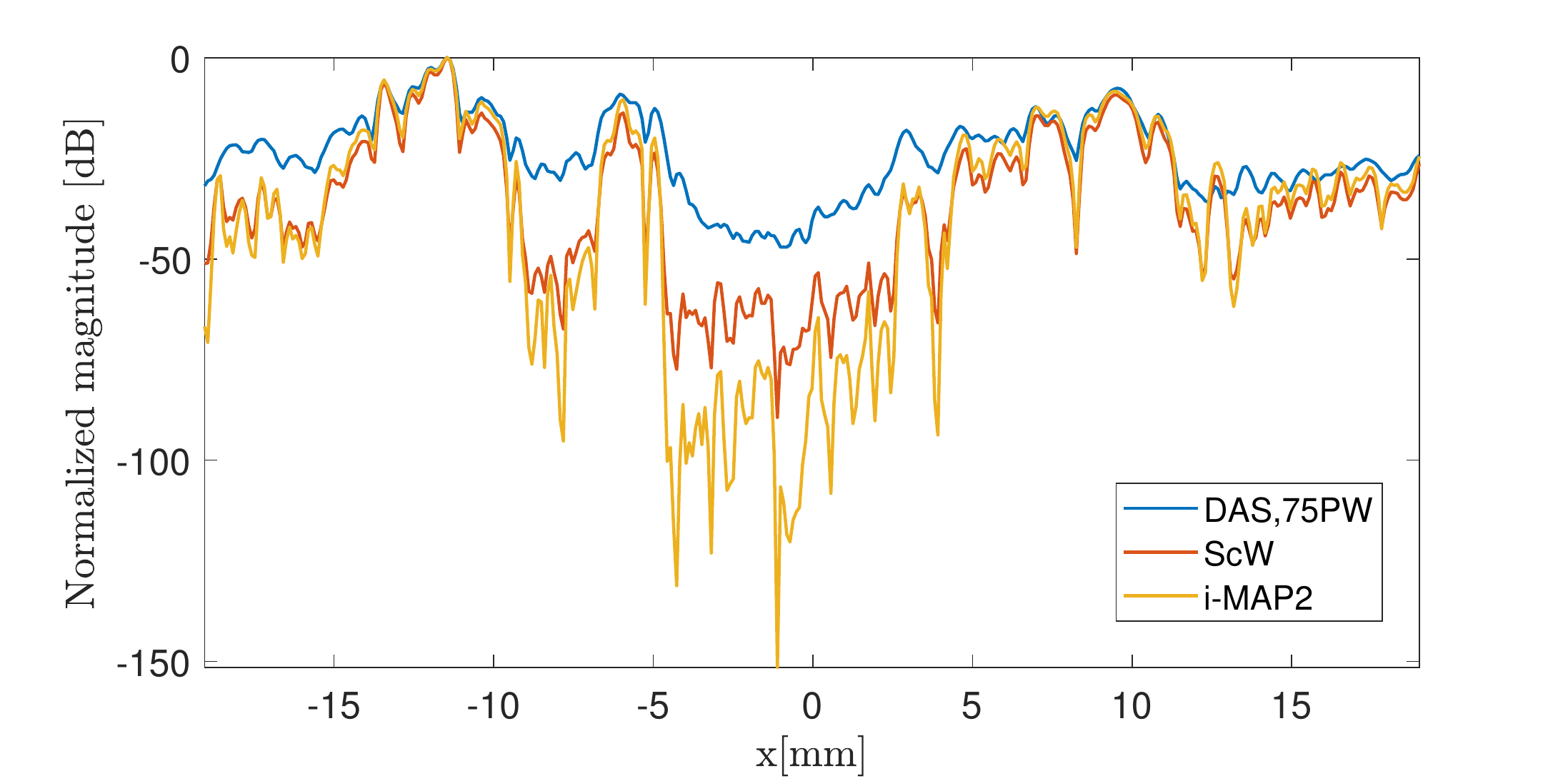}}%
  \centerline{(b)}\medskip
\end{minipage}
\caption{Lateral cross-section of the carotid from Fig. \ref{fig:carCross}, (a)  iMAP1 and ScW, (b) iMAP2 and ScW. iMAP1, iMAP2 and ScW are using 13 transmitted plane-waves. DAS is performed with 75 transmissions.}
\label{fig:carLatScanline}
\end{figure*}
%
\subsection{\textit{In-Vivo} Acquisition}
\label{ssec:in-vivo}

Finally, we apply the proposed beamformer to two \textit{in-vivo} datasets of a carotid artery. Based on the results of  Section~\ref{ssec:experiments}, we compare DAS obtained with 13 and 75 transmissions to all the other methods. Figures \ref{fig:carCross} and \ref{fig:carLong} present the cross section and the longitudinal scan of a carotid artery. The improvement of contrast is especially prominent in Fig.~\ref{fig:carCross}, where the cross section of the carotid artery is much cleaner with very sharp edge for iMAP1, iMAP2 and ScW. To demonstrate the difference in performance of the above three techniques, Fig. \ref{fig:carLatScanline} presents a lateral cross-section of the carotid from Fig. \ref{fig:carCross} at depth 1.8 cm. We note that all three methods with only 13 transmissions provide a sharper drop-off at the translation to the carotid cavity compared to DAS with 75 plane-waves. In terms of clutter suppression within the artery, iMAP2 outperforms ScW, however they both affect the speckle pattern. iMAP1 with 13 transmissions provides interference suppression comparable to DAS with 75 plane-waves with much less effect on speckle regions compared to iMAP2 and ScW.    

\section{Discussion and Conclusions}
\label{sec:discussion}
In this work we proposed a statistical interpretation of the beamforming process. We treat the signal of interest and the interference as uncorrelated Gaussian random variables and view beamforming as MAP estimation of the signal of interest.  
Implementation of MAP requires knowledge of the signal and interference variances. We propose using the current estimator of the signal of interest to improve the estimation of the distribution parameters and vice versa, leading to an iterative implementation of the MAP beamformer.
The resulting iterative scheme is simple computationally and does not require fine tuning of parameters.

The proposed method yields significant improvement of contrast compared to DAS as demonstrated using both simulated and experimental data. Only 13 plane-waves are required for iMAP1 to obtain the contrast of DAS with 75 transmissions. The second iteration, iMAP2, provides further improvement of contrast and leads to 80-100 dB noise suppression inside the cyst regions.
The comparison to Wiener postfilter, CF and ScW with signal model and parameter estimators proposed in previous works shows the superior performance of iMAP in terms of contrast and better preserved speckle patterns. 

In this work we proposed and implemented an algorithm based on a simple, yet, reasonable model. Our approach shows the ability to suppress interference without significantly increasing  complexity. Obviously, studying more involved models is of interest, since this can potentially further improve performance. We note, however, that the comparison of Wiener postfilter and iMAP performance points to a trade-off between the correctness of the model and the ability to estimate its parameters. We can consider iMAP and the Wiener postfilter as two extreme cases: iMAP is based on a simple model that requires estimation of only two parameters, the variances of the signal of interest and the noise. In contrast, Wiener postfilter does not make any assumptions on the noise distribution. As a result, the model is much more general, but it requires to estimate the entire noise covariance matrix. Empirical results presented in the paper show that iMAP outperforms Wiener postfilter in terms of contrast. This can be explained by poor estimation of the noise covariance matrix. 
A detailed study of more involved models and their effect on parameter estimation is left to future work.

In addition to quantitative measurements, based on simulated and experimental phantoms, two \textit{in-vivo} scans allow for visual assessment of the resulting image quality.
The  \textit{in-vivo} results are consistent with simulations and phantom scans in terms of contrast improvement. However, one can note that the background brightness is reduced in the vicinity of hyperechoic regions as appears in Figs. \ref{fig:carCross} (c-d) and \mbox{(f-g)} above the carotid cross-section on the right side of the hyperechoic structure. This artifact is seen in images obtained by iMAP, ScW and CF. All the above exploit the assumption of spatially white interference. This is not the case in a hyperechoic region that contaminates the signal detected by the elements in a highly correlated manner. To cope with this problem, the correlation between detected signals  should be taken into account leading to improved estimation of signal and noise variances.


To conclude, iMAP provides prominent improvement of contrast without affecting the resolution and results in a speckle patten comparable to that of DAS. Consequently, for plane-wave mode, fewer transmissions are required, maintaining high frame rate and reducing the amount of computations.
%

\bibliographystyle{IEEEtran}
\bibliography{IEEEfull,general}

\end{document}